\documentclass[final,11pt]{elsarticle}

\usepackage{amssymb}

\usepackage[margin=2.5cm]{geometry}
\usepackage{algorithm}
\usepackage{algpseudocode}

\newcounter{algsubstate}
\renewcommand{\thealgsubstate}{\alph{algsubstate}}
\newenvironment{algsubstates}
  {\setcounter{algsubstate}{0}%
   \renewcommand{\State}{%
     \stepcounter{algsubstate}%
     \Statex {\footnotesize\thealgsubstate:}\space}}
  {}

\usepackage{amsmath,bm}

\usepackage{physics}
\usepackage{subcaption,siunitx,booktabs}
\usepackage{graphicx,epstopdf}
\usepackage{graphics}
\usepackage{multirow}
\usepackage{fancyhdr, ifpdf}
\usepackage{amsxtra}
\usepackage{stmaryrd}
\usepackage{mathrsfs}
\usepackage{psfrag}
\usepackage{epsfig}
\usepackage{rotating}
\usepackage{setspace}
\usepackage{float}
\usepackage{amsfonts}
\usepackage{amsbsy}
\usepackage{amscd}
\usepackage{amsthm}
\usepackage{color}
\usepackage{tabularx}
\usepackage{caption}
\usepackage{subcaption}
\usepackage{sidecap}
\usepackage{dcolumn}
\usepackage{esint}
\usepackage{mathalfa}
\usepackage{cprotect}
\usepackage{ulem}
\usepackage[explicit]{titlesec}

\newcommand{\del}{\nabla}
\newcommand{\bPsi}{\boldsymbol{\Psi}}
\newcommand{\bvPsi}{\boldsymbol{\varPsi}}

\newcommand{\btau}{\boldsymbol{\tau}}
\newcommand{\bb}{\boldsymbol{b}}

\newcommand{\bff}{\boldsymbol{f}}

\newcommand{\bk}{\boldsymbol{k}}

\newcommand{\bx}{\boldsymbol{\textbf{x}}}
\newcommand{\by}{\boldsymbol{\textbf{y}}}

\newcommand{\bE}{\boldsymbol{\textbf{E}}}

\newcommand{\bH}{\boldsymbol{\textbf{H}}}
\newcommand{\bI}{\boldsymbol{I}}

\newcommand{\epsilonbar}{\bar{\epsilon}}
\newcommand{\fbar}{\bar{f}}
\newcommand{\bM}{\boldsymbol{\textbf{M}}}
\newcommand{\bL}{\boldsymbol{\textbf{L}}}

\newcommand{\bR}{\boldsymbol{\textbf{R}}}

\newcommand{\bPsibar}{\bar{\boldsymbol{\Psi}}}

\newcommand{\varphibar}{\bar{\varphi}}

\newcommand{\psibar}{\bar{\psi}}
\newcommand{\psitilde}{\widetilde{\psi}}
\newcommand{\Psibar}{\bar{\boldsymbol{\Psi}}}
\newcommand{\rhobar}{\bar{\rho}}

\newcommand{\Vbar}{\bar{V}}

\newcommand{\intomega}{\int_{\varOmega}}

\newcommand{\fintd}{\fint \displaylimits}

\newcommand{\Omegaper}{\Omega_\text{p}}
\newcommand{\dx}{\,d\bx}
\newcommand{\dy}{\,d\by}

\newcommand{\dk}{\,d\bk}
\newcommand{\btH}{\boldsymbol{\widetilde{\textbf{H}}}}

\newcommand{\Rthree}{ {\mathbb{R}^{3}} }

\renewcommand{\arraystretch}{1.2}
\newcommand{\chiepsx}{\,\btau^{\varepsilon}(\bx)\,}

\newcommand{\chieps}{\,\btau^{\varepsilon}}
\newcommand{\bdir}{\boldsymbol{\Upsilon}}
\newcommand{\dir}{\Upsilon}
\newcommand{\derveps}{\frac{d}{d\varepsilon}}
\newcommand{\atzero}{|_{\varepsilon = 0}}
\newcommand{\atzerob}{\bigg|_{\varepsilon = 0}}

\newcommand{\vself}[1]{\bar{V}^{#1}_{\tilde{\delta}}}

\newcommand{\dirac}{\tilde{\delta}}
\setcitestyle{citesep={,}}

\definecolor{hellgruen}{rgb}{0.2,0.7,0.2}

\makeatletter
\newcommand*{\rom}[1]{\expandafter\@slowromancap\romannumeral #1@}
\makeatother

\titleformat{\paragraph}[runin]
  {\normalfont\normalsize\bfseries}{}{11pt}{\theparagraph\hspace*{1em}#1:}
\titleformat{name=\paragraph,numberless}[runin]
  {\normalfont\normalsize\bfseries}{}{11pt}{#1:}
\bibpunct{[}{]}{;}{n}{}{}

\newcounter{bla}

\journal{Computer Physics Communications}

\begin{document}

\begin{frontmatter}



\title{DFT-FE -- A massively parallel adaptive finite-element code for large-scale density functional theory calculations}


\author[a]{Phani Motamarri\fnref{fn1}}
\author[a]{Sambit Das\fnref{fn1}}
\author[b]{Shiva Rudraraju}
\author[a]{Krishnendu Ghosh}
\author[c]{Denis Davydov}
\author[a,d]{Vikram Gavini\corref{author}}
\fntext[fn1]{Phani Motamarri and Sambit Das contributed equally to this work.}
\cortext[author] {Corresponding author.\\\textit{E-mail address:} vikramg@umich.edu}
\address[a]{Department of Mechanical Engineering, University of Michigan, Ann Arbor, MI 48109, USA}
\address[b]{Department of Mechanical Engineering, University of Wisconsin, Madison, WI 53706, USA}
\address[c]{Chair of Applied Mechanics, Friedrich-Alexander-Universit\"{a}t Erlangen-N\"{u}rnberg, Erlangen 91058, Germany}
\address[d]{Department of Materials Science \& Engineering, University of Michigan, Ann Arbor, MI 48109, USA}

\begin{abstract}
We present an accurate, efficient and massively parallel finite-element code, DFT-FE, for large-scale \textit{ab-initio} calculations (reaching $\sim 100,000$ electrons) using Kohn-Sham density functional theory (DFT).  DFT-FE is based on a local real-space variational formulation of the Kohn-Sham DFT energy functional that is discretized using a higher-order adaptive spectral finite-element (FE) basis, and treats pseudopotential and all-electron calculations in the same framework, while accommodating non-periodic, semi-periodic and periodic boundary conditions. We discuss the main aspects of the code, which include, the various strategies of adaptive FE basis generation, and the different approaches employed in the numerical implementation of the solution of the discrete Kohn-Sham problem that are focused on significantly reducing the floating point operations, communication costs and latency.  We demonstrate the accuracy of DFT-FE by comparing the energies, ionic forces and periodic cell stresses on a wide range of problems with popularly used DFT codes. Further, we demonstrate that DFT-FE significantly outperforms widely used plane-wave codes---both in CPU-times and wall-times, and on both non-periodic and periodic systems---at systems sizes beyond a few thousand electrons, with over $5-10$ fold speedups in systems with more than 10,000 electrons. The benchmark studies also highlight the excellent parallel scalability of DFT-FE, with strong scaling demonstrated on up to 192,000 MPI tasks.
\end{abstract}
 
\begin{keyword}
Electronic structure, real-space, spectral finite-elements, ionic forces, mixed-precision arithmetic, pseudopotential, all-electron, ab-initio molecular dynamics, band structure.
\end{keyword}

\end{frontmatter}
 


{\bf{Program summary}}

\begin{small}
\noindent
{\em Program Title:}        DFT-FE (https://github.com/dftfeDevelopers/dftfe)                           \\
{\em Journal Reference:}                                      \\
{\em Catalogue identifier:}                                   \\
{\em Licensing provisions:}  LGPL v3                          \\
{\em Programming language:}  C/C++                                 \\
{\em Computer:} Any system with C/C++ compiler and MPI library.     \\
{\em Operating system:}  Linux                                      \\
{\em RAM:} Ranges from few GBs for small problem sizes to around 50,000 GB for a system with 61,502 electrons.                                              \\
{\em Number of processors used:} Range from 64 to 192,000 MPI tasks (demonstrated).   \\
{\em Keywords:} Electronic structure, Real-space, Large-scale, Spectral finite-elements, Pseudopotential, All-electron.  \\
{\em Classification:}  0.7                                       \\
{\em External routines/libraries:} p4est (http://www.p4est.org/), deal.II (https://www.dealii.org/), \\ BLAS (http://www.netlib.org/blas/), LAPACK (http://www.netlib.org/lapack/), \\ELPA (https://elpa.mpcdf.mpg.de/), ScaLAPACK (http://www.netlib.org/scalapack/), \\Spglib (https://atztogo.github.io/spglib/), ALGLIB (http://www.alglib.net/), \\LIBXC (http://www.tddft.org/programs/libxc/), PETSc (https://www.mcs.anl.gov/petsc), \\ SLEPc (http://slepc.upv.es). \\ \\
{\em Nature of problem:} Density functional theory calculations.\\
   \\
{\em Solution method:} We employ a local real-space variational formulation of Kohn-Sham density functional theory that is applicable for both pseudopotential and all-electron calculations with arbitrary boundary conditions. Higher-order adaptive spectral finite-element basis is used to discretize the Kohn-Sham equations. Chebyshev polynomial filtered subspace iteration procedure (ChFSI) is employed to solve the nonlinear Kohn-Sham eigenvalue problem self-consistently. ChFSI in DFT-FE employs Cholesky factorization based orthonormalization, and spectrum splitting based Rayleigh-Ritz procedure in conjunction with mixed precision arithmetic. Configurational force approach is used to compute ionic forces and periodic cell stresses for geometry optimization.
 \\
   \\
{\em Restrictions:} Exchange correlation functionals are restricted to Local Density Approximation (LDA) and Generalized Gradient Approximation (GGA), with and without spin. The pseudopotentials available are optimized norm conserving Vanderbilt (ONCV) pseudopotentials and Troullier--Martins (TM) pseudopotentials. Calculations are non-relativistic.\\
   \\
{\em Unusual features:} DFT-FE handles all-electron and pseudopotential calculations in the same  framework, while  accommodating  periodic,  non-periodic  and  semi-periodic  boundary conditions.\\
   \\
{\em Running time:} This is dependent on problem type and computational resources used. Timing results for benchmark systems are provided in the paper.\\
   \\


\end{small}

\clearpage
\section{Introduction}
\label{sec:intro}
Quantum mechanical calculations based on Kohn-Sham density functional theory (DFT) occupy a substantial fraction of world's computational resources today, and have provided many important insights into a wide range of materials properties over the past decade. The Kohn-Sham approach~\cite{kohn65,kohn96} to DFT
provides an efficient formulation to compute the ground-state properties of materials systems by reducing the many-body Schr\"odinger problem of interacting electrons into an equivalent problem of non-interacting electrons in an effective mean field that is governed by the electron-density. While this effective single-electron formulation has no approximations, and is exact, in principle, the quantum mechanical interactions between electrons manifest in the form of an unknown exchange-correlation functional, which is modeled in practice. While improving the accuracy of the exchange-correlation functional is still an active area of research, the widely used models for the exchange-correlation functional~\cite{XCReview2005} have been shown to predict a range of materials properties across various materials systems with good accuracy.



Large-scale DFT calculations are crucial to improve the predictive capability of modeling materials systems, and enable computation based design of new materials in a variety of application areas. For example, accurate determination of dislocation core properties~\cite{rodney2017,Arias2000,Trinkle2005,Trinkle2008,Clouet2009,shin2011,shin2013,iyer2015,radhakrishnan2016,das2017} in metals and semiconductors, understanding ion conduction mechanisms and computing diffusivities in solid-state electrolytes~\cite{dawson2018,dive2018}, and large-scale bio-molecular simulations~\cite{cole2016}, all require the ability to conduct accurate and computationally efficient DFT calculations involving many thousands of atoms on both metallic and insulating systems. However, the solution to the governing equations in DFT demands significant computational resources, and accurate DFT calculations are routinely limited to materials systems with at most a few thousands of electrons restricting the system sizes to a few hundred atoms. This computational complexity is a serious bottleneck in the context of ab-initio molecular dynamics with long time-scales and atomic relaxations that require a large number of relaxation steps.

Traditionally, the widely used DFT codes employ either plane-waves~\cite{qe2009,gonze2002,VASP,exciting2014} or atomic-orbital type basis sets~\cite{Pople,jensen2002,cp2k2014,blum2009,nwchem} for DFT calculations. However, the use of plane-wave basis restricts simulation domains to be periodic. Further, these basis sets do not exhibit good parallel scalability, severely limiting the range of materials systems that can be studied. On the other hand, atomic orbital type basis sets are not systematically convergent for generic materials systems. Thus, to overcome the above limitations, there has been an increasing thrust in the development of systematically improvable and scalable real-space discretization techniques like finite-elements~\cite{tsuchida1995,tsuchida1996,tsuchida1998,pask1999,pask2005,sukumar2009,suryanarayana2010,zhou2011,motamarri2013,SCHAUER2013644,zhou2014,denis2016,kanungo2017,Davydov2018}, finite-difference~\cite{parsec2006,rescu2016,sparc2017a,sparc2017b,octopus2015,gpaw}, wavelets~\cite{luigi2008}, psinc functions~\cite{skylaris2005}, and other reduced order basis techniques ~\cite{dgdft2015,motamarri2015}.

We note that the traditional self-consistent approaches to solving the discretized nonlinear Kohn-Sham eigenvalue problem involves the diagonalization of the Kohn-Sham Hamiltonian to obtain the orthonormal eigenvectors, the computational complexity of which typically scales cubically with number of electrons. Hence, the computational cost of a Kohn-Sham DFT calculations becomes prohibitively expensive as the system size grows larger, and is another limiting factor in realizing large-scale DFT calculations. To circumvent this limitation, numerous efforts have focused on reducing the computational complexity by improving the system-size scaling of DFT calculations. These methods~\cite{godecker99,bowler2012,skylaris2005,fattebert2006,ls3df2008} rely on the exponential decay of the density matrix in real space, and the computational cost has been demonstrated  to linearly scale with number of atoms for systems with non-vanishing band gaps. Though reduced scaling approaches for metallic systems at moderate electronic temperatures have been developed~\cite{motamarri2014,skylaris2018,mohr2018}, they usually have a high computational prefactor. There are also other efforts in the literature which have focused on reducing the prefactor associated with the cubic computational complexity of the Kohn-Sham DFT calculations~\cite{dgdft2015,saad2006,saad2012} that are suited for both insulating and metallic systems. 

In this work, we focus on real-space adaptive spectral finite-element (FE) discretization of Kohn-Sham DFT which affords excellent parallel scalability. We employ Chebyshev filtering approach ~\cite{zhou2006} in conjunction with Cholesky factorization based orthonormalization procedure and spectrum splitting based Rayleigh-Ritz technique, along with mixed precision strategies, to reduce the computational prefactor of the cubic scaling steps.  These strategies enabled systematically convergent, computationally efficient and massively parallel DFT calculations (demonstrated up to $192,000$ MPI tasks) on material systems with tens of thousands of electrons for both metallic and insulating systems. The choice of FE discretization among other real space discretizations for DFT in this work is motivated by some key advantages it offers for electronic structure calculations. In particular, the FE basis naturally allows for arbitrary boundary conditions, provides good scalability on parallel computing platforms due to locality of the basis, and is amenable to adaptive spatial resolution, which can effectively be exploited for efficient solution of all electron DFT calculations as well as the development of coarse-graining techniques that seamlessly bridge electronic structure calculations with continuum. 

We present here DFT-FE, a massively parallel adaptive finite-element code for large-scale density functional theory calculations. The current endeavour extends the previous work ~\cite{motamarri2013}, which has demonstrated the advantage of higher-order spectral finite-elements in significantly improving the computational efficiency of DFT calculations. DFT-FE is based on local real-space variational formulation of Kohn-Sham DFT~\cite{motamarri2013,motamarri2018}, which handles all-electron and pseudopotential calculations in the same framework while accommodating periodic, non-periodic and semi-periodic boundary conditions. This unified local variational formulation is briefly discussed and the usefulness of this formulation to compute configurational forces~\cite{motamarri2018} for geometry optimization is also highlighted. These configurational forces correspond to generalized variational forces computed as the derivative of the Kohn-Sham energy functional with respect to the position of a material point $\bx$. These generalized forces that result from the inner variations of the Kohn-Sham energy functional inherently account for the Pulay corrections, and provide a unified framework to compute ionic forces as well as stress tensor for geometry optimization. We then introduce the FE discretization of Kohn-Sham DFT, and the resulting discretized Kohn-Sham equations are cast into a standard eigenvalue problem with modest computational cost by using a spectral FE basis in conjunction with the Gauss-Lobatto-Legendre (GLL) quadrature rules ~\cite{motamarri2013}. We subsequently discuss various adaptive mesh refinement procedures implemented in DFT-FE. Firstly, we employ an user defined adaptive mesh refinement procedure, involving user inputs for mesh sizes at different regions of interest in the simulation domain, which in turn can be estimated from a mesh size distribution $h(\bx)$ obtained from a constrained minimization of the discrete ground-state energy error estimate. Further, we also present an automatic mesh refinement procedure to construct an \textit{a priori} adaptive mesh, by employing numerically constructed single atom Kohn-Sham DFT wavefunctions. This approach is based on local error indicators obtained from the energy error estimates involving the discrete electronic wavefunctions.

We employ Chebyshev filtered subspace iteration technique (ChFSI)~\cite{zhou2006,motamarri2013} to self-consistently solve the discretized Kohn-Sham DFT problem. ChFSI technique involves three key computational steps: (i) Chebyshev filtering to compute a subspace that closely approximates the relevant eigensubspace; (ii) orthonormalization procedure to construct an orthonormal basis spanning the Chebyshev filtered subspace; and (iii) Rayleigh-Ritz procedure that involves projecting the Kohn-Sham Hamiltonian into the Chebyshev filtered subspace, followed by diagonalization of the projected Hamiltonian to compute the associated eigenpairs and a subspace rotation step. The computational complexity of Chebyshev filtering scales quadratically with system size, and is the dominant cost for small- to medium-scale system sizes. However, for large-scale systems, the cubic-scaling computational cost of orthonormalization and Rayleigh-Ritz procedure dominates. To this end, the numerical implementation in DFT-FE focuses on reducing the computational prefactor by using efficient numerical strategies, which include: (i) optimized FE cell level matrix operations; (ii) Cholesky factorization based Gram-Schmidt orthonormalization; (iii) mixed precision arithmetic, and (iv) spectrum splitting based Rayleigh-Ritz procedure. We note that the use of these techniques delays the onset of cubic computational complexity in DFT-FE to very large system sizes. Using benchmark problems, we demonstrate that DFT-FE exhibits close to quadratic scaling in computational complexity till $30,000$--$40,000$ electrons. 


In our numerical implementation of DFT-FE, we use two levels of parallelization: (i) domain-decomposition based on partitioning of the adaptive FE mesh, and (ii) parallelization over wavefunctions (band parallelization). Our numerical investigations show that best parallel scaling efficiency is achieved using domain-decomposition combined with moderate band parallelization. We assess the parallel scaling performance of DFT-FE on a range of system sizes, and observe excellent scalability on all systems. Notably, we obtain a scaling efficiency of $\sim50\%$ on 102,400 MPI tasks for a system containing $\sim40,000$ electrons. We note that DFT-FE's massive parallel scalability is a result of the locality of the FE basis as well as an effective parallel implementation of the various algorithms that reduce communication costs and latency. 

In order to assess the accuracy, computational efficiency and scalability of DFT-FE, we conduct a comprehensive comparison study with widely used plane-wave codes---QUANTUM ESPRESSO (QE)~\cite{qe2009,qe2017} and ABINIT~\cite{gonze2002}---on various pseudopotential DFT benchmark problems, which include periodic bulk metallic systems with a defect and non-periodic nano-particles. In the validation studies, we compare DFT-FE ground-state energies, ionic forces and cell stresses against QE results at different accuracy levels. We also consider all-electron periodic and non-periodic benchmark systems for assessing the accuracy of DFT-FE with \verb|exciting|~\cite{exciting2014} and NWChem~\cite{nwchem} codes. Next, we assess the computational efficiency and scalability of pseudopotential DFT-FE calculations with respect to QE and ABINIT. The FE and plane-wave discretizations in these benchmark calculations are chosen to be commensurate with chemical accuracy ($\sim10^{-4}$ Ha/atom and $\sim10^{-4}$ Ha/Bohr in ground-state energies and forces). The CPU-times on these benchmark systems show that DFT-FE is computationally more efficient than QE, even for periodic systems, beyond system sizes containing $3,000$ electrons, and significantly more efficient for larger system sizes by $4.5-12\times$. Comparing the minimum wall times, we find DFT-FE to be significantly faster than QE for all the systems considered here, with speedups reaching up to $16\times$ for larger systems. Finally, we also demonstrate the capability of DFT-FE to simulate very large systems reaching up to $\sim 60,000$ electrons, which are computationally prohibitive using plane-wave codes. For the largest benchmark system containing $\sim 60,000$ electrons, which is solved using a FE discretization  commensurate with chemical accuracy, we scale up to $192,000$ MPI tasks at $42\%$ efficiency and consume an average of $4.6$ minutes per SCF iteration.

Finally, we demonstrate the following capabilities in DFT-FE: i) ionic relaxations using an organometallic complex as the benchmark system, which is validated using QE ii) NVE \textit{ab-initio} molecular dynamics simulations using a representative bulk-metallic system containing $108$ atoms, which demonstrates energy conservation to very stringent accuracy, and iii) band-structure calculations of bulk Si and Si supercell with a vacancy, which are validated using QE. The accuracy, efficiency and scalability of the DFT-FE code demonstrated in this work, presents this as a useful code to conduct systematically convergent large-scale DFT calculations in an efficient manner, which can enable studies on complex materials systems that have not been possible heretofore. We note that the present release of DFT-FE is only a first step in a longer term effort, which includes future planned activities of incorporating advanced exchange-correlation functionals, Hamiltonians with spin-orbit coupling, dielectric calculations, electron-phonon coupling, and others capabilities into DFT-FE.    


The remainder of the paper is organized as follows. Section~\ref{sec:dftformul} discusses the governing equations in DFT, describes the local real space formulation implemented in DFT-FE that treats pseudopotential and all-electron calculations in the single framework, followed by a brief discussion on configurational forces used to evaluate ionic forces and stresses for geometry optimization.  FE discretization of the Kohn-Sham DFT problem is subsequently introduced with a discussion on the discrete Kohn-Sham equations. Section~\ref{sec:numImpl} describes the various aspects of numerical implementation in DFT-FE, which include adaptive mesh refinement strategies implemented in DFT-FE, and numerical implementation of the various algorithms employed in DFT-FE for the solution of the Kohn-Sham problem. Section~\ref{sec:results} presents the results on various benchmark systems, demonstrating the accuracy, computational efficiency and parallel scaling performance of the developed DFT-FE code in comparison with other widely used DFT codes. Section~\ref{sec:dftfeCapabilities} highlights other capabilities of the DFT-FE code, and we conclude with an outlook in Section~\ref{sec:conclusions}.

\section{Real space Kohn-Sham DFT formulation}
\label{sec:dftformul}
\subsection{Governing equations in DFT}
We consider a materials system with $N_e$ electrons and $N_a$ atoms whose position vectors are denoted by $\bR = \{\bR_1,\,\bR_2,\,\cdots \bR_{N_a}\}$. Neglecting spin, the variational problem  of evaluating the ground-state properties in density functional theory is equivalent to solving the $N$ lowest eigenvalues of the following non-linear eigenvalue problem~\cite{kohn65}:
\begin{equation}\label{contEigen}
\left(-\frac{1}{2} \del^{2} + V_{\text{eff}}(\rho,\bR)\right) {\psi}_{i} = \epsilon_{i} {\psi}_{i},\;\;\;\;\; i = 1,2,\cdots N\;\;\;\text{with}\;\;\;N > \frac{N_e}{2}\,,
\end{equation}
where
$\epsilon_{i}$ and $\psi_{i}$ denote the eigenvalues and corresponding eigenfunctions (also referred to as the canonical wavefunctions) of the Hamiltonian, respectively.  For clarity and notational convenience, the case of spin independent Hamiltonian is discussed here. However, extension to spin-dependent Hamiltonians~\cite{rmartin} is straightforward, and incorporated in DFT-FE. The electron density $\rho$ in equation ~\eqref{contEigen} can be expressed in terms of the orbital occupancy function  $f(\epsilon,\mu)$  and the canonical wavefunctions as
\begin{equation}\label{eq:elecdens}
\rho(\bx) = 2\sum_{i=1}^{N} f(\epsilon_i,\mu)|\psi_i(\bx)|^2 \,\,.
\end{equation}
The range of $f(\epsilon_i,\mu)$ lies in the interval $\left[0,1\right]$, and $\mu$ represents the Fermi-energy. In material systems with large number of eigenstates around the Fermi energy, the numerical instabilities  that may arise in the solution of the non-linear Kohn-Sham eigenvalue problem are avoided by using a smooth orbital occupancy function. In DFT-FE, $f$ is represented by the Fermi-Dirac distribution~\cite{VASP,godecker99} given by
\begin{equation}\label{fermidirac1}
f(\epsilon,\mu) = \frac{1}{1 + \exp\left(\frac{\epsilon - \mu}{\sigma} \right)}\,.
\end{equation}
In the above, $\sigma = k_{B} T$ denotes the regularization parameter with $k_B$ denoting the Boltzmann constant and $T$ representing the finite temperature. We note that as $\sigma\to 0$, the Fermi-Dirac distribution  tends to the Heaviside function. The constraint on the total number of electrons in the system ($N_e$) determines the Fermi-energy $\mu$, and is given by
\begin{equation}\label{cons}
\int \rho(\bx) \dx = 2\sum_i f(\epsilon_i,\mu) = N_e\,.
\end{equation}
We note that $f(\epsilon_i,\mu)$ is denoted as $f_i$ in the remainder of the manuscript. The effective single-electron potential, $V_{\text{eff}}(\rho,\bR)$, in the Hamiltonian in equation~\eqref{contEigen} is given by
\begin{equation}\label{veff}
 V_{\text{eff}}(\rho,\bR) = V_{\text{xc}}(\rho) + V_{\text{el}}(\rho,\bR) = \frac{\delta E_{\text{xc}}}{\delta \rho} + \frac{\delta E_{\text{el}}}{\delta \rho}
\end{equation}
In the above, $V_{\text{xc}}(\rho)$ denotes the exchange-correlation potential that accounts for quantum-mechanical interactions between electrons~\cite{XCReview2005}, and is given by the first variational derivative of the exchange-correlation energy $E_{\text{xc}}$. We adopt the generalized gradient approximation (GGA)~\cite{rmartin,gga1} for the exchange correlation functional description throughout the manuscript. However other forms of functionals involving local density (LDA, LSDA) are also incorporated in DFT-FE. In the case of GGA, the exchange-correlation energy is given by
\begin{equation}\label{exc}
E_{\text{xc}}(\rho) =\int \epsilon_{\text{xc}} (\rho,\del \rho) \rho(\bx) \dx.
\end{equation}
Numerous forms for $\epsilon_{\text{xc}} (\rho,\del \rho)$ have been proposed, and the three widely used forms are Becke (B88) ~\cite{becke}, Perdew and Wang (PW91) ~\cite{pw} and Perdew, Burke and Enzerhof (PBE) ~\cite{pbe}. 

The term $V_{\text{el}}(\rho)$, in the effective single-electron potential (equation~\eqref{veff}), accounts for the electrostatic interactions. In particular, it is the variational derivative of the classical electrostatic interaction energy between electrons and nuclei, $E_{\text{el}}$, which can further be decomposed as
\begin{equation}
E_{\text{el}}(\rho,\bR) = E_{\text{H}}(\rho) + E_{\text{ext}}(\rho,\bR) + E_{\text{zz}}(\bR)\,.
\end{equation}
In the above, $E_{\text{H}}$, $E_{\text{ext}}$ and $E_{\text{zz}}$ denote the electrostatic interaction energy between electrons (Hartree energy), interaction energy between nuclei and electrons, and repulsive energy between nuclei, respectively. These are given by
\begin{equation}\label{ElecEngy}
E_{\text{H}} = \frac{1}{2}\int\int\frac{\rho(\bx)\rho(\by)}{|\bx - \by|} \,\dx\,\dy ,\;\; E_{\text{ext}} = - \sum_{J}\int \rho(\bx)  \frac{Z_J}{|\bx-\bR_J|}\dx,\;\;E_{\text{zz}}= \frac{1}{2}\sum_{\substack{I,J \neq I}} \frac{Z_I Z_J}{|\bR_I-\bR_J|},\;\;
\end{equation}
where $Z_I$ denotes the charge on the $I^{th}$ nucleus. In the case of non-periodic boundary conditions representing an isolated atomic system, all integrals in equations~\eqref{ElecEngy} are over $\Rthree$, and the summations include all the atoms in the system. In the case of periodic boundary conditions representing an infinite periodic crystal, all integrals involving $\bx$ in equation~\eqref{ElecEngy} are over the periodic domain (supercell), whereas the integrals involving $\by$ are over $\Rthree$. Further, the summation over $I$ is on atoms in the periodic domain, and the summation over $J$ extends over all the lattice sites. Henceforth, unless otherwise specified, we will adopt this convention. Next, we define the nuclear charge distribution $b(\bx,\bR) = - \sum_{I}Z_I\tilde{\delta}(|\bx - \bR_I|)$ with 
$\tilde{\delta}(\bx - \bR_I)$ denoting a regularized Dirac distribution centered at $\bR_I$ (and similarly $b(\by,\bR) = - \sum_{J}Z_J\tilde{\delta}(|\by - \bR_J|)$) to reformulate the repulsive energy  $E_{\text{zz}}(\bR)$ as
\begin{equation}\label{repulsive}
\begin{split}
E_{zz} &= \frac{1}{2}\int \int \frac{b(\bx,\bR)\,b(\by,\bR)}{|\bx - \by|}\dx\dy\, - E_{\text{self}} \\&= \frac{1}{2}\int \int\left( \frac{b(\bx,\bR)\,b(\by,\bR)}{|\bx - \by|} - \sum_{I} \frac{Z_I^2\tilde{\delta}(|\bx-\bR_I|)\tilde{\delta}(|\by-\bR_I|)}{|\bx-\by|}\right) \dx\dy\,,
\end{split}
\end{equation}
where $E_{\text{self}}$ denotes the self energy of the nuclear charges which depends only on the nuclear charge distribution.

The tightly bound core electrons close to the nucleus of an atom do not influence chemical bonding in many materials systems, and, thus, may not play a significant role in governing many materials properties. Hence, it is a common practice to adopt the pseudopotential approach, where valence electronic wavefunctions are computed in an effective potential generated by the the nucleus and the core electrons. The pseudopotential is often defined by the  operator $\mathcal{V}_{\text{PS}} = \mathcal{V}_{\text{loc}} + \mathcal{V}_{\text{nl}}$ = $\sum_J (\mathcal{V}_{\text{loc}}^J + \mathcal{V}_{\text{nl}}^J)$, where $\mathcal{V}_{\text{loc}}^J$ and $\mathcal{V}_{\text{nl}}^J$ denote the local and non-local part of the pseudopotential operator for an atom $J$, respectively. Further, in the case of norm-conserving pseudopotentials, $\mathcal{V}_{\text{nl}}^J$ can be constructed as a separable pseudopotential operator ~\cite{rmartin,kb82} of the form $\sum_{lpm}\ket{\chi_{lpm}}h_{lp}\bra{\chi_{lpm}}$\,, with $\ket{\chi_{lpm}}$ denoting the pseudopotential projector. Here $l$ denotes the azimuthal quantum number, $p$ denotes the index corresponding to the projector component for a given $l$ while $m$ denotes the magnetic quantum number with $h_{lp}$ denoting the pseudopotential constant. Using the representation of the operator $\mathcal{V}_{PS}$ in the $\bx$ basis, $E_{\text{ext}}$ in pseudopotential Kohn-Sham DFT can be expressed as
\begin{equation}\label{eq:Eext_PS}
E_{\text{ext}}= 2 \sum_{i=1}^{N} \int \int f_i \,\psi^{*}_i(\bx)\, V_{\text{PS}}(\bx,\by,\bR)\, \psi_i(\by) \dy \dx\,.
\end{equation}
Norm conserving pseudopotentials are employed in DFT-FE, where the action of the nonlocal psuedopotential operator on a wavefunction is given by 
\begin{align}
V_{\text{nl}}\,\psi_i := \int V_{\text{nl}}(\bx,\by,\bR)\psi_i(\by)\dy 
&= \sum_{J} \sum_{lp} \sum_{m} C^{J,i}_{lpm}\,\,h^{J}_{lp}\,\,\chi_{lpm}^{J}(\bx,\bR_J)\,,\label{nlps}
\end{align}
\begin{equation}
\text{with}\;\;\;\;\;\;\;C^{J,i}_{lpm} = \int \chi_{lpm}^{J}(\bx,\bR_J)\psi_i(\bx)\dx, \;\;\;\;
\frac{1}{h^{J}_{lp}} = \ip{\xi_{lm}^J}{\chi_{lpm}^J} \,. \label{pspconst}
\end{equation}
In the above, $\ket{\xi_{lm}^J}$ denotes  the single atom pseudo-wavefunction.  Note that $h^J_{lp}$ does not depend on the magnetic quantum number $m$ as the spherical harmonics associated with angular variables in the inner product ~\eqref{pspconst} are normalized to unity. We remark that equation ~\eqref{nlps}  reduces to Troullier-Martins (TM) pseudopotential~\cite{tm91} in the Kleinman-Bylander form~\cite{kb82} for one projector component, i.e. $p=1$ for every $l$, while in the case of optimized norm conserving Vanderbilt pseudopotential (ONCV) ~\cite{oncv2013} there are two projector components ($p=1,\,2$) for every $l$. Both TM and ONCV norm-conserving pseudopotentials are implemented in DFT-FE. We further note that the accuracy of ONCV pseudopotentials are shown to be on par with PAW approaches widely employed in DFT codes~\cite{science2016}.

We note that the various components of the electrostatic interaction energy in ~\eqref{ElecEngy} and ~\eqref{repulsive} are non-local in real-space, and these extended interactions are reformulated as local variational problems as discussed in  ~\cite{gavini2007,motamarri2013}. To this end, we define the electrostatic potential corresponding to the $I^{th}$ nuclear charge $Z_I\tilde{\delta}(|\bx - \bR_I|)$ to be $\vself I(\bx)$ and the electrostatic potential corresponding to the total charge distribution $(\rho + b)$ to be $\varphi(\bx,\bR)$, and these potentials are given by:
\begin{equation}\label{pot}
\vself I(\bx) =\int \frac{-Z_I\tilde{\delta}(|\by-\bR_I|)}{|\bx-\by|} \dy,\;\;\;\; \varphi(\bx,\bR) = \int \frac{\rho(\by) + b(\by,\bR)}{|\bx - \by|} \dy \,. 
\end{equation}
Noting that the kernel corresponding to these extended interactions is the Green's function of the Laplace operator, these potentials can be efficiently computed by taking recourse to the solution of a Poisson problem. Using the potentials defined in ~\eqref{pot} and the expressions for different components of electrostatic energy in ~\eqref{ElecEngy}-\eqref{eq:Eext_PS}, we can rewrite the electrostatic energy $E_{\text{el}} = E_{H} + E_{\text{ext}} + E_{zz}$ as~\cite{das2015,motamarri2018}
\begin{equation}\label{eleclocal}
\begin{split}
  E_{\text{el}} &= \frac{1}{2} \int (\rho(\bx)\, +\, b(\bx,\bR))\, \varphi(\bx,\bR) \dx- \frac{1}{2}\sum_{I}\int -Z_I\tilde{\delta}(|\bx-\bR_I|)\vself I(\bx)\dx \\&+\sum_{J} \int \left(V^{J}_{\text{loc}}(|\bx-\bR_J|) - \vself J(|\bx-\bR_J|)\right)\rho(\bx)\dx + 2 \sum_{i=1}^{N} \int \int f_i \,\psi^{*}_i(\bx)\, V_{\text{nl}}(\bx,\by)\, \psi_i(\by) \dy \dx \,.
  \end{split}
\end{equation}
Finally, for given positions of nuclei, the reformulated governing equations for the Kohn-Sham DFT problem are:
\begin{subequations}\label{ksproblem}
\begin{gather}
\left(-\frac{1}{2} \del^{2} +  V_{\text{xc}} + \varphi + \sum_{J}(V^{J}_{\text{loc}} - \vself J) + V_{\text{nl}} \right){\psi}_{i} = \epsilon_{i} \,{\psi}_{i},\\
-\frac{1}{4\,\pi}\del^2\, \varphi(\bx,\bR) = \rho(\bx) +b(\bx,\bR)\,,\;\;\;\;-\frac{1}{4\,\pi}\del^2 \,\vself I(\bx,\bR_I) = -Z_I\tilde{\delta}(|\bx-\bR_I|)\,,\\[0.1in]
2\sum_{i}f(\epsilon_i,\mu) = N_e\,,\;\;\;\;
\rho(\bx) = 2\sum_{i}f(\epsilon_i,\mu)|\psi_{i}(\bx)|^2.
\end{gather}
\end{subequations}
Though, the above equations ~\eqref{eleclocal} and ~\eqref{ksproblem} represent a pseudopotential treatment, we note that an all-electron treatment can be realized by setting $V^{J}_{\text{loc}} = \vself J$ and $V_{\text{nl}} = 0$. We further remark that the equations ~\eqref{eleclocal} and ~\eqref{ksproblem} are equally valid for both periodic and non-periodic systems with appropriate boundary conditions. In a non-periodic setting, the simulation domain corresponds to a large enough domain, containing the compact support of the wavefunctions, with Dirichlet boundary conditions. In periodic calculations, it corresponds to a supercell with periodic boundary conditions.

For periodic systems, we now discuss the reduced Kohn-Sham equations on a periodic unit-cell. The Kohn-Sham eigenfunctions for infinite periodic crystals are given by the Bloch theorem ~\cite{rmartin,mermin}, and the Bloch-periodic Kohn-Sham problem on an infinite crystal reduces to a periodic problem on a unit-cell. In numerical simulations involving such periodic calculations, it is computationally efficient to deal with unit-cells that are much smaller than the supercells, and the computation of electron-density, kinetic energy and the electrostatic interaction energy involving the non-local pseudopotentials has an additional integration over the Brillouin zone (BZ). Using Bloch theorem, the Kohn-Sham eigenfunction $\psi_{n,\bk}(\bx)$ can be expressed as
\begin{equation}\label{bloch}
\psi_{n,\bk}(\bx) = e^{i\bk\cdotp\bx} u_{n,\bk}(\bx) \,,
\end{equation}
where $i = \sqrt{-1}$ and  $u_{n,\bk}(\bx)$ is a function that is periodic on the unit-cell while $\bk$ denotes a reciprocal space point in the Brillouin zone. Using ~\eqref{bloch}, the computation of electron-density in equation ~\eqref{eq:elecdens} is given by 
\begin{equation}
   \rho(\bx) = 2\sum_{n=1}^{N}\fintd_{BZ}f_{n,\bk} |u_{n,\bk}(\bx)|^2 \dk\,,
\end{equation}
where $\fintd_{BZ}$ denotes the volume average of the integral over the Brillouin zone corresponding to the periodic unit-cell $\Omega_p$ and $f_{n,\bk}$ denotes the orbital occupancy function corresponding to $u_{n,\bk}$. The contribution to the electrostatic interaction energy arising from the non-local pseudopotential using the Bloch theorem can be expressed as
\begin{equation}
    E_{\text{ext}}^{\text{nl}} = 2 \sum_{n=1}^{N} \fintd_{BZ} \: \int \displaylimits_{\Omegaper}\: \int\displaylimits_{\Rthree} f_{n,\bk}\; u^{*}_{n,\bk}(\bx)\;e^{-i\bk\cdotp\bx}\; V_{\text{nl}}(\bx,\by,\bR)\;e^{i\bk\cdotp\by}\;u_{n,\bk}(\by) \dy \dx \dk\,.
\end{equation}
Using the separable form of the nonlocal pseudopotential operator, we have
\begin{equation}
  V_{\text{nl}}\, u_{n,\bk} := e^{-i\bk\cdotp\bx}\int\displaylimits_{\Rthree} V_{\text{nl}}(\bx,\by,\bR) e^{i\bk\cdotp\by}\,u_{n,k}(\by) \dy = \sum_{a} \sum_{lpm} \sum_{r} e^{-i\bk.(\bx - \bL_r)} C_{lpm}^{a,n\bk}\,h^{a}_{lp}\,\,\chi_{lpm}^{a}(\bx,\bR_a+\bL_r)\,,
\end{equation}
where the summation over $r$ runs on all lattice points in the periodic crystal, and  $a$ runs on all the $N_a$ atoms in the unit-cell. We further note that $C_{lpm}^{a,n\bk}$ in the above equation is given by
\begin{equation}
    C_{lpm}^{a,n\bk} =  \int \displaylimits_{\Omegaper}\sum_{r}e^{i\bk.(\bx - \bL_{r})}\; \chi^{a}_{lpm}(\bx,\bR_a+\bL_{r})\; u_{n,\bk}(\bx) \dx \,.
\end{equation}
Finally, using the local formulation of the extended electrostatic interaction energy discussed previously, the computation of the electronic ground-state, for a given position of atoms, in the context of unit-cell periodic DFT calculations is given by the following equations:
\begin{subequations}\label{ksproblemPer}
\begin{gather}
\left(-\frac{1}{2} \del^{2} - i \,\bk.\del  + \frac{1}{2} |\bk|^2 +  V_{\text{xc}} + \varphi + \sum_{J}(V^{J}_{\text{loc}} - \vself J) + V_{\text{nl}} \right){u}_{n,\bk} = \epsilon_{n,\bk} \,{u}_{n,\bk} \;\;\text{on}\;\; \Omega_p, \label{evp}\\
-\frac{1}{4\,\pi}\del^2\, \varphi(\bx,\bR) = \rho(\bx) +b(\bx,\bR)\;\;\text{on}\;\; \Omega_p\,,\;\;\;\;-\frac{1}{4\,\pi}\del^2 \,\vself I(\bx,\bR_I) = -Z_I\tilde{\delta}(|\bx-\bR_I|)\;\;\text{on}\;\; \Rthree \,, \\[0.1in]
2\sum_{n=1}^{N}\fintd_{BZ}f_{n,\bk}\dk = N_e\,,\;\;\;\;
\rho(\bx) = 2\sum_{n=1}^{N}\fintd_{BZ}f_{n,\bk} |u_{n,\bk}(\bx)|^2 \dk\,.
\end{gather}
\end{subequations}
In the above, periodic boundary conditions are imposed on $\Omega_p$ for the fields $u_n(\bx,\bk)$ and $\varphi(\bx)$. Furthermore, we remark that the equations ~\eqref{ksproblemPer} involve an additional integration over the Brillouin zone (BZ) and is evaluated using numerical quadratures. These numerical quadratures replace the integration over BZ by a weighted sum  over $\bk$  points in the first BZ, and this sampling of the BZ is done using the Monkhorst-pack (MP) grid ~\cite{mpgrid} in DFT-FE. The symmetry operations of the underlying Bravais lattice is used to extract the reduced number of k-points sampling the irreducible Brillouin zone (IBZ). The origin of this MP grid could either coincide with the origin of the BZ or can be shifted by half of the grid spacing in order to maximize the benefit from crystal symmetry mediated BZ reduction ~\cite{rmartin}. Finally, the set of equations ~\eqref{ksproblemPer} is solved self-consistently, with the eigenvalue problem ~\eqref{evp} being solved for $N$ lowest bands for every $\bk$-point in IBZ. We remark that, as noted previously, periodic all-electron calculations can be realized by setting $V^{J}_{\text{loc}} = \vself J$ and $V_{\text{nl}} = 0$.
\subsection{Variational formulation}
The Kohn-Sham governing equations discussed in the previous subsection are the Euler-Lagrange equations of a local variational Kohn-Sham problem that corresponds to the computation of the electronic ground-state free energy for a given position of atoms. The variational problem can be formulated in terms of wavefunctions, fractional occupancies and the electrostatic potentials as given by~\cite{motamarri2018}:
\begin{equation}\label{saddlepoint}
\mathcal{F}_0(\bR) = \min_{\bff \in [0,1]^{N}}\min_{\bPsi \in (\mathcal{Y})^{N}} \;\max_{\varphi \in \mathcal{Y}}\; \mathcal{L}(\bff,\bPsi,\varphi;\bR)\;\;\text{such that}\;\;\int \psi^{*}_{i}\psi_j \dx = \delta_{ij},\;\;2\sum_{i}f_i = N_e\,,
\end{equation}
\vspace{-0.25in}
\begin{align*}
\text{where}\;\;\;\mathcal{L}(\bff,\bPsi,\varphi;\bR) = \widetilde{\mathcal{L}}(\bff,\bPsi) 
+ \min_{\mathcal{V} \in (H^{1}(\Rthree))^{N_a}}\mathcal{L}_{\text{el}}(\bff,\bPsi,\varphi,\mathcal{V};\bR)\,,
\end{align*}
\vspace{-0.1in}
\begin{equation}
\text{with}\;\;\widetilde{\mathcal{L}}(\bff,\bPsi) = T_{\text{s}}(\bff,\bPsi)+ E_{\text{xc}}(\rho) + E_{\text{ent}}(\bff).
\end{equation}
We note that $\bPsi = \{\psi_1(\bx),\psi_2(\bx),\psi_3(\bx), \cdots, \psi_N(\bx)\}$, and $\bff = \{f_1,f_2,f_3\cdots f_N\}$ denotes the vector of orbital occupancy factors, while
$\mathcal{V} = \{V^{1},V^{2},\cdots,V^{N_a}\}$ denotes the vector containing the trial electrostatic potentials corresponding to all nuclear charges in the simulation domain. Here, $T_{\text{s}}(\bff,\bPsi)$ denotes the kinetic energy of non-interacting electrons and $E_{\text{ent}}(\bff)$ denotes the electronic entropy contribution, and the corresponding expressions are given by
\begin{gather}
 T_{\text{s}}(\bff,\bPsi) = 2\,\sum_{i=1}^{N}\int f_i\, \psi_{i}^{*}(\bx) \left(-\frac{1}{2} \del^{2} \right)\,\psi_i(\bx) \dx \,,\\
 E_{\text{ent}} = -2\,\sigma\; \sum_{i=1}^{N} \left[f_i \ln f_i + (1- f_i)\ln(1 - f_i)\right]\,.
\end{gather}
The energy functional corresponding to electrostatic energy, $\mathcal{L}_{\text{el}}$, can be expressed in the local form as~\cite{motamarri2018}
\begin{equation}
\begin{split}
&\mathcal{L}_{\text{el}}(\bff,\bPsi,\varphi,\mathcal{V},\bR) = \int\left[-\frac{1}{8\pi} |\del \varphi(\bx)|^2 + (\rho(\bx) + b(\bx,\bR))\varphi(\bx)\right]\dx \\
&+ \sum_{I} \int \left[\frac{1}{8\pi} |\del V^{I}(\bx)|^2  +  Z_I \tilde{\delta}(|\bx - \bR_I|) V^{I}(\bx) \right]\dx\, + \sum_{J} \int \left(V^{J}_{\text{loc}}(|\bx-\bR_J|) - \vself J(|\bx-\bR_J|)\right)\rho(\bx)\dx \\
 & + 2 \sum_{i=1}^{N} \int \int f_i \,\psi_{i}^{*}(\bx)V_{\text{nl}}(\bx,\by,\bR)\, \psi_i(\by)\dy\dx\,,
 \end{split}
\end{equation}
where $\vself J$ denotes the electrostatic potential corresponding to the $J^{th}$ nuclear charge (see equation~\eqref{pot}), or analogously $\bar{\mathcal{V}}_{\delta} = \{\vself 1,\vself 2,\cdots,\vself {N_a}\}=arg \, \min_{\mathcal{V} \in (H^{1}(\Rthree))^{N_a}}\mathcal{L}_{\text{el}}(\bff,\bPsi,\varphi,\mathcal{V};\bR) $. Further, we note that, $\mathcal{Y}$ in equation ~\eqref{saddlepoint} denotes a suitable function space that guarantees the existence of minimizers. We remark that numerical computations involve the use of bounded domains, which in non-periodic calculations correspond to a large enough domain containing the compact support of the wavefunctions, and, in periodic calculations, correspond to the super-cell\footnote{while the variational problem in equation~\eqref{saddlepoint} is presented for super-cells in the case of periodic calculations, it can be extended to periodic unit-cells using the Bloch Ansatz as discussed in ~\cite{motamarri2018}.}. Denoting such an appropriate bounded domain by $\Omega$ subsequently, $\mathcal{Y}=H^{1}_{0}(\Omega)$ in the case of non-periodic problems, and $\mathcal{Y}=H^{1}_{per}(\Omega)$ in the case of periodic problems.
\subsection{Configurational Forces}\label{sec:configForce}
We employ configurational forces approach~\cite{motamarri2018} to evaluate ionic forces and periodic unit-cell stresses in DFT-FE. These configurational forces correspond to the generalized variational force computed as the derivative of the Kohn-Sham energy functional ~\eqref{saddlepoint} with respect to the position of a material point $\bx$. This approach 
provides a unified framework to compute ionic forces as well as stress tensor for geometry optimization, and inherently accounts for the Pulay corrections owing to the variational nature of the formulation. For the sake of completeness, we present here the expressions for the configurational forces, in terms of the Kohn-Sham eigenfunctions, as the derivative of the energy functional ~\eqref{saddlepoint} with respect to $\bx$. We refer to~\cite{motamarri2018} for the derivation of configurational forces, and a comprehensive discussion on this topic. 

Let $\chieps : \Rthree \rightarrow \Rthree$ represent the infinitesimal perturbation of the underlying space, mapping a material point $\bx$ to a new point $\bx'$ such that ${\btau}^0 = \textbf{I}$. Further, let the generator of this mapping be $\bdir = \derveps\chiepsx\atzero$ such that $\chieps$ is constrained to rigid body deformations in the compact support of the regularized nuclear charge distribution $\bb(\bx)$ in order to preserve the integral constraint $\int \tilde{\delta}(\bx - \bR_I) \dx = 1$. Denoting $\mathcal{F}_0(\chieps)$ to be the ground-state free energy in the perturbed space, the configurational force is evaluated by computing the G\^ateaux derivative of  $\mathcal{F}_0(\chieps)$ given by
\begin{equation}\label{Eshelbyforce}
\frac{d\mathcal{F}_0(\chieps)}{d\varepsilon}\atzerob = \int\displaylimits_{\Omega} \bE:\del \bdir (\bx) \dx + \sum_{I} \int\displaylimits_{\Rthree} \bE'^I:\del \bdir (\bx) \dx \,+\, \text{F}^{\text{PSP}}\,,
\end{equation}
where $\bE$ and  $\bE'^I$ denote the Eshelby tensors whose expressions in terms of the solutions of the saddle point problem ~\eqref{saddlepoint} on the original space are provided below.  If $\bPsibar = \{\psibar_1,\psibar_2,\psibar_3 \cdots \psibar_N \}$ denote the Kohn-Sham eigenfunctions corresponding to the lowest $N$ eigenvalues $\bar{\boldsymbol{\epsilon}}=\{\bar{\epsilon}_1,\bar{\epsilon}_2,\cdots,\bar{\epsilon}_N\}$ with occupancies $\bar{\bff}=\{\fbar_1,\fbar_2 \cdots \fbar_N\}$, and $\varphibar$ denotes the electrostatic potential (all solutions of the saddle point problem ~\eqref{saddlepoint}), the expressions for the Eshelby tensor $\bE$ and $\text{F}^\text{PSP}$ in equation ~\eqref{Eshelbyforce} are given by
\begin{align*}\label{EshelbyforceOrth}
&\bE = \Biggl(\sum_{i=1}^{N}\Bigl(\fbar_i\del \psibar_{i}^{*}(\bx)\cdot\del \psibar_i(\bx) - 2 \, \fbar_i\, \bar{\epsilon}_i\,\psibar^{*}_i(\bx)\; \psibar_i(\bx)\Bigr)  + \varepsilon_{\text{xc}}(\rhobar,\del \rhobar)\rhobar(\bx)  -\frac{1}{8\pi} |\del \varphibar(\bx)|^2 + \rhobar(\bx) \varphibar(\bx) \,\notag\\
&+ \sum_J (V_{\text{loc}}^{J} - \vself J)\rhobar(\bx) \,+  \text{E}^{\text{nl}} \,+ {\text{E}^{\text{nl}}}^{*} \Biggr)\bI - \sum_{i=1}^{N}\fbar_i\Bigl[\del \psibar_{i}^{*}(\bx) \otimes \del \psibar_i(\bx)
 + \del {\psibar_i}(\bx) \otimes \del \psibar_{i}^{*}(\bx)\Bigr] \,\notag\\
& - \frac{\partial}{\partial \del \rho}(\varepsilon_{\text{xc}}(\rhobar,\del \rhobar)\rhobar(\bx)) \otimes \del \rhobar  + \frac{1}{4\pi}\del \varphibar(\bx) \otimes \del \varphibar(\bx)\,,\\
&\bE'^{I} = \frac{1}{8\pi} |\del \Vbar_{\dirac}^{I}(\bx)|^2 \bI - \frac{1}{4\pi}\del \Vbar_{\dirac}^{I}(\bx) \otimes \del \Vbar_{\dirac}^{I} (\bx)\,,
\end{align*}
where
\vspace{-0.1in}
\begin{gather}
\rhobar(\bx) = 2\sum_{i} \fbar_{i}\,\psibar^{*}_i(\bx)\,\psibar_i(\bx)\,,\;\;\;\;\fbar_i = \frac{1}{1 + \exp(\frac{\bar{\epsilon}_i -\mu}{k_B\,T})} \,,\notag \\
\text{E}^{\text{nl}} = 2\;\sum_{i=1}^{N} \sum_{J} \sum_{lpm} \fbar_i\; h_{lp}^{J} \,\psibar_{i}^{*}(\bx)\;\chi^{J}_{lpm}(\bx,\bR_J) {\int\displaylimits_{\Omega} \chi^{J}_{lpm}(\by,\bR_J)\, \psibar_i(\by) \dy} \,.\notag 
\end{gather}
Further,
\begin{equation}
\text{F}^{\text{PSP}} = \sum_{J} \int\displaylimits_{\Omega} \rhobar(\bx)\left(\del \left(V^{J}_{\text{loc}}(|\bx-\bR_J|) - \vself J(|\bx-\bR_J|)\right)\right)\cdot \left(\bdir(\bx) - \bdir(\bR_J) \right)\dx  \,+\, \text{F}^\text{nl} \,+\,  {\text{F}^\text{nl}}^{*}\,,\notag
\end{equation}
where
\begin{equation*}
\text{F}_\text{nl} = 2\;\sum_{i=1}^{N} \sum_{J} \sum_{lpm} \fbar_i h_{lp}^{J}\Bigg[\int\displaylimits_{\Omega} \psibar_{i}^{*}(\bx) \del \chi^{J}_{lpm}(\bx,\bR_J)\cdotp\left(\bdir(\bx) - \bdir(\bR_J)\right)\dx\Biggr]
\Biggl[\int\displaylimits_{\Omega}\chi^{J}_{lpm}(\by,\bR_J)\psibar_{i}(\by) \dy\Biggr] \,.
\end{equation*}
Though equation~\eqref{EshelbyforceOrth} represents the pseudopotential case, the expression for an all-electron case is realized by setting $E^{\text{nl}} = 0$, $\text{F}^{\text{PSP}} = 0$ and  $V^{J}_{\text{loc}}(|\bx-\bR_J|) = \vself J(|\bx-\bR_J|)$. Finally, we remark that the configurational forces provide a unified expression for computing both the ionic forces as well as periodic unit-cell stress by using an appropriate choice of the generator $\bdir (\bx)$. In particular, the force on any given atom is computed via equation ~\eqref{Eshelbyforce} by choosing the compact support of $\bdir (\bx)$ to contain the atom of interest, and the periodic unit-cell stress tensor is evaluated by choosing $\dir_{i} = C_{ij}x_j$ in  ~\eqref{Eshelbyforce} as explained in ~\cite{motamarri2018}.


\subsection{Discrete Kohn-Sham DFT equations}
We introduce here the finite-element (FE) discretization of the Kohn-Sham DFT problem by representing various electronic fields in the FE basis, a piece-wise polynomial basis generated from the FE discretization~\cite{brenner2002}. In particular, we employ $C^{0}$ continuous Lagrange polynomial basis interpolated over Gauss-Lobatto-Legendre nodal points.  The FE discretization of the Kohn-Sham DFT problem described here is along the lines of our prior work~\cite{motamarri2013} and is briefly discussed here, highlighting the important differences in this work. We specifically note here that the real-space formulation of Kohn-Sham DFT as presented in equation ~\eqref{saddlepoint} results in a saddle point problem (min-max problem) in the electronic fields. Thus, it is possible that the electronic ground-state energy obtained from a single FE discretization of all the solution fields in the Kohn-Sham DFT problem can be non-variational. To address this, we seek to solve the electrostatic problem to a more stringent accuracy than the Kohn-Sham eigenvalue problem. To this end, we consider two FE triangulations for representing the wavefunctions and the electrostatic potentials, namely $\mathcal{T}^h$ and $\mathcal{T}^{h_{el}}$ with the characteristic mesh-sizes denoted by $h$ and $h_{el}$, respectively. We consider $\mathcal{T}^{h_{el}}$ to be a uniform subdivision of $\mathcal{T}^h$. Denoting the subspaces spanned by the FE basis corresponding to triangulations $\mathcal{T}^h$ and $\mathcal{T}^{h_{el}}$ to be $\mathbb{V}_h^{M}$ (with dimension $M$) and $\mathbb{V}_{h_{el}}^{M_{el}}$ (with dimension $M_{el}>M$), we note that $\mathbb{V}_h^{M} \subset \mathbb{V}_{h_{el}}^{M_{el}}$. Finally, the representation of the various fields in the Kohn-Sham problem~\eqref{ksproblem}---the wavefunctions and the electrostatic potentials---in the FE basis is given by
\begin{equation}\label{fem}
\psi^{h}_{i}(\bx) = \sum_{j=1}^{M} N^{h}_j(\bx) \psi^{j}_{i}\,\,,\;\;\;\;\varphi^{h_{el}}(\bx) = \sum_{j=1}^{M_{el}} N^{h_{el}}_j(\bx) \varphi^{j}\,,\;\;\;\;{\vself {J^{h_{el}}}}(\bx) = \sum_{j=1}^{M_{el}} N^{h_{el}}_j(\bx)\vself {J^{j}} \,,
\end{equation}
where $N^{h}_{j}:1\leq j \leq M$ denotes the FE basis spanning $\mathbb{V}_h^{M}$ and $N^{h_{el}}_{j}:1\leq j \leq M_{el}$ denotes the FE basis spanning $\mathbb{V}_{h_{el}}^{M_{el}}$ . We note that $\psi^{h}_{i}$, $\varphi^{h_{el}}$ and $\vself {J^{h_{el}}}$ denote the FE discretized fields, with $\psi^{j}_{i}$, $\varphi^{j}$ and $\vself {J^{j}}$ denoting the coefficients in the expansion of the $i^{th}$ discretized wavefunction and the electrostatic potentials, which also correspond to the nodal values of the respective fields at the $j^{th}$ node on the FE mesh. 

The FE discretization of the Kohn-Sham eigenvalue problem~\eqref{ksproblem} results in a generalized eigenvalue problem given by
$\bH \hat{\bvPsi}_{i} = \epsilon^{h}_{i} \bM \hat{\bvPsi}_{i}$
where $\bH$  denotes the discrete Hamiltonian matrix with matrix elements $\text{H}_{jk}$, $\bM$ denotes the overlap matrix (or commonly referred to as the mass matrix in finite element literature) with matrix elements $\text{M}_{jk}$, and $\epsilon^{h}_{i}$ denotes the $i^{th}$ eigenvalue corresponding to the discrete eigenvector $\hat{\bvPsi}_i$. The expression for the discrete Hamiltonian matrix, $\text{H}_{jk} = \text{H}_{jk}^{\text{loc}} + \text{H}_{jk}^{\text{nl}}$, is given in terms of  
\begin{equation}\label{discreteHam}
\text{H}_{jk}^{\text{loc}} = \frac{1}{2} \int_{\Omega} \del N^h_{j} (\bx) .\del N^h_k(\bx) \dx +\int_{\Omega} V^{h}_{\text{eff,loc}}(\bx,\bR)\, N^h_{j}(\bx) N^h_{k}(\bx) \dx\,.
\end{equation}
%
In the above, $V_{\text{eff,loc}}^{h}$ denotes the local part of the effective single-electron potential computed in the FE basis as the sum of discretized exchange-correlation potential $V_{\text{xc}}^{h}$, total electrostatic potential $\varphi^{h_{el}}(\bx)$ and the local pseudopotential term as follows:
\begin{equation}
 V_{\text{eff,loc}}^{h}(\bx,\bR) = V_{\text{xc}}^{h}(\bx) + \varphi^{h_{el}}(\bx) + \sum_J \left(V^{J}_{\text{loc}}(|\bx-\bR_J|) - {\vself {J^{h_{el}}}}(|\bx-\bR_J|)\right)\,.
\end{equation}
In the case of all-electron calculations, $V_{\text{eff,loc}}^{h}(\bx,\bR) = V_{\text{xc}}^{h}(\bx) + \varphi^{h_{el}}(\bx)$ and $\text{H}_{jk}^{\text{nl}}$ is zero. In the case of pseudopotential calculations, $H_{jk}^{\text{nl}}$ is given by
\begin{equation}\label{eq:nonlocalHamDiscrete}
\text{H}_{jk}^{\text{nl}} = \sum_{J = 1}^{N_a} \sum_{lpm} C^{J}_{lpm,j}h^{J}_{lp}C^{J}_{lpm,k}\,,\;\;\text{where}\;\;\;C^{J}_{lpm,j} =  \int_{\Omega} \chi^{J}_{lpm}(\bx,\bR_J) N^h_j(\bx) \dx\,.
\end{equation}
Finally, the matrix elements of the overlap matrix $\bM$ are given by $\text{M}_{jk}=\int_{\Omega} N^h_j(\bx) N^h_k(\bx) \dx$.
We note that the matrices $\bH^{\text{loc}}$ and $\bM$ are sparse as the FE basis functions are local in real space and have a compact support (a finite region where the function is non-zero). Further, the vectors $C^{J}_{lpm,j}$ in $\bH^{\text{nl}}$ are also sparse since the projectors $\chi^{J}_{lpm}(\bx,\bR_J)$ have a compact support, thus rendering a sparse structure to the discrete Hamiltonian $\bH$.

In order to explore efficient solution strategies, it is desirable to transform the generalized eigenvalue problem into a standard eigenvalue problem. Since the matrix $\bM$ is positive definite symmetric, there exists a unique positive definite symmetric square root of $\bM$, and is denoted by $\bM^{1/2}$. Hence, the following holds true:
\begin{equation}
\bH \hat{\bvPsi}_{i} = \epsilon^{h}_{i} \bM \hat{\bvPsi}_{i} \;\;
\Rightarrow\qquad \bH \hat{\bvPsi}_{i} = \epsilon^{h}_{i} \bM^{1/2} \bM^{1/2} \hat{\bvPsi}_{i} \;\;
\Rightarrow\qquad \btH \widetilde{\bvPsi}_{i} = \epsilon^{h}_{i} \widetilde{\bvPsi}_{i} \label{hep}\,,
\end{equation}
\begin{equation}
\text{where} \;\;\; \widetilde{\bvPsi}_{i} = \bM^{1/2} \hat{\bvPsi}_{i}\,,\;\;\;\;
\btH = \bM^{-1/2}\bH\bM^{-1/2}\,.
\end{equation}
We note that $\btH$ is a Hermitian matrix, and~\eqref{hep} represents a standard Hermitian eigenvalue problem. The actual eigenvectors are recovered by the transformation $\hat{\bvPsi}_{i} = \bM^{-1/2} \widetilde{\bvPsi}_{i}$. 
Furthermore, we note that the matrix $\bM^{-1/2}$ can be evaluated with modest computational cost by using a spectral FE basis in conjunction with the use of Gauss-Lobatto-Legendre (GLL) quadrature for the evaluation of integrals in the overlap matrix, that renders the overlap matrix diagonal~\cite{motamarri2013}. This renders the matrix $\btH$  the same sparsity structure as the matrix $\bH$.

Finally, for the given positions of nuclei, the discrete Kohn-Sham eigenvalue problem along with the discretized Poisson equations for the electrostatic potentials ($\varphi^{h_{el}}$ and $\vself {J^{h_{el}}}$) are to be solved self-consistently, and are given by:
\begin{subequations}\label{eq:discreteSolve}
\begin{gather}
\bM^{-1/2}\bH\bM^{-1/2} \widetilde{\bvPsi}_{i} = \epsilon^{h}_{i} \widetilde{\bvPsi}_{i} \,,\label{eq:kohnShamSolve} \\
\sum_{j=1}^{M_{el}}\left[\frac{1}{4\pi}\int_{\Omega} \nabla N_i^{h_{el}}({\bx}). \nabla N_j^{h_{el}}({\bx})\,{\dx}  \right]\varphi^j
=\int_{\Omega}\left(\rho^h({\bx}) + b^{h_{el}}(\bx,\bR)\right)N_i^{h_{el}}({\bf x}) \,{\rm d}{\bf x} \,, \quad \label{eq:phiTotDiscreteSolve}\\
\sum_{j=1}^{M_{el}}\left[\frac{1}{4\pi}\int_{\Omega_J} \nabla N_i^{h_{el}}({\bx}). \nabla N_j^{h_{el}}({\bx})\,{\rm d} {\bf x} \right]\bar{V}^{J^j}_{\tilde{\delta}}
=\int_{\Omega_J}\left( b^{h_{el}}_J(|\bx - \bR_J|) \right)N_i^{h_{el}}({\bf x}) \,{\dx} \,, \quad \forall J \,, \label{eq:vselfDiscreteSolve}\\
2\sum_{i}f(\epsilon^{h}_i,\mu) = N_e\,,\;\;\;\;
\rho^{h}(\bx) = 2\sum_{i}f(\epsilon^h_i,\mu)|\psi^h_{i}(\bx)|^2 \,.
\end{gather}
\end{subequations}
We note that the nuclear charges in DFT-FE implementation are located on the nodes of the FE triangulation, and are treated as point charges. Thus, the nuclear charge distribution in the discrete setting $b^{h_{el}}(\bx,\bR)$ in equation~\eqref{eq:phiTotDiscreteSolve} is given by $b^{h_{el}}(\bx,\bR) = \sum_I b_{I}^{h_{el}}(|\bx - \bR_I|)$ with $b_{I}^{h_{el}} = - Z_I{\delta}(|\bx - \bR_I|)$ where $\delta(|\bx - \bR_I|)$ denotes the Dirac-delta distribution centered at the position of the atom $\bR_I$.  The boundary conditions used for the computation of the discrete potential field $\varphi^{h_{el}}(\bx)$ in equation ~\eqref{eq:phiTotDiscreteSolve} are either homogeneous Dirichlet boundary conditions or periodic boundary conditions depending on whether the problem is non-periodic or periodic. Further, the discrete self potential $\bar{V}^{J^{h_{el}}}_{\tilde{\delta}}$ associated with individual nuclear charge $J$  is solved using the discrete Poisson equation ~\eqref{eq:vselfDiscreteSolve} subject to Dirichlet boundary conditions  with prescribed Coulomb potential applied on a domain $\Omega_J$ enclosing the atom $J$.  After obtaining the electronic ground-state from the solution of the discrete Kohn-Sham problem (equations~\eqref{eq:discreteSolve}), we compute the discrete total ground-state energy  $E^h$  in terms of the discrete solution fields ($\bar{\epsilon}^h_i, \rhobar^h, \varphibar^{h_{el}},\vself {I^{h_{el}}}$) as follows:
\begin{equation}\label{discreteenergy}
E^h = E^h_{\text{band}} - E^{h}_{\text{pot}} +  E^h_{\text{xc}}(\rhobar^h,\del \rho^h) + E^{h_{el}}_{\text{el}}\,,
\end{equation}
where, 
\begin{gather*}
 E^h_{\text{band}} = 2 \sum_i f(\epsilonbar^h_i,\mu) \epsilonbar^h_i\,, \;\;\;
 E^h_{\text{pot}}  = \int_\Omega \rhobar^h(\bx) \left(V^h_{\text{xc}}(\bx) + \varphibar^{h_{el}}(\bx)\right) \dx \,,\\
\begin{split}
E^{h_{el}}_{\text{el}} &=  \int_\Omega \left[-\frac{1}{8\pi} |\del \varphibar^{h_{el}}(\bx)|^2 + (\rhobar^{h_{el}}(\bx) + b^{h_{el}}(\bx,\bR))\varphibar^{h_{el}}(\bx)\right]\dx \\
&+ \sum_{I} \int_\Rthree \left[\frac{1}{8\pi} |\del \vself {I^{h_{el}}}(\bx)|^2  -  b^{h_{el}}_{I}(|\bx - \bR_I|) \vself {I^{h_{el}}}(\bx) \right]\dx \,.
\end{split}
\end{gather*}

\section{Numerical implementation}
\label{sec:numImpl}
DFT-FE is built over the deal.II open-source finite-element library~\cite{dealII90}, and uses its underlying finite element constructs, adaptive mesh refinement architecture and efficient parallel vector objects.
In this section, we discuss various aspects of numerical implementation of the Kohn-Sham DFT problem within the framework of spectral finite-element discretization in DFT-FE. We first begin with a discussion on the strategies implemented for adaptive mesh refinement in DFT-FE, followed by a detailed discussion on the various steps involved in the implementation of the Kohn-Sham self-consistent field iteration procedure. 
\subsection{Adaptive mesh refinement}
One of the significant strengths of the FE basis is that it can accommodate adaptive spatial resolution, which in turn can be effectively exploited for the efficient solution of DFT calculations ~\cite{motamarri2013,bylaska,lehtovaara,tsuchida1996,denis2016} as well as the development of coarse-graining techniques that seamlessly bridge electronic structure calculations with continuum~\cite{qcofdft,choly2005,gang2006}. In DFT-FE, adaptive mesh refinement is carried out using octree-based hexahedral mesh generator based on the  `Parallel AMR on Forests of Octrees' (p4est)  library ~\cite{p4est2011} via deal.II open-source finite-element library~\cite{dealII90}. Spatial discretization via this octree refinement produces a non-conforming mesh (cf. Fig.~\ref{fig:udamr}), resulting in a potentially discontinuous function approximation. However, the continuity of a function across refinement interfaces is enforced by algebraic constraints that require the value of the function on the `hanging node' to be consistent with the approximation along the neighboring element face or edge.  We discuss the two adaptive mesh refinement strategies implemented in DFT-FE: (i) user-defined adaptive mesh refinement (UDAMR) procedure, involving user inputs for mesh sizes at different regions of the simulation domain, and (ii) automatic adaptive mesh refinement procedure (AAMR) guided by \textit{a priori} error estimates. We note that the simulation domain encloses the given materials system for a non-periodic problem, while for a periodic problem, the simulation domain is a periodic unit-cell.

\subsubsection{User defined adaptive mesh refinement (UDAMR)}
\begin{figure}[htp]
 \centering
\includegraphics[scale=0.5]{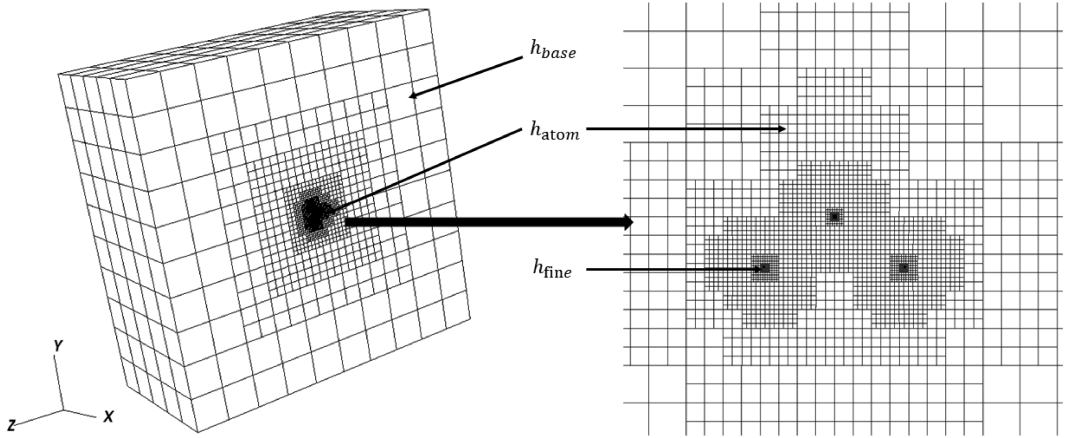}
\caption{User defined adaptive mesh schematic for all-electron DFT calculation. Case study:  SiF$_4$ molecule. The schematic shows the slice of the 3D FE mesh cut with a plane normal to x-axis. The three locally refined mesh regions with mesh size $h_{\text{fine}}$ indicate the vicinity of one Silicon and two Fluorine atoms on the sliced plane. The schematic also shows the coarse-graining from $h_{\text{fine}}$ to $h_{\text{base}}$.}
\label{fig:udamr}
\end{figure}

The user defined adaptive mesh in DFT-FE is constructed to have a refined mesh around the atom up to a certain radius $r_{\text{atom}}$ and coarsens away.  This is accomplished by discretizing the given simulation domain using an initial coarse uniform mesh with mesh size $h_{\text{base}}$.  Subsequently, the FE cells whose centroids are within a distance of $r_{\text{atom}}$ units from each of the atomic positions are marked for refinement. These marked cells are refined until a target mesh size $h_{\text{atom}}$ is achieved. Hence, the mesh size parameters $h_{\text{base}}$,  $h_{\text{atom}}$  and the parameter $r_{\text{atom}}$ form the user defined input to generate the adaptive mesh.  Though these parameters are adequate for generating adaptive meshes in the case of pseudopotential DFT calculations involving smooth solution fields, all-electron DFT calculations require much finer meshes in the close vicinity of the nuclear position due to the highly oscillatory nature of the wavefunctions near the nuclei. Hence, we introduce another mesh parameter $h_{\text{fine}}$ which prescribes the mesh-size of the elements that share the vertex at the nuclear position. A user defined adaptive mesh depicting the above parameters in the case of SiF$_4$ molecule for all-electron DFT calculation is shown in Fig.~\ref{fig:udamr} .

We note that the mesh parameters $h_{\text{base}}$, $h_{\text{atom}}$, $h_{\text{fine}}$ for a given materials system can be estimated from the optimal mesh size distribution $h(\bx)$ obtained by minimizing the discretization error in the ground-state energy for a fixed number of FE cells $N_E$. To this end, we recall the procedure described in ~\cite{motamarri2013} to determine this mesh-size distribution. Let $E$ be the ground-state energy for the continuous problem ~\eqref{ksproblem} with $\Psibar = \{\psibar_{1}\,,\,\psibar_2 \,\cdots\,\psibar_{N}\}$ representing the electronic wavefunction solutions of the continuous problem ~\eqref{ksproblem}, and, further, let $E_h$ denote the discretized ground-state energy with  $\Psibar^{h} = \{\psibar_{1}^{h}\,,\,\psibar_2^{h} \,\cdots\,\psibar_{N}^{h}\}$ representing the electronic wavefunction solutions of the discrete problem \eqref{eq:discreteSolve}. As discussed in the previous section, we note that the triangulation $\mathcal{T}^{h_e}$ employed in the solution of the electrostatics problems ~\eqref{eq:phiTotDiscreteSolve} and ~\eqref{eq:vselfDiscreteSolve} is more resolved than the triangulation $\mathcal{T}^{h}$ used for solving Kohn-Sham eigenvalue problem ~\eqref{eq:kohnShamSolve}, and is usually chosen to be the uniform subdivision of the triangulation $\mathcal{T}^{h}$. Thus, the dominant discretization error will correspond to the error in the wavefunctions, and as derived in equation (47) of ~\cite{motamarri2013}, the discretization error $|E - E_h|$ (retaining only the leading order terms) is given by 
\begin{equation}\label{boundsdom}
|E - E^h| \leq C  \Bigl(\sum_{i}{|\psibar_{i} - \psibar^{h}_{i}|}^{2}_{1,\Omega}\Bigr) \leq  \mathcal{C}\sum_{e}h_e^{2k}\left[\sum_{i}{|\psibar_{i}|}_{k+1,\Omega_e}^{2} \right]\,,
\end{equation}
where $e$ denotes an element in the FE discretization, with mesh size $h_e$ covering the domain $\Omega_e$, and $|\,.\,|_{k+1,\Omega_e}$ denotes the $(k+1)$ semi-norm over $\Omega_e$. Using the definition of semi-norms, equation~\eqref{boundsdom} can be written as
\begin{equation}\label{meshdistribbound}
|E_h - E| \leq \mathcal{C} \sum_{e} \Bigl[ h_{e}^{2k}\int_{\Omega_{e}} \Bigl[\sum_{i} {|D^{k+1}\psibar_{i}(\bx)|}^{2}\Bigr] \dx\Bigr] \leq \mathcal{C'} \intomega h^{2k}(\bx)\Bigl[\sum_{i}{|D^{k+1}\psibar_{i}(\bx)|}^{2}\Bigr] \dx\,,
\end{equation}
where $h(\bx)$ denotes the element size distribution function defining the target element size at point $\bx$ in the simulation domain. The mesh distribution $h(\bx)$ can be estimated by minimizing the approximation error in energy in ~\eqref{meshdistribbound} subject to fixed number of elements, which is given by
\begin{equation}
\min_{h} \intomega \Bigl\{ h^{2k}( \bx )\Bigl[\sum_{i}{|D^{k+1}\psibar_{i}(\bx)|}^{2} \Bigr] \Bigr\}\dx \quad \mbox{subject to} \intomega\frac{\dx}{h^{3}(\bx)}=N_{E}\,.
\end{equation}
The solution to this variational problem is given by
\begin{equation}\label{optimmesh}
h(\bx) = A \Bigl(\sum_{i}{|D^{k+1}\psibar_{i}(\bx)|}^{2}\Bigr)^{-1/(2k+3)}\,,
\end{equation}
where the constant $A$ is computed from the constraint that the total number of elements is $N_E$. We note that the mesh size distribution $h(\bx)$ in equation ~\eqref{optimmesh} involves the knowledge of $\psibar_{i}(\bx)$, which is not known $\textit{a priori}$. However, from a practical standpoint, single atom DFT wavefunctions constructed numerically from the solution of 1D radial Kohn-Sham DFT problem can be used. Thus, the mesh parameters $h_{\text{base}}$, $h_{\text{atom}}$, $h_{\text{fine}}$ can be estimated from $h(\bx)$, which provides a systematic procedure to prescribe these quantities. 

We note that the adaptive mesh refinement procedure via the octrees employed in DFT-FE produces a non-conforming mesh such that the ratio of edge lengths of two neighboring cells is at most 2:1. Due to this constraint the target mesh sizes of $h_{\text{atom}}$ and $h_{\text{fine}}$ can only be approximately realized, and may also lead to more degrees of freedom than required. This increase in degrees of freedom can be more pronounced in the case of all-electron DFT calculations. Thus, DFT-FE also provides an automatic adaptive mesh generation strategy, guided by local error indicators involving the FE discretized solution fields, and is discussed subsequently. 

\subsubsection{Automatic adaptive mesh refinement (AAMR)}
A number of recent works~\cite{zhou2008,chen2011,bao2012,chen2014,denis2016,shen2018} have been devoted to adaptive mesh refinement strategies employing local error indicators expressed in terms of the solution fields of the discrete Kohn-Sham problem. Local error indicators relying on eigenvalue problem residuals as well as jump in the derivative of solution fields across the face of FE cells (Kelly error indicators) have been employed in many of the recent works ~\cite{zhou2008,chen2011,chen2014,denis2016}. Further, error indicators based on $H^{1}$ semi-norms of the wavefunctions~\cite{bao2012} and those based on coarsening mesh approaches~\cite{shen2018} have also been employed. Most of the above methods start with an initial coarse triangulation on which the discrete Kohn-Sham problem is solved, and a local error indicator in terms of the discrete solution fields is subsequently employed to mark the cells to be refined ({\it a posteriori} mesh adaption). This procedure is usually repeated till convergence, and the process generates a sequence of increasingly refined adaptive FE approximations.

We note that many of the aforementioned {\it a posteriori} adaptive mesh refinement strategies require the solution of the Kohn-Sham problem during the course of the adaptive refinement procedure, which can be very expensive to employ for large scale calculations \footnote{computational complexity of the Kohn-Sham problem is cubic-scaling with number of electrons, thus making the repeated solution of a large-scale problem very expensive during the course of adaptive mesh refinement procedure.}. To this end, we present an efficient strategy to construct an adaptive mesh \textit{a priori}, before beginning the SCF procedure, by making use of numerically computed single-atom Kohn-Sham DFT wavefunctions. The approach is based on a local error indicator obtained from an error estimate on the energy involving the discrete wavefunctions, in contrast to the error estimate in equation~\eqref{boundsdom} that involves the continuous wavefunctions. We first derive this energy error estimate following the mesh adaption ideas in ~\cite{radio}, and subsequently present the algorithm implemented in DFT-FE for automatic adaptive mesh refinement.

Let $\psitilde^h(\bx)$  be the $(k-1)^{\text{th}}$ interpolant of $\psibar^{h}(\bx)$ with $k\geq 2$ denoting the FE interpolating polynomial. As opposed to the approximation error $|E-E^h|$, we will work with $|E-\widetilde{E}^h|$ where $\widetilde{E}^h$ denotes the discrete Kohn-Sham ground-state energy obtained using the $(k-1)^{\text{th}}$ interpolants of the FE solution of the Kohn-Sham problem (equation~\eqref{eq:discreteSolve}). To this end, following along similar lines as the derivation of equation ~\eqref{boundsdom}, the dominant term in $|E-\widetilde{E}^h|$ can be derived to be
\begin{equation}\label{boundsdominter}
|E - \widetilde{E}^h| \leq \widetilde{C}  \Bigl(\sum_{i}{|\psibar_{i} - \psitilde^{h}_{i}|}^{2}_{1,\Omega}\Bigr)\,.
\end{equation}
We recall, in the above expression, $|\,.\,|_{1,\Omega}$ denotes the $H^{1}$ semi-norm over $\Omega$. Using triangle inequality, we have
\begin{equation}\label{triainequal}
{|\psibar_{i} - \psitilde^{h}_{i}|}_{1,\Omega} = {|(\psibar_{i} - \psibar^{h}_{i}) + (\psibar^{h}_{i} - \psitilde^{h}_{i})|}_{1,\Omega} \leq {|\psibar_{i} - \psibar^{h}_{i} |}_{1,\Omega} + {|\psibar^{h}_{i} - \psitilde^{h}_{i}|}_{1,\Omega}
\end{equation}
We note that, asymptotically, as $h \rightarrow 0$, the first term,  ${|\psibar_{i} - \psibar^{h}_{i}|}_{1,\Omega}$\,, in the above inequality~\eqref{triainequal} is $\order{h^k}$ ~\cite{motamarri2013}, while the second term, ${|\psibar^{h}_{i} - \psitilde^{h}_{i}|}_{1,\Omega}$\,, is of order $\order{h^{k-1}}$. Hence, the error is dominated by ${|\psibar^{h}_{i} - \psitilde^{h}_{i}|}_{1,\Omega}$\,, which in turn can be bounded in terms of the FE mesh size as~\cite{ciarlet2002}
\begin{equation}
{|\psibar_{i} - \psitilde^{h}_{i}|}_{1,\Omega} \leq  \hat{C} {|\psibar^{h}_{i} - \psitilde^{h}_{i}|}_{1,\Omega} \leq \hat{C}^{'} \sum_{e} (h_e)^{k-1} {|\psibar^{h}_{i}|}_{k,\Omega_e}\,,
\end{equation}
where $e$ denotes an element in the FE discretization, with mesh size $h_e$ covering the domain $\Omega_e$. Hence, the energy error estimate $|E - \widetilde{E}^h|$ in equation ~\eqref{boundsdominter} can be written as
\begin{equation}\label{errorBoundApost}
|E - \widetilde{E}^h| \leq C  \sum_e  (h_e)^{2(k-1)} \Bigl(\sum_{i} |\psibar^{h}_{i}|^2_{k,\Omega_e}\Bigr)\,.
\end{equation}

The energy error estimate ~\eqref{errorBoundApost} motivates $r_e= (h_e)^{2(k-1)} \sum_i {|D^{k}\psibar_i^{h}(\bx)|}^2$ as a useful local error indicator to be used for the automatic adaptive mesh refinement procedure. This error indicator is ideally suited for {\it a-posteriori} mesh adaption, i.e., use the solution from the current mesh to conduct mesh adaption. However, such a procedure will be impractical for large-scale calculations, where the Kohn-Sham problem has to be solved for every iterate of the refinement procedure. Thus, in DFT-FE, we adopt the strategy to use this error indicator to conduct mesh refinement {\it a-priori}. To this end, we use single-atom wavefunctions interpolated onto the FE mesh (denoted as $\psi_i^{h}(\bx)$) as good approximations to $\psibar_i^{h}(\bx)$ for the purpose of mesh-refinement, and use these in the computation of the local error indicator.  

Thus, the automatic adaptive mesh refinement procedure in DFT-FE starts with an initial triangulation $\mathcal{T}^{h^{(0)}}$, and a sequence of nested triangulations $\mathcal{T}^{h^{(n)}}$ are generated using the following iterative procedure (Algorithm~\ref{alg:meshadapt}): \textbf{Interpolate single-atom DFT wavefunctions} $\rightarrow$ \textbf{Estimate local error} $\rightarrow$ \textbf{Mark for refinement} $\rightarrow$ \textbf{Refine} $\rightarrow$ \textbf{Check convergence}.
In particular, the triangulation $\mathcal{T}^{h^{(n+1)}}$ is constructed from $\mathcal{T}^{h^{(n)}}$ by interpolating the single-atom DFT wavefunction data onto the current mesh $\mathcal{T}^{h^{(n)}}$. The local error indicator ${r_e}^{(n)}$ is computed for each FE cell, and the cells are ranked in the order of decreasing error. A predefined fraction ($\beta$) of cells are marked for refinement, by selecting those at the top of the ordered cells (corresponding to the highest local error), and the refinement procedure is carried out to generate the triangulation $\mathcal{T}^{h^{(n+1)}}$. As the AAMR is executed as an {\textit {a-priori}} adaption scheme, a key aspect of AAMR is to devise a stopping criterion for the refinement algorithm. To this end, as the sequence of nested triangulations get generated in AAMR, we examine the convergence of kinetic energy term, i.e, $\sum_i \int |\del \psi_i^{h^{(n)}}(\bx)|^2 \,\dx$.  This is motivated from ~\eqref{boundsdominter}, as the leading order error in the ground-state energy results from the error in the kinetic energy. The refinement algorithm is terminated when the kinetic energy is converged to within a prescribed tolerance.

\begin{algorithm}
\begin{algorithmic}[1]
\caption{ Automatic Adaptive Mesh Refinement Procedure (AAMR) in DFT-FE }
\label{alg:meshadapt}
\State Generate an initial coarse mesh $\mathcal{T}^{h^{(0)}}$ for the given configuration of atoms.
\For{\texttt{ n = 0 to $n_{\text{max}}$}}
\State {Interpolate single atom Kohn-Sham DFT wavefunctions on to the mesh $\mathcal{T}^{h^{(n)}}$:   $\psi_i^{h^{(n)}}(\bx)$.}
\State {Compute the quantity $q^{(n)} = \sum_i \int |\del \psi_i^{h^{(n)}}(\bx)|^2 \dx$ on $\mathcal{T}^{h^{(n)}}$.}
\If{$|q^{(n)} - q^{(n-1)}| < tol_q$}
   \State break;
\Else
   \State continue;
\EndIf
\State {Compute the local error indicator $r_{e}^{(n)}= (h_e)^{2(k-1)} \sum_i {|D^{k}\psi_i^{h^{(n)}}(\bx)|}^2$ in each FE cell}
\State {Rank the FE cells in the order of decreasing error $r_e^{(n)}$, and refine the top fraction $\beta$ of the cells with the maximum error to generate a new mesh $\mathcal{T}^{h^{n+1}}$.}
\EndFor
\end{algorithmic}
\end{algorithm}

\begin{figure}[htp]
 \centering
\includegraphics[scale=0.5]{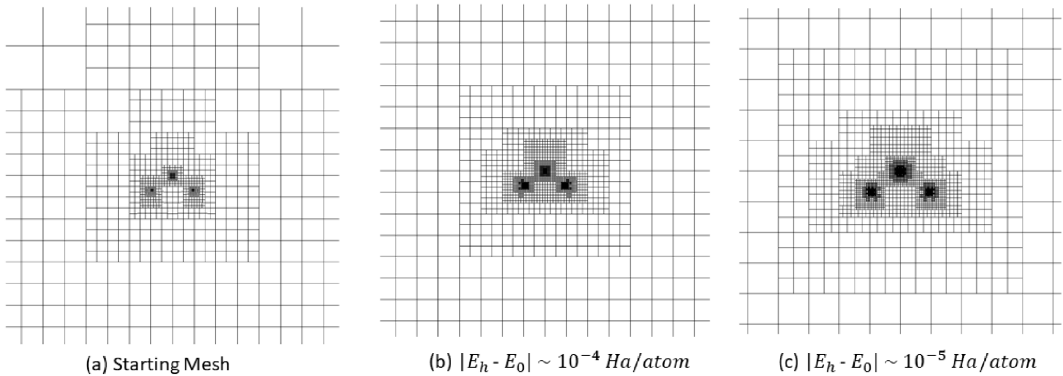}
\caption{Slices of 3D FE meshes obtained using AAMR scheme at different discretization errors. Case Study: All-electron DFT calculation on SiF$_4$ molecule.}
\label{fig:aamr}
\end{figure}

In order to compare AAMR with UDAMR, we conduct a comparative study between these mesh adaption schemes. To this end, we choose representative benchmark examples involving non-periodic norm conserving (ONCV) pseudopotential, all-electron DFT calculations on SiF$_4$ molecule, and a periodic all-electron DFT calculation on Si diamond unit-cell. The mesh attributes, degrees of freedom per atom, and the discretization errors in ground-state energies, atomic forces and periodic unit-cell hydrostatic stresses are tabulated for the above benchmark systems (cf. Tables~\ref{tab:AMRPSP}--\ref{tab:AMRAEPer}). We note that the reference data for the ground-state energy per atom ($E_0$), force vector ($\bff_0$) and hydrostatic stress ($\sigma_0$) to measure the discretization errors are obtained using highly refined FE calculations. These results indicate that AAMR procedure as described in Algorithm~\ref{alg:meshadapt} provides FE meshes with significantly lesser degrees of freedom than the UDAMR procedure for a range of discretization errors. In the case of pseudopotential DFT calculations on SiF$_4$ molecule, we observe that AAMR scheme resulted in a FE mesh with $3$ times lesser degrees of freedom than the UDAMR scheme for discretization errors of the $\order{10^{-5}}$ in ground-state energies and forces. While in the case of all-electron DFT calculations on the same benchmark system, AAMR scheme resulted in a FE mesh with $20$ times lesser degrees of freedom than the UDAMR scheme for discretization errors of the $\order{10^{-4}}$ in ground-state energies and forces. Fig.~\ref{fig:aamr} shows the slices of a 3D FE meshes obtained using AAMR scheme at discretization errors of around $10^{-4}$ Ha/atom and $10^{-5}$ Ha/atom in the case of SiF$_4$ all-electron DFT calculations. Further, in the case of all-electron DFT periodic calculations on Si diamond unit-cell, AAMR scheme resulted in a FE mesh with $3$ times lesser degrees of freedom for discretization errors of the $\order{10^{-4}}$ in ground-state energies.

\begin{figure}[!]
    \centering
    \begin{subfigure}{0.48\textwidth}
        \centering
        \includegraphics[scale=0.4]{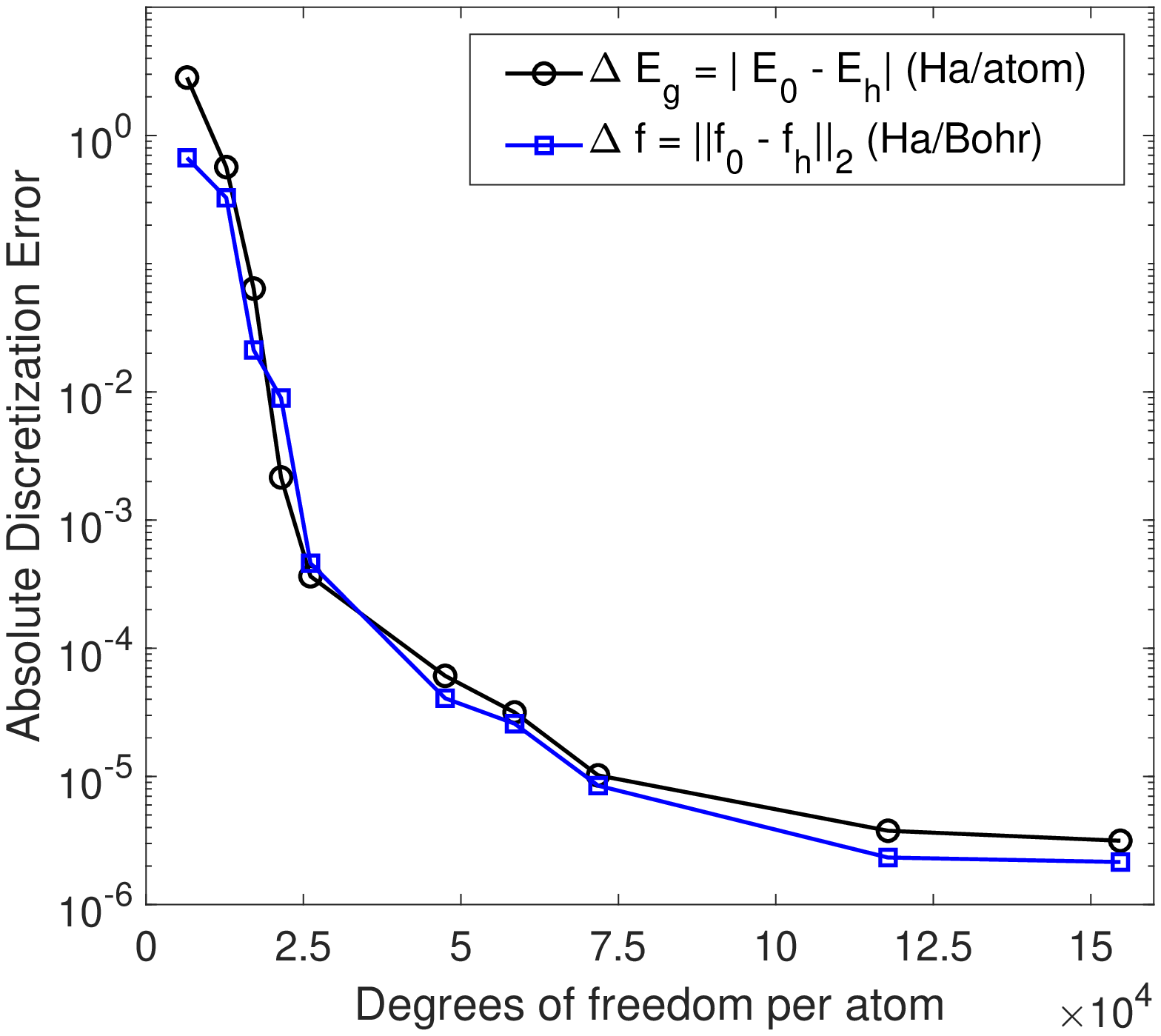}
        \caption{SiF$_4$ ONCV Pseudopotential Calculation}
        \label{fig:SiF4ConvPSP}
    \end{subfigure}
    ~ 
    \begin{subfigure}{0.48\textwidth}
        \centering
        \includegraphics[scale=0.4]{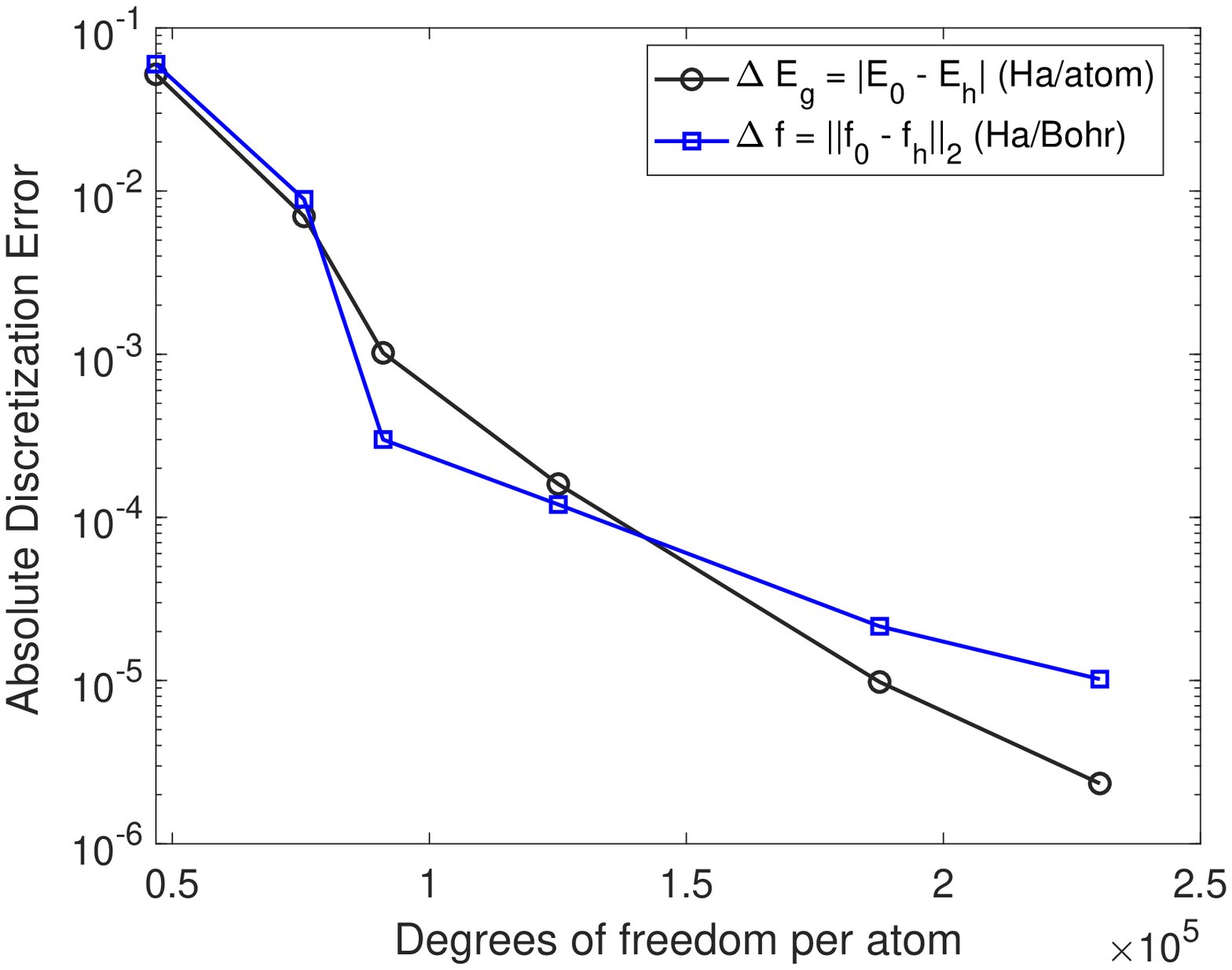}
        \caption{SiF$_4$ All-electron Calculation}
        \label{fig:SiF4ConvAE}
    \end{subfigure}
    \caption{Systematic convergence of FE basis in DFT-FE. Case study in SiF$_4$ molecule. Discretization errors in ground-state energy and force on one of the F atoms are plotted against degrees of freedom obtained at different stages of the AAMR procedure}
    \label{fig:SysConv}
\end{figure}

We also demonstrate the systematic convergence of energies and forces obtained from the sequence of FE meshes generated using AAMR. To this end, we plot in Fig. ~\ref{fig:SiF4ConvPSP} and  Fig.~\ref{fig:SiF4ConvAE} the absolute discretization errors in ground-state energy and force vector (on a specific atom) for the sequence of increasingly refined meshes obtained from AAMR. Fig.~\ref{fig:SysConv} demonstrates that absolute discretization errors as low as $\order{10^{-5}}$ Ha/atom in the ground-state energy and $\order{10^{-5}}$ Ha/Bohr in the forces can be attained using AAMR, which are significantly better than chemical accuracy. We note that the discretization errors stagnate for increasingly finer meshes in Fig.~\ref{fig:SysConv} and this behavior is a consequence of using \textit{a priori} error indicator involving single atom wavefunctions in AAMR.

\begin{table}
\centering
\small
\caption{Comparison between user defined adaptive mesh refinement strategy (UDAMR) and automatic adaptive mesh generation strategy (AAMR). \textbf{Case Study: Pseudopotential non-periodic DFT calculation on SiF$_4$ molecule.} SiF$_4$ details: Si-F bondlength = $2.91$ Bohr, F-Si-F Tetrahedral angle = ${109.47}^{0}$,  $E_0 = -19.94369693$ Ha/atom, $\bff_0$ on F atom = $(0.0288669, 0.0, -0.0204124)$ Ha/Bohr. $\beta = 0.03$ in Algorithm ~\ref{alg:meshadapt}. $|\Delta E_g| = |E_0 - E_h|$ and $|\Delta f| = ||\textbf{f}_0 - \textbf{f}_h||_2$ are the discretization errors in ground-state energy and force vector, respectively.}\label{tab:AMRPSP}
\begin{tabular}{|c|c|c|c|c|}
\hline \hline
 \small{Adaptive mesh} & $h_{\text{min}}$, $h_{\text{max}}$ (a.u.), FE$_{ord}$ ; & \small{FE basis per atom} & $|\Delta E_g|$  & $|\Delta f|$  \\
 \small{strategy} & (\small{tol}$_{q}$\,(Ha/atom)) &  &  &\\
\hline \hline
UDAMR & 0.72, 12.5, 4  & 24,965 & 1.5 $\times$ 10$^{-3}$ & 1.5 $\times$ 10$^{-4}$\\\hline
UDAMR & 0.36, 12.5, 4 & 193,108 & 1.1 $\times$ 10$^{-5}$ & 1.72 $\times$ 10$^{-5}$\\\hline
AAMR & 0.28, 20, 4 ; (1 $\times$ 10$^{-3}$) & 26,115 & 3.6 $\times$ 10$^{-4}$ & 4.6 $\times$ 10$^{-4}$ \\ \hline 
AAMR & 0.28, 20, 4 ; (5 $\times$ 10$^{-5}$) & 47,495 & 6 $\times$ 10$^{-5}$ &  4.1 $\times$ 10$^{-5}$ \\ \hline
AAMR & 0.15, 20, 4 ; (2 $\times$ 10$^{-5}$) & 58,520 & 3 $\times$ 10$^{-5}$ & 2.6 $\times$ 10$^{-5}$\\\hline\hline
\end{tabular}
\end{table}

\begin{table}
\centering
\small
\caption{Comparison between user defined adaptive mesh refinement strategy (UDAMR) and automatic adaptive mesh generation strategy (AAMR). \textbf{Case Study: All-electron non-periodic calculation on SiF$_4$ molecule}. Details: Si-F bondlength = 2.91 Bohr, F-Si-F Tetrahedral angle = 109.47$^{0}$,  $E_0 = -137.76973324$ Ha/atom, $f_0$ on $\bff_0$ on F atom $= (0.02595397, 0.0, -0.01834901)$ Ha/Bohr. $\beta = 0.03$ in Algorithm ~\ref{alg:meshadapt}. $|\Delta E_g| = |E_0 - E_h|$ and $|\Delta f| = ||\textbf{f}_0 - \textbf{f}_h||_2$ are the discretization errors in ground-state energy and force vector, respectively.}\label{tab:AMRAE} 
\begin{tabular}{|c|c|c|c|c|}
\hline \hline
 \small{Adaptive mesh} & $h_{\text{min}}$, $h_{\text{max}}$ (a.u.), FE$_{ord}$ ; & \small{FE basis per atom} & $|\Delta E_g|$  & $|\Delta f|$  \\
 strategy & (\small{tol}$_{q}$\,(Ha/atom))  &  &  &\\
\hline \hline
UDAMR & 0.024, 12.5, 4  & 382,705 & 2 $\times$ 10$^{-3}$ & 1.8 $\times$ 10$^{-3}$\\\hline
UDAMR & 0.012, 12.5, 4 & 2,493,531 & 3.9 $\times$ 10$^{-4}$ & 6.8 $\times$ 10$^{-4}$\\\hline
AAMR & 0.019, 20, 4 ; (1 $\times$ 10$^{-2}$) & 91,002 & 1.02 $\times$ 10$^{-3}$ & 3.3 $\times$ 10$^{-4}$\\\hline 
AAMR & 0.0097, 20, 4 ; (1 $\times$ 10$^{-3}$) & 125,083 & 1.6 $\times$ 10$^{-4}$ &  1.2 $\times$ 10$^{-4}$\\\hline
AAMR & 0.0024, 20, 4 ; (5 $\times$ 10$^{-5}$) & 187,591 & 9.8 $\times$ 10$^{-6}$ & 2.15 $\times$ 10$^{-5}$\\\hline\hline
\end{tabular}
\end{table}
%
\begin{table}
\centering
\small
\caption{Comparison between user defined adaptive mesh refinement strategy (UDAMR) and automatic adaptive mesh generation strategy (AAMR). \textbf{Case Study: All-electron periodic calculation on Si diamond unit-cell}. Si unit-cell details: Lattice constant $= 10.0065~Bohr$, $E_0 = -289.350382455$ Ha/atom,  $\sigma_0$ on unit-cell $= -0.00077999$ Ha/Bohr$^3$. $\beta = 0.03$ in Algorithm ~\ref{alg:meshadapt}. $|\Delta E_g| = |E_0 - E_h|$ and $|\Delta \sigma| = |\sigma_0 - \sigma_h|$ are the discretization errors in ground-state energy and hydrostatic stress, respectively.}\label{tab:AMRAEPer}
\begin{tabular}{|c|c|c|c|c|}
\hline \hline
 \small{Adaptive mesh} & $h_{\text{min}}$, $h_{\text{max}}$ (a.u.), FE$_{ord}$;& \small{FE basis per atom} & $|\Delta E_g|$  & $|\Delta \sigma|$  \\
 strategy & (tol$_{q}$\,(Ha/atom))  &  &  &\\
\hline \hline
UDAMR & 0.013, 1.67, 4 & 188,724 & 2.0 $\times$ 10$^{-3}$ & 6.9 $\times$ 10$^{-5}$ \\\hline
UDAMR & 0.013, 1.67, 5 & 360,695 & 3.6 $\times$ 10$^{-4}$ & 4.3  $\times$ 10$^{-5}$ \\\hline
AAMR & 0.0097, 1.25, 4 ; (1 $\times$ 10$^{-2}$) & 75,563 & 1.75 $\times$ 10$^{-3}$    & 5.48 $\times$ 10$^{-5}$ \\\hline 
AAMR & 0.0048, 1.25, 4 ; (1 $\times$ 10$^{-3}$) & 111,860 & 8.0 $\times$ 10$^{-5}$ & 2.67 $\times$ 10$^{-5}$ \\\hline
AAMR & 0.0024, 1.25, 4 ; (5 $\times$ 10$^{-5}$) & 283,238 & 6.7 $\times$ 10$^{-6}$ & 1.5 $\times$ 10$^{-5}$ \\\hline\hline
\end{tabular}
\end{table}


\newpage
\subsection{SCF Algorithm}
\label{sec:scfalgo}
The discrete nonlinear Hermitian eigenvalue problem is solved self-consistently along with Poisson equations (see equation ~\eqref{eq:discreteSolve}) to compute the Kohn-Sham ground-state solution. Algorithm~\ref{alg:scf} lists all the steps in the SCF procedure followed in DFT-FE. We use adaptive higher order spectral finite-elements in conjunction with computationally efficient and scalable Chebyshev filtered subspace iteration technique (ChFSI)~\cite{saad2006,motamarri2013} to evaluate the occupied eigenspace of the discrete Kohn-Sham Hamiltonian. We further employ Anderson and Broyden schemes~\cite{anderson1965, broyden1965} for electron-density mixing, and the finite-temperature Fermi-Dirac smearing~\cite{VASP} to avoid the charge sloshing associated with metallic systems. 

The ChFSI procedure in Algorithm~\ref{alg:scf} involves the Chebyshev filtering (CF), orthonormalization (CholGS), and the Rayleigh-Ritz procedure (RR). We note that CF scales quadratically with number of atoms, while CholGS and RR scale cubically with number of atoms. Thus, for small to medium scale system sizes CF is the dominant computational cost, while for larger system sizes the computational cost of CholGS and RR dominates. To this end, the numerical implementation in DFT-FE focuses on reducing the prefactor and improving scalability of the ChFSI procedure by exploiting efficient methods and cache-friendly data-structures like FE cell level matrix-matrix multiplications, mixed precision strategies and spectrum splitting approach, as will be discussed subsequently. Furthermore, the electrostatic potentials are computed by solving a Poisson problem, which employs a matrix-free framework of the deal.II finite-element library~\cite{kronbichler2012,dealII90} in conjunction with a Jacobi preconditioned conjugate gradient solver. We note that the above matrix-free framework computes the matrix-vector product of the FE operator on the fly without ever storing it as a sparse matrix. Such on the fly computations benefit from significantly lower memory access costs and have been demonstrated to outperform global sparse-matrix based methods on modern computing architectures~\cite{kronbichler2012}.

\begin{algorithm}
\caption{Self Consistent Field (SCF) iteration in DFT-FE}
\label{alg:scf}
\begin{algorithmic}[1]

\State Compute the self-potentials (${\vself {J^{h_{el}}}}$) corresponding to the nuclear charges by solving the discrete Poisson equations~\eqref{eq:vselfDiscreteSolve}.

\State Compute the discrete pseudopotential projector matrices $C^{J}_{lpm,j}$ (see equation~\eqref{eq:nonlocalHamDiscrete}).

\State Start with an initial guess for $\rho^h_{\rm in}({\bf x})$,  obtained from the superposition of single atom charge densities, and an initial guess for $\bm{\widetilde{\Psi}}$ using single-atom Kohn-Sham DFT wavefunctions.

\State [ES] Get the total electrostatic potential $\varphi^h({\bf x},{\bf R})$ by solving the discrete Poisson equation ~\eqref{eq:phiTotDiscreteSolve}.

\State Get effective potential, $V_{\rm eff,loc}^{h}(\rho^h_{\rm in}({\bf x}),{\bf R}) = V_{\text{xc}}^h + \varphi^{h_{el}} + \sum_{J}(V^{J^{h}}_{\text{loc}} - {\vself {J^{h_{el}}}})$ (see equation~\eqref{ksproblem}).

\State Compute the FE cell level Hamiltonian matrices corresponding to $H_{jk}^{\text{loc}}$ (see equation~\eqref{discreteHam}).  

\State Employ Chebyshev-filtered subspace iteration (ChFSI) method to get the occupied subspace spanning the $N (N > N_{e}/2)$ lowest eigenvectors of ${\bf \widetilde{H}}$ (see equation ~\eqref{hep}).
 \begin{algsubstates}
    \State [CF] Chebyshev filtering of $\bm{\widetilde{\Psi}}$ (see Section~\ref{sec:cheby}).
    \State [CholGS] Orthonormalize the Chebyshev filtered basis $\bm{\widetilde{\Psi}}$ (call Algorithm~\ref{alg:CholGS} in Section~\ref{sec:CholGS}).
    \State [RR] Perform the Rayleigh-Ritz procedure (call Algorithm~\ref{alg:rr} in Section~\ref{sec:rr}).
\end{algsubstates} 

\State [DC] Compute new output electron density, $\rho^h_{\rm out}({\bf x})$ (call Algorithm~\ref{alg:dc} in Section~\ref{sec:rr}).
   
\State    If $\norm{\rho^h_{\rm out}({\bf x}) - \rho^h_{\rm in}({\bf x})} \leq \textrm{tolerance}$, \textit{stop}; Else, compute new $\rho^h_{\rm in}({\bf x})$ using a mixing scheme (see Section \ref{sec:mixing}) and go to step 4.
\end{algorithmic}
\end{algorithm}

\subsection{Chebyshev filtering}\label{sec:cheby}
In DFT-FE, Chebyshev polynomial filtering technique~\cite{zhou2006} is used to adaptively approximate the wanted eigenspace (the lowest $N$ occupied eigenfunctions) of the FE discretized Hamiltonian ${\bf \widetilde{H}}$~\cite{motamarri2013}. In practice, $N$ is typically chosen as $N_e/2+ b$ to allow for finite-temperature Fermi-Dirac smearing, where $b$ is usually $\rm{(5-10) \%}$ of $N_e/2$. In a given SCF iteration step, a scaled Hamiltonian ${\bf \bar{H}}$ is obtained by scaling and shifting ${\bf \widetilde{H}}$  such that the unwanted spectrum of ${\bf \widetilde{H}}$ is mapped on to $[-1, 1]$, and the wanted spectrum is mapped on to $\left(-\infty,-1\right)$ to exploit the fast growth property of Chebyshev polynomials in this region. Subsequently, the action of a degree $m$ Chebyshev polynomial filter, $T_m({\bf \bar{H}})$, on the input subspace, $\bm{\widetilde{\Psi}}$, is computed recursively as
\begin{equation}\label{chebFilter}
T_m({\bf \bar{H}})  \bm{\widetilde{\Psi}}=\left[2{\bf \bar{H}} T_{m-1}({\bf \bar{H}})-T_{m-2}({\bf \bar{H}})\right]\bm{\widetilde{\Psi}}\,.  
\end{equation}
We use an adaptive filtering strategy in which multiple sweeps of ChFSI procedure are performed till the residual norm of the eigenpair closest to the Fermi energy reaches below a specified tolerance $\delta$, chosen to between $1 \times 10^{-2} - 5 \times 10^{-2}$. Our numerical experiments in the case of pseudopotential electronic ground-state calculations show that while multiple calls to ChFSI are triggered in the first few SCF iterations, there is an overall reduction in the number of ChFSI calls (due to reduced number of SCF iterations) when employing the adaptive filtering strategy in comparison to employing a single sweep in all SCF iterations.
We remark that despite using the adaptive filtering strategy, for atomic relaxations or molecular dynamics simulations, multiple Chebyshev filtering calls are typically not triggered as the wavefunctions from the previous electronic ground-state calculation are reused as a starting guess. We note that the choice of the Chebyshev polynomial degree $m$ in equation~\eqref{chebFilter} is based on the upper bound of the spectrum of ${\bf \widetilde{H}}$, which is governed by the smallest mesh size employed in the finite element discretization. A Chebyshev polynomial degree between 20--50 is typically used in DFT-FE for pseudopotential calculations, whereas significantly higher Chebyshev polynomial degrees ($\sim 500-1000$) are required for all-electron calculations.   

\subsubsection{Practical implementation aspects of Chebyshev filtering}
The computational complexity of Chebyshev filtering scales as $\mathcal{O}(MN)$, where $M$ is the size of the discretized Hamiltonian ${\bf \widetilde{H}}$ and $N$ is the number of occupied states. Since Chebyshev filtering is the dominant computational cost in DFT-FE for small to medium sized systems (up to 20,000 electrons), we optimize the core kernel in the Chebyshev filtering procedure, which involves the computation of ${\bf \bar{H}} {\bf X}$ in equation~\eqref{chebFilter}, with ${\bf X}$ denoting a trial subspace in the course of the Chebyshev recursive iteration.  To this end, we first explicitly compute and store the FE
Hamiltonian matrices (cell level Hamiltonian matrices), and subsequently extract the cell level wavefunction matrices from the global wavefunction vectors ${\bf X}$. We then employ BLAS \verb|Xgemm| routines to compute the  matrix-matrix products involving cell Hamiltonian and wavefunction matrices, and assemble them to get the global wavefunction vectors. We note that global FE sparse matrix approaches, particularly when dealing with large number of wavefunction vectors, are more memory-bandwidth limited\footnote{The cell level matrix approach is similar in spirit to matrix-free based approaches, which have been demonstrated to have lower memory access costs than global FE sparse-matrix based methods~\cite{kronbichler2012}.} and incur a higher communication cost\footnote{The global FE sparse matrix framework in deal.II library currently does not take advantage of performing MPI communication of multiple vectors in a single communication call.} than the cell level matrix approach employed above. 

In case of large problems with many thousands of wavefunction vectors, the peak memory during Chebyshev filtering can be quite high if implemented naively by filtering all the wavefunction vectors simultaneously, as multiple temporary memories of size ${\bf X}$ are needed in the course of the Chebyshev recursive iteration. Hence, to reduce the peak memory, we use a blocked approach by filtering blocks of wavefunction vectors, ${\bf X}_b$  with block size denoted by $B^f$, based on the rationale that Chebyshev filtering can be performed on each wavefunction vector independently. Further, the blocked approach also allows us to take advantage of batched \verb|Xgemm| \footnote{Batched operations are efficient for performing many small matrix-matrix multiplications concurrently on multiple threads. Currently such routines are available in vendor optimized BLAS libraries such as Intel MKL.} routines in ${\bf \widetilde{H}}{\bf X}_b$ to perform the aforementioned cell level matrix-matrix products concurrently on multiple threads, which we found to be faster than using multiple threads on standard \verb|Xgemm| calls involving very skewed matrix dimensions when blocked approach is not used. Additionally, we use a single contiguous memory block to store the global wavefunction vectors as well as the block wavefunction vectors, where the data layout is such that for each degree of freedom the corresponding wavefunction values are stored contiguously. This leads to more cache-friendly data access while copying the data between the global wavefunction vectors and the cell wavefunction matrices. Furthermore, we exploit the fact that all wavefunction vectors have identical communication pattern to minimize the total number of MPI point-to-point communication calls in  ${\bf \widetilde{H}}{\bf X}_b$, which reduces the network latency.

The optimal value of the Chebyshev filtering block size, $B^f$, depends on two competing factors---very small sizes lead to higher memory access overheads and communication latency, whereas very large sizes increase peak memory and reduce the efficiency of batched \verb|Xgemm| routines.  Based on numerical experiments, we find the optimal range of $B^f$ to be between 300--400, which is set as the default in DFT-FE. 

\subsection{Cholesky factorization based Gram-Schmidt orthonormalization}\label{sec:CholGS}
ChFSI involves orthonormalization procedure after the Chebyshev filtering step to prevent the ill-conditioning of the filtered vectors in the course of the subspace iteration procedure. This procedure scales cubically with number of electrons and becomes one of the dominant computational costs in large-scale problems (greater than 20,000 electrons). To this end, we employ Cholesky factorization based Gram-Schmidt (CholGS) orthonormalization technique in DFT-FE. This is shown to be more efficient and scalable~\cite{luigi2008,bekas2010} than the commonly used classical Gram-Schmidt procedure. Algorithm~\ref{alg:CholGS} shows the steps involved in the CholGS procedure. The $\mathcal{O}(N^2)$ dot products involved in classical Gram-Schmidt are replaced by more cache-friendly matrix-matrix multiplications in CholGS (steps 1 and 4). Furthermore, the single communication call involved in the computation of overlap matrix ${\bf S}$ in CholGS has a much lower communication latency in comparison to $\mathcal{O}(N^2)$ communication calls in classical Gram-Schmidt.

\begin{algorithm}
\caption{Cholesky-Gram-Schimdt (CholGS) orthonormalization}
\label{alg:CholGS}
\begin{algorithmic}[1]
\State Compute overlap matrix, ${\bf S}=\bm{\widetilde{\Psi}^{\dagger}}\bm{\widetilde{\Psi}}$.\quad ($\mathcal{O}(MN^2)$)

\State Perform Cholesky factorization of the overlap matrix, ${\bf S}={\bf L}{\bf L}^{\dagger}$.\quad ($\mathcal{O}(N^3)$)

\State Compute ${\bf L}^{-1}$.\quad $\mathcal{O}(N^3)$

\State Construct orthonormal basis: ${\bm{{\widetilde{\Psi}}}^{{\rm o}}}={\bm{\widetilde{\Psi}}}{\bf {L^{-1}}^{\dagger}}$. \quad $(\mathcal{O}(MN^2))$
\end{algorithmic}
\end{algorithm}

\subsubsection{Parallel implementation aspects of CholGS in Algorithm~\ref{alg:CholGS}}
\begin{figure}
\includegraphics[width=0.9\textwidth]{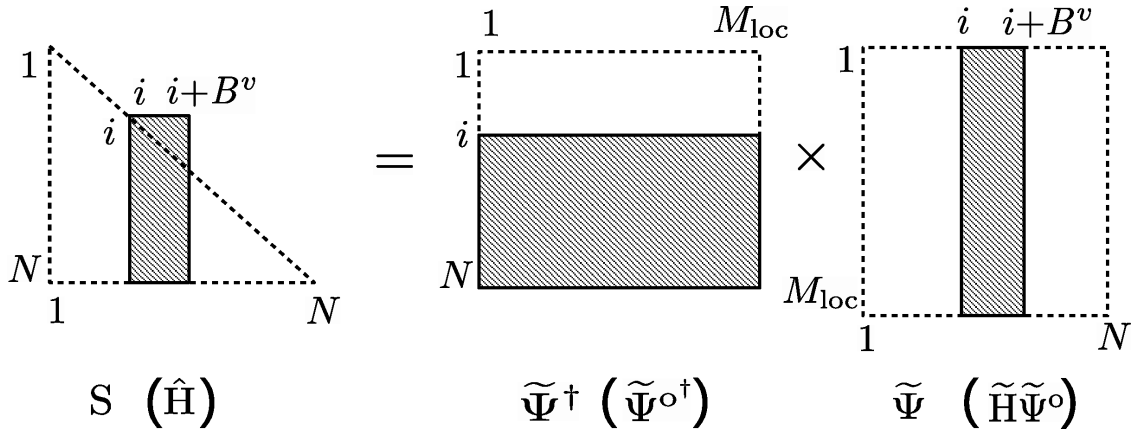}
 \centering
\caption{Blocked approach computation of lower triangular part of the Hermitian overlap matrix, ${\bf S}=\bm{\widetilde{\Psi}^{\dagger}}\bm{\widetilde{\Psi}}$ in Algorithm~\ref{alg:CholGS}, and of the Hermitian projected Hamiltonian, ${\bf \hat{H}}=\bm{{\widetilde{\Psi}}^{{\rm o}^{\dagger}}} {\bf \widetilde{H}} \bm{{\widetilde{\Psi}}^{{\rm o}}}$ in Algorithm~\ref{alg:rr}.}
\label{fig:XtHX}
\end{figure}

\begin{figure}
\includegraphics[width=0.95\textwidth]{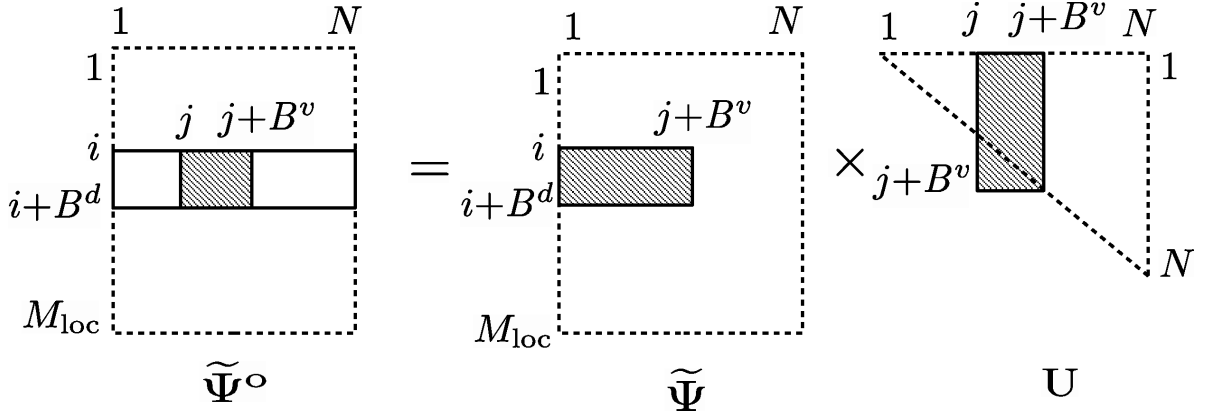}
 \centering
\caption{Two level blocked approach computation of $\bm{{\widetilde{\Psi}}^{\rm o}}=\bm{\widetilde{\Psi}}{\bf U}$, where ${\bf U}$ is an upper triangular matrix.}
\label{fig:LX}
\end{figure}

\paragraph{\small Computation of overlap matrix}
We first note that $\bm{\widetilde{\Psi}}$ is stored in parallel as a $M_{\rm loc} \times N$ matrix, where $M_{\rm loc}$ is the number of FE nodes owned locally by a given MPI task. Accordingly, a straightforward approach to compute the overlap matrix  ${\bf S}$ in step 1 involves the evaluation of local contributions of $\bm{\widetilde{\Psi}^{\dagger}}\bm{\widetilde{\Psi}}$ (a $N \times N$ matrix) on each MPI task, and then accumulating the local contributions to  ${\bf S}$ using the \verb|MPI_Allreduce| collective routine. However, this approach requires memory corresponding to a $N \times N$ matrix on each MPI task, and hence is not practically applicable for large-scale problems $\sim(N>20,000)$. To avoid this large memory footprint in both storage of ${\bf S}$ as well as computation of the local contributions, we use the popular 2D cyclic block grid distribution of ScaLAPACK library~\cite{scalapack1997} to distribute the memory of ${\bf S}$, and use a blocked approach to compute the local contributions of $\bm{\widetilde{\Psi}^{\dagger}}\bm{\widetilde{\Psi}}$ to ${\bf S}$. Further in the blocked approach, we also exploit the Hermiticity of ${\bf S}$, by computing only the lower triangular portion of ${\bf S}$. Fig.~\ref{fig:XtHX} shows the schematic of the blocked approach with block size $B^v$, where $[i,N] \times [i,i+B^v]$ sub-matrices of ${\bf S}$ are computed successively one after another. Computation of each sub-matrix first involves computation of the local contribution in each MPI task by performing matrix-matrix multiplication between $[i,N] \times M_{\rm loc}$ block of $\bm{\widetilde{\Psi}^{\dagger}}$ and $M_{\rm loc} \times [i,i+B^v]$ of $\bm{\widetilde{\Psi}}$ using BLAS \verb|Xgemm| routine, followed by accumulation of the local contributions using the \verb|MPI_Allreduce| collective. Subsequently, the corresponding sub-matrix entries of the ScaLAPACK parallelized ${\bf S}$ are filled. Overall, the above blocked approach combined with ScaLAPACK parallelization of ${\bf S}$ provides both memory optimization and efficiency improvements. 
\vspace{-0.15in}
\paragraph{\small Computation of inverse of Cholesky factor}
Cholesky factorization of ${\bf S}$ in step 2 and inversion of the Cholesky factor ${\bf L}$ in step 3 are performed using ScaLAPACK routines \verb|pXpotrf| and \verb|pXtrtri|, respectively. Based on the numerical experiments conducted on a large benchmark systems, we find that the steps 2 and 3 are a minor cost compared to other steps in CholGS. For instance, the cost of steps 2 and 3 combined contributed to about 7\% of the total wall time for CholGS for a system containing 61,502 electrons (see Fig.~\ref{fig:mixedPrecCholGSTimings}). 
\vspace{-0.1in} 
\paragraph{\small Construction of orthonormal vectors}
Similar to step 1, computation of the orthonormalized basis ${\bm{{\widetilde{\Psi}}}^{{\rm o}}}$ in step 4 also has a large memory footprint when performed simply as a matrix-matrix multiplication between the local portion ($M_{\rm loc} \times N$ matrix) of the parallel distributed ${\bm{\widetilde{\Psi}}}$ and the full ${\bf {L^{-1}}^{\dagger}}$ ($N \times N$ matrix) on every MPI task. For large-scale problems this leads to a high peak memory due to storage of the full ${\bf {L^{-1}}^{\dagger}}$ on every MPI task, and also to store the computed ${\bm{{\widetilde{\Psi}}}^{{\rm o}}}$, which requires the same memory size as ${\bm{\widetilde{\Psi}}}$. Hence, we compute ${\bm{{\widetilde{\Psi}}}^{{\rm o}}}$ using two blocked levels to address both of these memory issues, as shown schematically in Fig.~\ref{fig:LX}. First, we employ an outer blocked level over $M_{\rm loc}$ with block size $B^d$, which allows reuse of the memory of $\bm{{\widetilde{\Psi}}}$ to store ${\bm{{\widetilde{\Psi}}}^{{\rm o}}}$. In particular, we compute $[i,i+B^d] \times [1,N]$ sub-matrices of ${\bm{{\widetilde{\Psi}}}^{{\rm o}}}$ one after the other and copy the orthonormalized sub-matrices back on to $\bm{{\widetilde{\Psi}}}$, thereby requiring only an additional $B^d \times N$ memory. Secondly, we employ an inner blocked level where each $[i,i+B^d] \times [1,N]$ sub-matrix of ${\bm{{\widetilde{\Psi}}}^{{\rm o}}}$ is further divided into $[i,i+B^d] \times [j,j+B^v]$ sub-matrices and successively computed. Similar to the blocked approach used in step 1, this inner blocked level removes the requirement to store the full ${\bf {L^{-1}}^{\dagger}}$ while also exploiting the triangular matrix property of ${\bf {L^{-1}}^{\dagger}}$. Each $[i,i+B^d] \times [j,j+B^v]$ sub-matrix in the inner blocked level is computed by performing a matrix-matrix multiplication between a $[i,i+B^d] \times [1,j+B^v]$ sub-matrix of ${\bm{\widetilde{\Psi}}}$ and a $[1,j+B^v] \times [j,j+B^v]$ sub-matrix of ${\bf {L^{-1}}^{\dagger}}$. We note that ${\bf {L^{-1}}^{\dagger}}$ is stored in a ScaLAPACK parallel format after the end of step 3. Thus to obtain the $[1,j+B^v] \times [j,j+B^v]$ sub-matrix of ${\bf {L^{-1}}^{\dagger}}$ in each MPI task, we first use the local portion of the parallel ${\bf {L^{-1}}^{\dagger}}$ to fill the corresponding entries in the sub-matrix and the rest as zeros, and subsequently use the \verb|MPI_Allreduce| collective to gather and communicate the filled sub-matrix to all MPI tasks. 
\vspace{-0.15in}
\paragraph{\small Remarks on block sizes}
We now discuss few considerations regarding the choice of optimal values for the block sizes $B^{v}$ (used above in steps 1 and 4) and $B^{d}$ (used above in step 4). Too small values of $B^v$ will lead to computational overheads in the \verb|Xgemm| calls due to the highly skewed matrix dimensions which are not cache-friendly, and, further, the total number of MPI collective communication calls will increase leading to higher communication latency. On the other hand too large values of $B^v$ will deprecate the efficiency benefit of exploiting the Hermiticity of ${\bf S}$ in step 1 and the triangular matrix nature of ${\bf {L^{-1}}^{\dagger}}$ in step 4. Based on numerical experiments, we find that value of $B^v$ between 350--500 is optimal. Similarly, the choice of $B^d$ is based on two competing factors--- too small values of $B^d$ incur higher computational and communication overheads due to repeated access of ${\bf {L^{-1}}^{\dagger}}$ for every outer level block computation, whereas larger values increase the peak memory required in step 4. We find that $B^d$ values between 2000--3000 have very negligible overhead costs while still providing memory efficiency when $M_{\rm loc}$ is much larger than $B^d$.

\subsubsection{Mixed precision approaches in CholGS} To further reduce the prefactor of the CholGS algorithm, we make use of mixed precision arithmetic in steps 1 and 4 of Algorithm~\ref{alg:CholGS}, which are the dominant costs in the CholGS algorithm. Mixed precision approaches for orthonormalization in the context of electronic-structure calculations have been explored previously by~\cite{tsuchida2012}. We first develop a mixed precision approach for step 1, where the computation of the overlap matrix, {\bf S} can be split into computation of the diagonal and the off-diagonal parts:
\begin{equation}\label{eq:overlapmixedprec}
  {\bf S}={\bf S_{d}} + {\bf S_{od}},
\end{equation}
where ${\bf S_d}$ is a matrix containing the diagonal entries of ${\bf S}$. We take advantage of the fact that ${\bf S_{od}} \to \bf{0}$ as the SCF approaches convergence and hence compute ${\bf S_{od}}$ using single precision BLAS \verb|Xgemm| routines, while the computation of diagonal entries of ${\bf S_d}$ is performed using double precision BLAS routines at negligible computational cost.  Similarly, step 4 can be split into
\begin{equation}
\bm{{\widetilde{\Psi}}^{\rm o}}={\bm{\widetilde{\Psi}}}{\bf {L^{-1}_d}^{\dagger}}+{\bm{\widetilde{\Psi}}}{\bf {L^{-1}}_{od}^{\dagger}},
\end{equation}
where ${\bf {L^{-1}_d}^{\dagger}}$ is a matrix containing the diagonal entries of ${\bf {L^{-1}}^{\dagger}}$.  Taking advantage of the fact that ${\bf {L^{-1}_{od}}^{\dagger}} \to \bf{0}$ as the SCF approaches convergence, we compute ${\bm{\widetilde{\Psi}}}{\bf {L^{-1}_{od}}^{\dagger}}$
using single precision BLAS \verb|Xgemm| routines, while the computation of ${\bm{\widetilde{\Psi}}}{\bf {L^{-1}_d}^{\dagger}}$ is performed as a double precision scaling operation at negligible computational cost. We remark that, in addition to the reduction of computational costs, the use of mixed precision also reduces the communication costs in steps 1 and 4 as the \verb|MPI_Allreduce| collectives employed in these steps communicate the relevant single precision data with half the MPI message size (bytes), in comparison to their double precision counterparts. 

\begin{figure}
\includegraphics[scale=0.5]{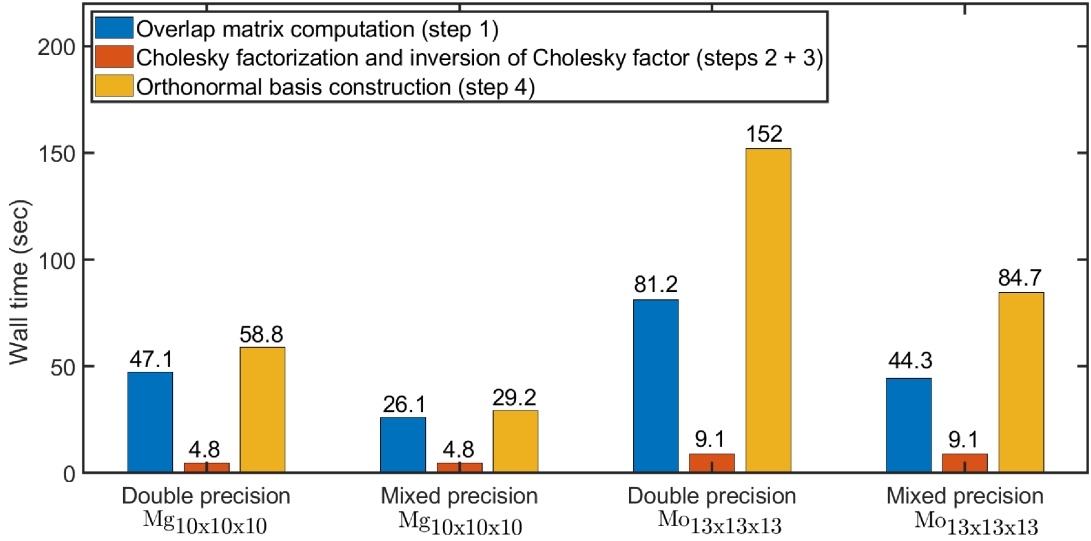}
 \centering
\caption{Comparison of CholGS algorithm (Algorithm~\ref{alg:CholGS})  wall times for a single SCF step in using mixed precision arithmetic in steps 1 and 4. Case studies: (i) $\textrm{Mg}_{\textrm{10x10x10}}$  with 39,990 electrons run on 51,200 MPI tasks and (ii) $\textrm{Mo}_{\textrm{13x13x13}}$ with 61,502 electrons run on 64,000 MPI tasks.}
\label{fig:mixedPrecCholGSTimings}
\end{figure}

The computational cost reduction in steps 1 and 4 of the mixed precision approach is demonstrated in Fig.~\ref{fig:mixedPrecCholGSTimings} for large-scale benchmark problems involving 39,900 and 61,502 electrons. We find this approach to be around 2 times faster in comparison to double precision approach. Furthermore, we also examine the accuracy and robustness of the mixed precision algorithm in the overall SCF convergence in Section~\ref{sec:mixedPrecRR}, and is discussed in detail subsequently.

\subsection{Rayleigh-Ritz procedure and electron-density computation}\label{sec:rr}
Rayleigh-Ritz (RR) procedure in ChFSI involves the following steps: i) computation of the projected Hamiltonian, ${\bf \hat{H}}=\bm{{\widetilde{\Psi}}^{{\rm o}^{\dagger}}} {\bf \widetilde{H}} \bm{{\widetilde{\Psi}}^{{\rm o}}}$ into the space spanned by the orthonormalized wavefunctions $\bm{{\widetilde{\Psi}}^{{\rm o}}}$, ii) diagonalization of ${\bf \hat{H}}$: ${\bf \hat{H}} \bm{{\rm Q}}=\bm{{\rm Q}}\bm{{\rm D}}$, where $\bm{{\rm D}}$ contains all the eigenvalues of ${\bf \hat{H}}$ in ascending order and $\bm{ {\rm Q}}$ contains the corresponding eigenvectors, iii) subspace rotation of $\bm{{\widetilde{\Psi}}^{{\rm o}}}$: $\bm{{\widetilde{\Psi}}}^{\bf R}=\bm{{\widetilde{\Psi}}^{{\rm o}}} \bm{{\rm Q}}$. 
Subsequently, the output electron-density at a point ${\bf x}$ belonging to a FE cell $e$ is computed as
\begin{equation}\label{eq:elecDensSums}
\rho^h_{\rm out}({\bf x})= 2\sum_{i=1}^{N} f(\epsilon^h_i,\mu)|\psi^h_i(\bx)|^2
=2\sum_{i=1}^{N} \left[f(\epsilon^h_i,\mu) \left(\sum_{j=1}^{M_e} {{\psi}}_{i}^{e,j} N^e_j({\bf x}) \right) \left( \sum_{k=1}^{M_e} {{{\psi}}^{{e,k}^*}_{i}} N^e_k({\bf x}) \right) \right],
\end{equation}
where ${\left\{N^e_1({\bf x}) ,\,N^e_2({\bf x}), \, \cdots \, N^e_{M_e}({\bf x})\right\}}$ denote the FE basis functions associated with the given cell ($M_e$ denoting the number of nodes in the cell),  and  ${\left\{{\psi}_i^{e,1}, \,{\psi}_i^{e,2}, \, \cdots \, {\psi}_i^{e,M_e}\right\}}$ denote the corresponding nodal values of the $i^{\rm th}$ wavefunction, $\bm{{\psi}}^h_i({\bf x})$ in FE cell $e$. 
Using the subspace rotated wavefunctions $\bm{{\widetilde{\Psi}}}^{\bf R}$, Equation~\ref{eq:elecDensSums} can be re-written as
\begin{equation}\label{eq:elecDens}
\rho^h_{\rm out}({\bf x})=2 \,\bm{n^{{e}^{T}}}({\bf x})
 \bm{{\hat{\Psi}}^{\bf R}_e} f \left( {\bf D}, \mu \right) \bm{{\hat{\Psi}}_e^{{\bf R}^{\dagger}}}
\bm{n^e}({\bf x}),
\end{equation}
where  \begin{equation}
\bm{n^e}({\bf x})={\left[N^e_1({\bf x}) \,N^e_2({\bf x}) \, \cdots \, N^e_{M_e}({\bf x})\right]}^T,
\end{equation}
and the matrix $\bm{{\hat{\Psi}}^{\bf R}_e}$ contains the FE cell column vectors extracted from $\bm{\hat{\Psi}}^{\bf R}$ which is given by
\begin{equation}\label{eq:scaling}
\bm{{\hat{\Psi}}^{\bf R}}= {\bf M}^{-1/2}\bm{{\widetilde{\Psi}}}^{\bf R}. 
\end{equation}

In the above, the computational complexity of steps i), ii) and iii) of the Rayleigh-Ritz procedure scales as $\mathcal{O}(MN^2)$, $\mathcal{O}(N^3)$, and $\mathcal{O}(MN^2)$, respectively, while the electron-density computation scales as $\mathcal{O}(MN)$. Rayleigh-Ritz procedure is, thus, one of the significant bottlenecks for large-scale problems. To this end, we employ two strategies in DFT-FE: spectrum-splitting and mixed precision, to reduce the prefactor of Rayleigh-Ritz procedure, as discussed below.

\subsubsection{Spectrum-splitting in RR}
\begin{algorithm}
\caption{Spectrum-splitting based Rayleigh Ritz procedure (RR)}
\label{alg:rr}
\begin{algorithmic}[1]
\State Compute ${\bf \hat{H}}=\bm{{\widetilde{\Psi}}^{{\rm o}^{\dagger}}} {\bf \widetilde{H}} \bm{{\widetilde{\Psi}}^{{\rm o}}}$.
\State Compute  $N_{\rm fr}$ largest eigenstates of ${\bf \hat{H}}$: ${\bf \hat{H}}\bm{{\rm Q}_{\rm fr}}=\bm{{\rm Q}_{\rm fr}}\bm{{\rm D}_{\rm fr}}$.
\State Subspace rotation to compute fractionally occupied eigenstates: $\bm{{\widetilde{\Psi}}_{\rm fr}}^{\bf R}=\bm{{\widetilde{\Psi}}^{{\rm o}}} \bm{{\rm Q}_{\rm fr}}$.
\end{algorithmic}
\end{algorithm}

The key idea behind spectrum-splitting is that the eigenvalues and eigenvectors of the projected Hamiltonian ${\bf \hat{H}}$ with orbital occupancy function $f_i=1$ are not explicitly necessary for the computation of the electron-density in equation~\eqref{eq:elecDens}. This can be exploited to achieve significant computational savings when most of the Kohn-Sham states are fully occupied as is the case for typically used Fermi-Dirac smearing temperatures $\sim 500 \,{\rm K}$.  Such methods have been developed in previous works in the context of both pseudopotential~\cite{rescu2016,amartya2018} and all-electron DFT~\cite{motamarri2017} calculations, which we have adapted in DFT-FE. Furthermore, we additionally take advantage of spectrum-splitting to develop a mixed precision technique to reduce the computational cost of the projected Hamiltonian computation. We discuss below the implementation of the spectrum-splitting algorithm in DFT-FE.

Let $N_{\rm oc}$ denote the number of Kohn-Sham eigenstates with full occupancies ($f_i=1$), and $N_{\rm fr} = N - N_{\rm oc}$ denote the number of remaining states with partial occupancies. We consider the following split in the diagonalization of ${\bf \hat{H}}$:
\begin{equation}
{\bf \hat{H}}
=
\left[
\begin{array}{c|c}
{\bf Q_{oc}} & {\bf Q_{fr}}
\end{array}
\right] 
\left[
\begin{array}{c|c}
{\bf D_{oc}} & {\bf 0}  \\
\hline
{\bf 0} & {\bf D_{fr}}
\end{array}
\right]
\left[
\begin{array}{c}
{\bf Q_{oc}^{\dagger}} \\
\hline
{\bf Q_{fr}^{\dagger}}
\end{array}
\right],
\end{equation}
where ${\bf Q_{oc}}$ contains the eigenvectors corresponding to $N_{\rm oc}$ eigenvalues of ${\bf \hat{H}}$, which are stored as the diagonal entries of ${\bf D_{oc}}$. On the other hand, ${\bf Q_{fr}}$ contains the eigenvectors corresponding to remaining $N_{\rm fr}$ eigenvalues of ${\bf \hat{H}}$, which are stored as the diagonal entries of ${\bf D_{fr}}$. Similarly $f \left( {\bf D}, \mu \right)$ can be split as
\begin{equation}\label{eq:fSplit}
 f \left( {\bf D}, \mu \right)
 =
  \left[
 \begin{array}{c|c}
 f \left({\bf D_{oc}},\mu\right) & {\bf 0}  \\
\hline
{\bf 0} & f \left({\bf D_{fr}},\mu \right)
\end{array}
\right].
\end{equation}
Using the above equation~\eqref{eq:fSplit} along with the scaling step in equation~\eqref{eq:scaling} and subspace rotation: $\bm{{\widetilde{\Psi}}}^{\bf R}=\bm{{\widetilde{\Psi}}^{{\rm o}}} \bm{{\rm Q}}$, equation~\eqref{eq:elecDens} can be written as
\begin{equation}\label{eq:rhoSpectrumSplitIntermediate}
\rho^h_{\rm out}({\bf x})=2\, \bm{n^{{e}^{T}}}({\bf x})
   \bm{{\hat{\Psi}}_{\bm e}^{{\rm o}}}
 \left[
\begin{array}{c|c}
{\bf Q_{oc}} & {\bf Q_{fr}}
\end{array}
\right] 
  \left[
 \begin{array}{c|c}
 f \left({\bf D_{oc}},\mu\right) & {\bf 0}  \\
\hline
{\bf 0} & f \left({\bf D_{fr}},\mu \right)
\end{array}
\right]
\left[
\begin{array}{c}
{\bf Q_{oc}^{\dagger}} \\
\hline
{\bf Q_{fr}^{\dagger}}
\end{array}
\right]
\bm{{\hat{\Psi}}_e^{{{\rm o}}^{\dagger}}}
\bm{n^e}({\bf x}),    
\end{equation}
where $\bm{{\hat{\Psi}}^{{\rm o}}_{\bm e}}$ denotes the FE cell level vectors of ${\bm{{\hat{\Psi}}^{\rm o}}}={\bf M}^{-1/2} \bm{{\widetilde{\Psi}}^{{\rm o}}}$.
We note that $f \left({\bf D_{oc}},\mu\right)={\bf I_{oc}}$, an $N_{\rm oc} \times N_{\rm oc}$ identity matrix and hence  equation~\eqref{eq:rhoSpectrumSplitIntermediate} can be recast in the following way:
\begin{align}
\rho^h_{\rm out}({\bf x})= & 2 \,\bm{n^{{e}^{T}}}({\bf x})
 \bm{{\hat{\Psi}}_{\bm e}^{{\rm o}}}
 \left[
\begin{array}{c|c}
{\bf Q_{oc}} & {\bf Q_{fr}}
\end{array}
\right] 
  \left[
  {\bf I}+
  \left(
 \begin{array}{c|c}
 {\bf 0} & {\bf 0}  \\
\hline
{\bf 0} & f \left({\bf D_{fr}},\mu \right)-{\bf I_{fr}}
\end{array}
\right)
\right]
\left[
\begin{array}{c}
{\bf Q_{oc}^{\dagger}} \\
\hline
{\bf Q_{fr}^{\dagger}}
\end{array}
\right]
\bm{{\hat{\Psi}}_e^{{{\rm o}}^{\dagger}}}
\bm{n^e}({\bf x})\notag\\
=& 2 \,\bm{n^{{e}^{T}}}({\bf x})\left[  \bm{{\hat{\Psi}}_{\bm e}^{{\rm o}}} \bm{{\hat{\Psi}}_{\bm e}^{{\rm o}^{\dagger}}}
+ \bm{{\hat{\Psi}}^{{\rm o}}_{\bm e}} {\bf Q_{fr}} \left(f\left(\bm{{\rm D}_{\rm fr}},\mu \right) -{\bf I_{fr}} \right){\bf Q_{fr}^{\dagger}}\bm{{\hat{\Psi}}_{\bm e}^{{\rm o}^{\dagger}}}\right]\bm{n^e}({\bf x})\notag\\
=& 2 \,\bm{n^{{e}^{T}}}({\bf x})\left[  \bm{{\hat{\Psi}}_{\bm e}^{{\rm o}}} \bm{{\hat{\Psi}}_{e}^{{{\rm o}}^{\dagger}}}
+ \bm{{\hat{\Psi}}_{{\rm fr},e}}^{\bf R} \left(f\left(\bm{{\rm D}_{\rm fr}},\mu \right) -{\bf I_{fr}} \right)\bm{{\hat{\Psi}}_{{\rm fr},e}^{{\bf R}^{\dagger}}}\right]\bm{n^e}({\bf x}),
\label{eq:elecDensSP}
\end{align}
where $\bm{{\hat{\Psi}}_{{\rm fr},e}}^{\bf R}$ denotes the FE cell level vectors of $\bm{{\hat{\Psi}}_{\rm fr}}^{\bf R}= {\bf M}^{-1/2}\bm{{\widetilde{\Psi}}^{{\rm o}}} \bm{{\rm Q}_{\rm fr}}$.
\begin{algorithm}
\caption{Electron-density computation (DC)}
\label{alg:dc}
\begin{algorithmic}[1]
\State Compute Fermi-energy ($\mu$) using the constraint:
\begin{equation*}
2\left(N_{\rm oc}+\sum_{i=N_{\rm oc}}^{N}f(\epsilon_i^h,\mu)\right)=N_e.
\end{equation*}
\State Scale $\bm{{\widetilde{\Psi}}^{{\rm o}}}$ and  $\bm{{\widetilde{\Psi}}_{\rm fr}}^{\bf R}$: $\bm{{\hat{\Psi}}^{{\rm o}}}= {\bf M}^{-1/2}\bm{{\widetilde{\Psi}}^{{\rm o}}},\,
 \bm{{\hat{\Psi}}_{\rm fr}}^{\bf R}= {\bf M}^{-1/2}\bm{{\widetilde{\Psi}}_{\rm fr}}^{\bf R}$.
\State Compute electron density using equation~\eqref{eq:elecDensSP}:
\begin{equation*}
\rho^h_{\rm out}({\bf x})=2 \,\bm{n^{{e}^{T}}}({\bf x})\left[ {\bm{{\hat{\Psi}}_{\bm e}^{{\rm o}}}}\bm{{\hat{\Psi}}_{e}^{{{\rm o}}^{\dagger}}}
+\bm{{\hat{\Psi}}_{{\rm fr},e}}^{\bf R}\left(f\left(\bm{{\rm D}_{\rm fr}},\mu \right) -{\bf I_{fr}} \right)\bm{{\hat{\Psi}}_{{\rm fr},e}^{{\bf R}^{\dagger}}}\right]\bm{n^e}({\bf x}).
\end{equation*}
\end{algorithmic}
\end{algorithm}

\begin{figure}
\includegraphics[scale=0.5]{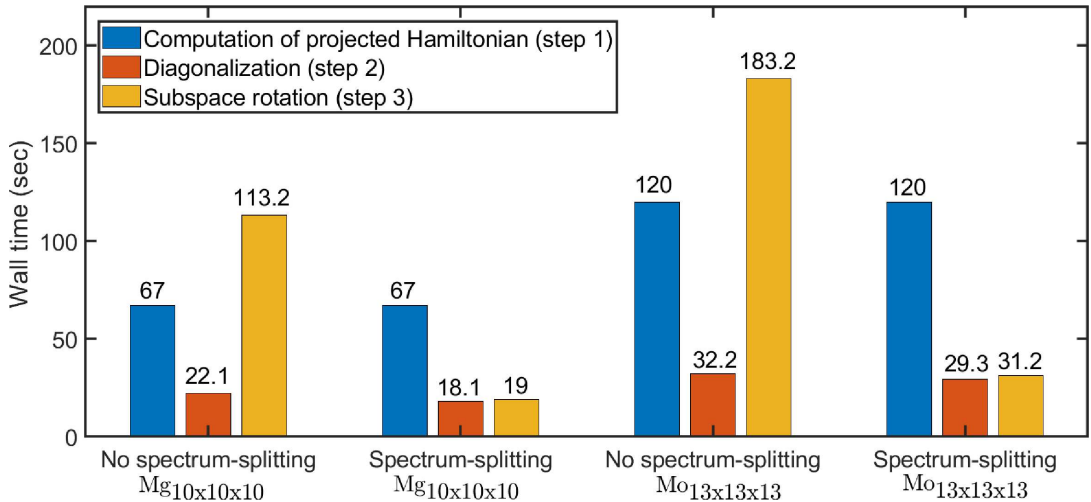}
 \centering
\caption{Comparison of Rayleigh-Ritz procedure wall times for a single SCF step by using spectrum-splitting (Algorithm~\ref{alg:rr}). Case studies: (i) $\textrm{Mg}_{\textrm{10x10x10}}$  with 39,990 electrons run on 51,200 MPI tasks and (ii) $\textrm{Mo}_{\textrm{13x13x13}}$ with 61,502 electrons run on 64,000 MPI tasks. $N_{\rm fr}$ for both case studies is 15\% of $N$}
\label{fig:specSplitRRTimings}
\end{figure}

In the above, it is evident that the electron-density computation requires only the $N_{\rm fr}$ largest eigenstates of ${\bf \hat{H}}$. Accordingly, the spectrum-splitting based algorithms for the Rayleigh-Ritz procedure and electron-density computation in DFT-FE are given in Algorithm~\ref{alg:rr} and Algorithm~\ref{alg:dc}, respectively. Even with finite-temperature Fermi-Dirac smearing, $N_{\rm fr}$ is usually a small fraction of $N$. From our numerical experiments, we find that $N_{\rm fr}$ is 10--15\% of $N$ for metallic systems, and much smaller percentage ($<5\%$) for insulating and semi-conducting systems. This translates to significant cost savings in the subspace rotation step as shown in Fig.~\ref{fig:specSplitRRTimings}. This is because the usual full subspace rotation: $\bm{{\widetilde{\Psi}}}^{\bf R}=  \bm{{\widetilde{\Psi}}^{{\rm o}}} \bm{{\rm Q}}$, which scales as $\mathcal{O}(MN^2)$ is now replaced by a significantly cheaper partial subspace rotation step: $\bm{{\widetilde{\Psi}}_{\rm fr}}^{\bf R}= \bm{{\widetilde{\Psi}}^{{\rm o}}} \bm{{\rm Q}_{\rm fr}}$ (step 3 of Algorithm~\ref{alg:rr}), which scales as $\mathcal{O}(MNN_{\rm fr})$. Furthermore, step 2, which now amounts to a partial diagonalization of ${\bf \hat{H}}$ to compute the $N_{\rm fr}$ largest eigenstates, can be exploited to reduce diagonalization cost. In the literature, iterative approaches like LOBPCG~\cite{rescu2016}, and inner Chebyshev filtering~\cite{amartya2018} are shown to be better than ScaLAPACK's direct eigensolver for partial diagonalization. However, iterative approaches may not be robust for metallic systems in the limit of vanishing band gaps. Hence in DFT-FE, we perform partial diagonalization using the ELPA library's~\cite{elpa2014,elpaCray, elpaOpt} direct eigensolver, which is more scalable than ScaLAPACK's eigensolver and competes with the aforementioned iterative approaches with respect to minimum solution time. Fig.~\ref{fig:specSplitRRTimings} shows the direct diagonalization times\footnote{The ELPA diagonalization times quoted here are run on NERSC Cori Intel KNL nodes which have 1.4 GHz clock frequency. On a higher clock frequency machine (eg: IBM Power and Intel Skylake architectures), these diagonalization timings are faster by a factor of 2--3~\cite{elpaOpt}.} (step 2) for very large system sizes with 39,990 electrons and 61,502 electrons.

\subsubsection{Mixed precision in RR}\label{sec:mixedPrecRR}
\begin{figure}
\includegraphics[scale=0.5]{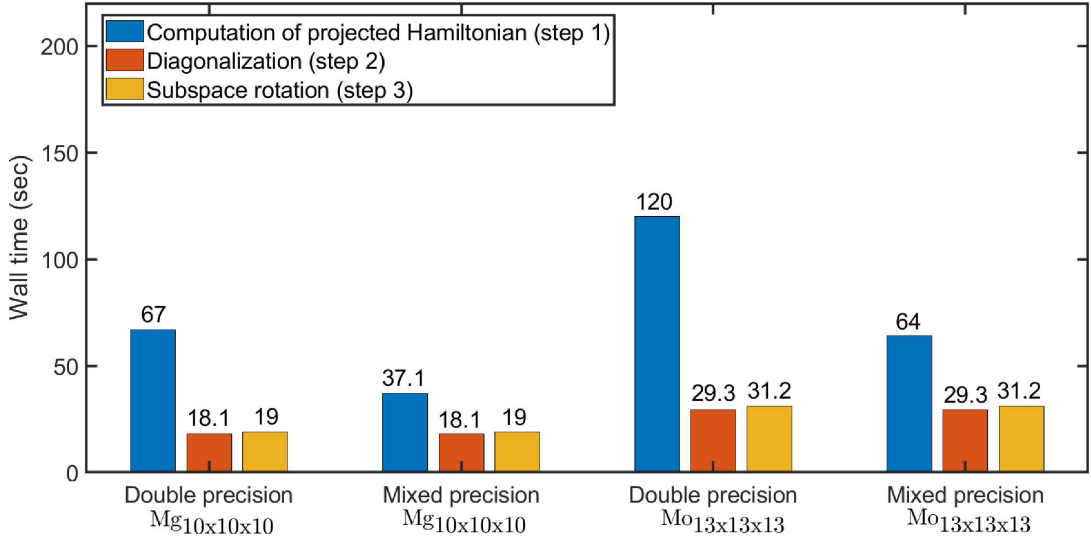}
 \centering
\caption{Comparison of Rayleigh-Ritz procedure (Algorithm~\ref{alg:rr}) wall times for a single SCF step by using mixed precision arithmetic in the computation of projected Hamiltonian. Case studies: (i) $\textrm{Mg}_{\textrm{10x10x10}}$  with 39,990 electrons run on 51,200 MPI tasks and (ii) $\textrm{Mo}_{\textrm{13x13x13}}$ with 61,502 electrons run on 64,000 MPI tasks.}
\label{fig:mixedPrecRRTimings}
\end{figure}
We observe that the computation of the projected Hamiltonian, ${\bf \hat{H}}=\bm{{{\widetilde{\Psi}}^{{\rm o}}}^{\dagger}} {\bf \widetilde{H}} \bm{{\widetilde{\Psi}}^{{\rm o}}}$ is the most dominant cost in the Rayleigh Ritz procedure using the spectrum-splitting technique (see Fig.~\ref{fig:specSplitRRTimings}). To this end, we develop a mixed precision  algorithm to reduce the prefactor of the computation of ${\bf \hat{H}}$ and illustrate the procedure here. 
We first consider the following split of the orthonormalized wavefunctions  $\bm{{\widetilde{\Psi}}^{{\rm o}}}$
\begin{equation}
\bm{{\widetilde{\Psi}}^{{\rm o}}}=\left[\bm{{\widetilde{\Psi}}_{\rm oc}^{{\rm o}}} \, \bm{{\widetilde{\Psi}}_{\rm fr}^{{\rm o}}} \right],    
\end{equation}
where the columns $\bm{{\widetilde{\Psi}}_{\rm oc}^{{\rm o}}}$ and $\bm{{\widetilde{\Psi}}_{\rm fr}^{{\rm o}}}$ contain the first $N_{\rm oc}$ and the remaining $N_{\rm fr}$ wavefunctions, respectively. We next rewrite the partial eigendecomposition of  ${\bf \hat{H}}:{\bf \hat{H}}\bm{{\rm Q}_{\rm fr}}=\bm{{\rm Q}_{\rm fr}}\bm{{\rm D}_{\rm fr}}$ (see step 2 of Algorithm~\ref{alg:rr}) as
\begin{equation}\label{eq:partialDiagMixedPrec1}
\left[
\begin{array}{c|c}
{\bf \hat{H}}_{\bf oc-oc} & {\bf \hat{H}}_{\bf oc-fr}  \\
\hline
{\bf \hat{H}}_{\bf fr-oc} & {\bf \hat{H}}_{\bf fr-fr}
\end{array}
\right]
\left[
\begin{array}{c}
\bm{{\rm Q}_{\rm fr}^{a}}  \\
\hline
\bm{{\rm Q}_{\rm fr}^{b}}  
\end{array}
\right]
=
\left[
\begin{array}{c}
\bm{{\rm Q}_{\rm fr}^{a}}  \\
\hline
\bm{{\rm Q}_{\rm fr}^{b}}  
\end{array}
\right]
\bm{{\rm D}_{\rm fr}},
\end{equation}
where ${\bf \hat{H}}_{\bf oc-oc}=\bm{{\widetilde{\Psi}}_{\rm oc}^{{\rm o}^{\dagger}}} {\bf \widetilde{H}} \bm{{\widetilde{\Psi}}_{\rm oc}^{{\rm o}}}$,  ${\bf \hat{H}}_{\bf fr-fr}=\bm{{\widetilde{\Psi}}_{\rm fr}^{{\rm o}^{\dagger}}} {\bf \widetilde{H}} \bm{{\widetilde{\Psi}}_{\rm fr}^{{\rm o}}}$,  ${\bf \hat{H}}_{\bf fr-oc}=\bm{{\widetilde{\Psi}}_{\rm fr}^{{\rm o}^{\dagger}}} {\bf \widetilde{H}} \bm{{\widetilde{\Psi}}_{\rm oc}^{{\rm o}}}$, and ${\bf \hat{H}}_{\bf oc-fr}=\bm{{\widetilde{\Psi}}_{\rm oc}^{{\rm o}^{\dagger}}} {\bf \widetilde{H}} \bm{{\widetilde{\Psi}}_{\rm fr}^{{\rm o}}}$. As the SCF approaches convergence, $\bm{{\widetilde{\Psi}}^{{\rm o}}}$ tends to the eigenfunctions of ${\bf \widetilde{H}}$, and hence the limiting behaviour of equation~\eqref{eq:partialDiagMixedPrec1} can be written as
\begin{equation}\label{eq:partialDiagLim1}
\left[
\begin{array}{c|c}
{\bf \hat{H}}_{\bf oc-oc} \to {\bf D_{oc}} & {\bf \hat{H}}_{\bf oc-fr} \to {\bf 0} \\
\hline
{\bf \hat{H}}_{\bf fr-oc} \to {\bf 0} & {\bf \hat{H}}_{\bf fr-fr} \to {\bf D_{fr}}
\end{array}
\right]
\left[
\begin{array}{c}
\bm{{\rm Q}_{\rm fr}^{a}} \to {\bf 0} \\
\hline
\bm{{\rm Q}_{\rm fr}^{b}}  \to {\bf I_{fr}}
\end{array}
\right]
= 
\left[
\begin{array}{c}
\bm{{\rm Q}_{\rm fr}^{a}} \to {\bf 0} \\
\hline
\bm{{\rm Q}_{\rm fr}^{b}}  \to {\bf I_{fr}}
\end{array}
\right]
\bm{{\rm D}_{\rm fr}}.
\end{equation}
Using equation~\eqref{eq:partialDiagLim1} the limiting behaviour of the partial eigendecomposition of  ${\bf \hat{H}}:{\bf \hat{H}}\bm{{\rm Q}_{\rm fr}}=\bm{{\rm Q}_{\rm fr}}\bm{{\rm D}_{\rm fr}}$ is written as
\begin{equation}\label{eq:partialDiagLim2}
 {\bf \hat{H}}_{\bf fr-fr} \bm{{\rm Q}_{\rm fr}^{b}} = \bm{{\rm Q}_{\rm fr}^{b}}  \bm{{\rm D}_{\rm fr}}.
\end{equation}
Equation~\eqref{eq:partialDiagLim2} provides the rationale to design a mixed precision algorithm to compute ${\bf \hat{H}}$ by employing double precision BLAS \verb|Xgemm| routine to compute the ${\bf \hat{H}}_{\bf fr-fr}$ sub-matrix, while all the other sub-matrices: ${\bf \hat{H}}_{\bf oc-oc}$, ${\bf \hat{H}}_{\bf fr-oc}$ and ${\bf \hat{H}}_{\bf oc-fr}$  are computed using single precision BLAS \verb|Xgemm| routine. Since $N_{\rm fr}$ is typically less than 15\% of $N$, the computation of ${\bf \hat{H}}_{\bf fr-fr}$  using double precision is a very small computational cost in this approach. This leads to overall computational savings by a factor of around 2 in computation of ${\bf \hat{H}}$ as shown in Fig.~\ref{fig:mixedPrecRRTimings}. Additionally, in Table~\ref{tab:mixedPrecAccuracy}, we examine the accuracy and robustness in employing mixed precision algorithms for both Rayleigh-Ritz and orthonormalization (see Section~\ref{sec:CholGS}) steps on various benchmark systems in DFT-FE. These benchmark system are chosen such that the FE discretization errors are $\sim 10^{-4}$ Ha/atom in ground-state energy and $\sim 10^{-4}$ Ha/Bohr in ionic forces. The results in Table~\ref{tab:mixedPrecAccuracy} show that number of SCFs do not change between mixed precision and double precision approaches, and further the mixed precision algorithms incur negligible errors in both energies and forces in comparison to the double precision calculations. Notably, these errors are about two orders of magnitude lower than the discretization errors.

In addition to using mixed precision in computation of ${\bf \hat{H}}$, we also use a blocked approach for memory and computational efficiency improvements. The computational efficiency improvement in using the blocked approach arises from exploiting the Hermiticity of ${\bf \hat{H}}$ as shown in Fig.~\ref{fig:XtHX}. The implementation of the blocked approach used here is similar to the implementation of the blocked approach in the overlap matrix computation (see Section~\ref{sec:CholGS}). Finally, we remark that the use of spectrum splitting technique in conjunction with the mixed precision algorithm in the Rayleigh-Ritz procedure provides efficiency gains by a factor of around 3 for the large benchmark systems considered in Fig.~\ref{fig:specSplitRRTimings} and ~\ref{fig:mixedPrecRRTimings}.

\begin{table}[htbp]
\centering
\small
\caption{\label{tab:mixedPrecAccuracy}Accuracy and robustness study of mixed precision computations in CholGS orthonormalization and Rayleigh-Ritz procedure on benchmark systems. Energy difference, maximum atomic force difference magnitude ($\max\limits_{1\le i \le N_{a}} \left\|{\bf f}^i_{\rm dp} -{\bf f}^i_{\rm sp}\right\|$) and total number of SCFs are reported with respect to double precision calculations. ${\bf f}^i_{\rm dp}$ and ${\bf f}^i_{\rm sp}$ denote atomic force on ${i}^{\rm th}$ atom for double precision and single precision calculations respectively. Discretization errors for the benchmark systems are $\sim 10^{-4}$ Ha/atom in ground-state energy and $\sim 10^{-4}$ Ha/Bohr in ionic forces. More details about the benchmark systems are given in Section~\ref{sec:timing}.}
\begin{tabular}{|c|c|c|c|c|}
\hline
System & Energy difference & Maximum force difference & Total SCFs \\
       &  (Ha/atom)  & magnitude (Ha/Bohr)            & (Double, Mixed) \\
\hline
$\textrm{Mg}_{\textrm{6x6x6}}$ &  $7 \times 10^{-12}$ & $2 \times 10^{-6}$ &  (49, 49)\\
\hline
$\textrm{Cu}_{\textrm{4-shell}}$& $5 \times 10^{-12}$ & $3 \times 10^{-6}$  & (46, 46)\\
\hline
$\textrm{Mo}_{\textrm{6x6x6}}$&  $3 \times 10^{-12}$ & $7 \times 10^{-7}$  & (49, 49)\\
\hline
\end{tabular}
\end{table}

\subsection{Mixing schemes}\label{sec:mixing}
The SCF iteration procedure for solving the Kohn-Sham eigenvalue problem can be viewed as a fixed point iteration in electron density or effective potential. In terms of electron density, this fixed point problem can be written as $\rho=F[\rho]$,
where $F[\rho]$ involves computing the occupied eigenspace for a given $\rho$. This fixed point iteration can be accelerated by mixing the electron density using an appropriate mixing scheme~\cite{anderson1965, broyden1965, Kerker1981, eyert1996, kudin2002, linlin2013b, zhou2018}. In the present DFT-FE software release, we implement two kinds of mixing schemes: n-stage Anderson mixing scheme \cite{anderson1965} and Broyden mixing scheme~\cite{broyden1965}. We have used Anderson mixing scheme in all the simulations conducted in the present work. In a future release, we plan to implement more advanced mixing strategies~\cite{Kerker1981,linlin2013b}, which provide improved SCF convergence rate independent of system size. 
\subsection{Parallelization}
\label{sec:paral}
The primary level of parallelization in the DFT--FE code is based on domain decomposition of the adaptive FE mesh into partitions and distributing them to different MPI tasks. This is accomplished through the deal.II finite element library with p4est~\cite{BangerthBursteddeHeisterEtAl11}. We note that the FE basis is localized with a compact support on the cells shared by a FE node. Hence only the FE nodes on the processor boundaries need to be communicated, which has a significantly smaller communication cost in comparison to the all-to-all communication required in global basis sets like plane-waves. This allows for excellent parallel scalability of DFT--FE, which we demonstrate subsequently in Section~\ref{sec:scaling}. 

To further improve scalability, we implement a second level of parallelization over wavefunctions (band parallelization) in each of the key computational steps of the SCF iteration: Chebyshev filtering, CholGS (steps 1 and 4 of Algorithm~\ref{alg:CholGS}), Rayleigh--Ritz procedure (steps 1 and 3 of Algorithm~\ref{alg:rr}), and electron density computation. In particular, computations over the total number of wavefunctions ($\bm{\widetilde{\Psi}}$)  are divided into groups of wavefunctions (band groups) and distributed among a group of MPI sub--communicators, with each sub--communicator doing computations on a single band group of size $N_{b_i}$, denoted by $\bm{\widetilde{\Psi}_{b_i}}$, where $i$ denotes the band group index. The use of band parallelization in Chebyshev filtering and electron density computation does not involve any inter band group communication of $\bm{\widetilde{\Psi}_{b_i}}$'s. However, the computation of the electron density requires accumulation of the electron density contribution from each band group incurring a very small communication cost. Further, we exploit band parallelization in the computation of overlap matrix and orthonormal basis construction in CholGS as ${\bf S}_{\bm{b_i}}=\bm{\widetilde{\Psi}^{\dagger}}\bm{\widetilde{\Psi}_{b_i}}$ and ${\bm{{\widetilde{\Psi}}_{b_i}^{\rm o}}}={\bm{\widetilde{\Psi}}}{ {{\bf L^{-1}_{\bm{b_i}}}}^{\bf \dagger}}$, respectively. We note that ${ {{\bf L^{-1}_{\bm{b_i}}}}^{\bf \dagger}}$ denotes $N \times N_{b_i}$ sub-matrix of ${\bf {L^{-1}}^{\dagger}}$. Similarly, band parallelized computation of projected Hamiltonian in Rayleigh--Ritz procedure is performed as ${\bf \hat{H}}_{\bm{b_i}}=\bm{{\widetilde{\Psi}}^{{\rm o}^{\dagger}} {\bf \widetilde{H}}} \bm{{\widetilde{\Psi}}_{b_i}^{{\rm o}}}$. We note that all-to-all communications of $\bm{\widetilde{\Psi}_{b_i}}$'s and $\bm{\widetilde{\Psi}^{\rm o}_{b_i}}$'s across band groups are performed before beginning the orthonormalization and Rayleigh-Ritz procedure, respectively. Furthermore, all-to-all communications of ${\bf S}_{\bm{b_i}}$'s and ${\bf \hat{H}}_{\bm{b_i}}$'s across band groups are also performed to compute the matrices ${\bf S}$ and ${\bf \hat{H}}$ ($N \times N$ dimensions). The above all-to-all communications involve large memory sizes, and hence the communication cost can increase significantly with increase in band parallelization groups, thus affecting parallel scaling efficiency. However, a modest amount of band parallelization can be combined with domain decomposition parallelization to extend the parallel scalability in DFT-FE, as discussed below. Additionally, we have also implemented parallelization over k points for problems involving multiple k-point sampling over the Brillouin zone in periodic calculations.



\begin{figure}
\centering
\includegraphics[scale=0.3]{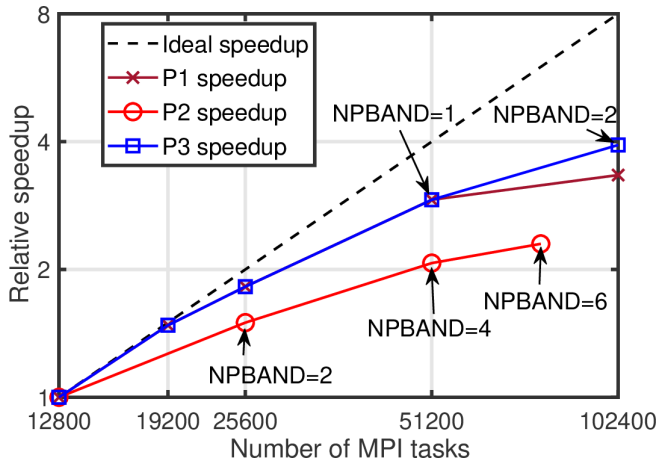}
\caption{Comparison of parallel scalability of $\textrm{Mg}_{\textrm{10x10x10}}$ (39,990 electrons) using three different parallelization approaches: (P1) only domain decomposition parallelization, (P2) primarily band parallelization with minimal domain decomposition parallelization, and (P3) domain decomposition parallelization till parallel scaling efficiency of $\sim70\%$ followed by moderate band parallelization. The number of band parallelization groups used in approaches P2 and P3 are denoted by NPBAND. This benchmark study comprised of $~\sim94\,{\rm million}$ $4^{\rm th}$ order FE basis functions.}
\label{fig:scalabilityStrategy}
\end{figure}

 We now compare the scalability of three different parallelization approaches in DFT-FE: (i) only domain decomposition parallelization (P1), (ii) primarily band parallelization with just enough domain decomposition parallelization to fit the memory (P2), and (iii) domain decomposition parallelization till parallel scaling efficiency of $\sim70\%$ followed by moderate band parallelization (P3). We conduct comparative studies on a large benchmark system containing 3999 atoms (39,990 electrons) using the above three approaches, and the results are shown in Fig.~\ref{fig:scalabilityStrategy}. We observe that only domain decomposition parallelization (P1) provides better parallel scalability than the primarily band parallelization approach (P2)---$73\%$ efficiency vs. $52\%$ efficiency at 51,200 MPI. This is attributed to the significant increase in MPI collective communication cost of wavefunctions as the number of band parallelization groups increase. However, the use of band parallelization is beneficial for medium--large system sizes when appreciable scaling from domain decomposition parallelization has already been extracted, as is evident from Fig.~\ref{fig:scalabilityStrategy}, where the best parallel scaling efficiency is obtained for the combined parallelization approach (P3). In particular, we use domain decomposition parallelization till 51,200 MPI tasks ($73\%$ efficiency), and band parallelization using two band parallelization groups to achieve $49\%$ efficiency at 102,400 MPI tasks. We remark that only $41\%$ efficiency is achieved for the same 102,400 tasks by solely using domain decomposition parallelization approach (P1). Based on the above comparison, in the remainder of this work, we primarily use the combined parallelization approach (P3) to scale DFT-FE calculations, particularly for large system sizes.
 

\subsection{Geometry optimization algorithm}
\label{sec:geoopt}
Geometry optimization involves relaxation of ionic forces and periodic unit-cell stresses, which are computed using the configurational forces approach as discussed in Section~\ref{sec:configForce}. In DFT-FE, geometry optimization is performed using Polak--Ribiere-Polyak non-linear conjugate gradient algorithm (PRP-CG)~\cite{shewchuk1994} with a secant line search. Furthermore, in the case of atomic force relaxation, we regenerate the finite-element mesh corresponding to the new atomic positions for every geometry update step. We note that the parallel adaptive mesh generation in DFT-FE, which is performed via the deal.II package is a very minor cost in comparison to the electronic ground-state solve. Additionally, we have also implemented an interface to checkpoint and restart the non-linear conjugate gradient solver for the geometry relaxation.

In the future, we plan to improve the secant line search in our non-linear conjugate gradient algorithm implementation with the more robust Brent's method~\cite{brent1971}. We also plan to implement alternate geometry optimization algorithms like Limited-memory Broyden-Fletcher-Goldfarb-Shanno (L-BFGS)~\cite{liu1989} and Fast Inertial Relaxation Engine (FIRE)~\cite{bitzek2006}, which demonstrate faster convergence than PRP-CG in many cases.
\section{Results and discussion}
\label{sec:results}
In this section, we demonstrate the accuracy, parallel scaling performance and computational efficiency of the developed DFT-FE code on various benchmark systems involving both pseudopotential and all-electron DFT calculations. GGA~\cite{gga1} exchange correlation of the PBE form~\cite{pbe} is employed in all the calculations, and additionally ONCV~\cite{oncv2013} pseudopotentials from the SG15 database~\cite{oncv2015} are employed in all the pseudopotential DFT calculations. Further, we use Fermi-Dirac smearing with temperature $T = 500$ $K$, and the $n$-stage Anderson mixing scheme~\cite{anderson1965} for mixing the electron-density in the SCF iteration procedure. 

We first validate the accuracy of DFT-FE with widely used DFT basis sets on benchmark materials systems involving periodic and non-periodic pseudopotential and all-electron DFT calculations. Second, we demonstrate the parallel scalability of DFT-FE on pseudopotential benchmark systems with sizes ranging from $2,550$ electrons to $39,990$ electrons. Third, we assess the computational efficiency of the DFT-FE code by comparing to popular plane-wave based codes---QUANTUM ESPRESSO (QE)~\cite{qe2009,qe2017}, and ABINIT~\cite{gonze2002}---on periodic and non-periodic pseudopotential benchmark systems with sizes ranging from $2,550$ to $20,470$ electrons. Finally, we also conduct large-scale DFT calculations on sizes ranging from $27,986$ to $61,502$ electrons using DFT-FE that are computationally prohibitive using plane-wave codes. The discretization parameters in the above computational efficiency studies are chosen to be commensurate with chemical accuracy (discretization errors of $\sim 10^{-4}$ Ha and $\sim 10^{-4}$ Ha/Bohr in energy per atom and ionic forces respectively), based on the validation studies on the same benchmark systems at smaller sizes.

All the numerical simulations with computational times reported in this work were executed on the Cori supercomputer at the National Energy Research Scientific Computing (NERSC) center. In particular, we used Cori's Phase \rom{2} partition containing 9,688 compute nodes based on  Intel Xeon Phi processors. Each compute node has the following specifications:  single-socket Intel Xeon Phi $7250$ (``Knights Landing'') processor with $68$ physical cores per node @ $1.4$ GHz and $96$ GB of memory per node. Cori uses a Cray Aries with Dragonfly topology for inter-node communication with  $45.0$ TB/s global peak bisection bandwidth. 

The numerical simulations for validation of DFT-FE with respect to QE were performed on the Theta supercomputer at Argonne Leadership Computing Facility (ALCF). Theta comprises 4,392 compute nodes, with each compute node having the following specifications: single 1.3 GHz Intel Xeon Phi 7230 SKU chip (``Knights Landing'') with 64 cores, 16 GB Multi-Channel DRAM (MCDRAM) and 192 GB DDR4 memory. The interconnect topology is a dual place Dragonfly with ten groups. Each group consists of two cabinets or racks. The total bi-section bandwidth is 7.2 TB/sec.

The DFT-FE simulations reported in this section were run using $32$ MPI tasks per node and $2$ OpenMP threads (for \verb|BLAS| operations), except for a few small system sizes (less than $3,000$ electrons) where using $64$ MPI tasks per node and a single OpenMP thread was found to be more efficient. Similarly, QE and ABINIT simulations were run using optimal MPI tasks-OpenMP threads combinations based on the problem size--- $32$ MPI tasks per node and $2$ OpenMP threads for smaller problem sizes (less than $3,000$ electrons), and $16$ MPI tasks per node and $4$ OpenMP threads for larger problem sizes where more memory per MPI task is required. Furthermore, MPI task to core binding was also appropriately set for all the above combinations. Our numerical experiments showed that using more than $4$ OpenMP threads provided negligible performance gains in DFT-FE as well as in QE and ABINIT.
\subsection{Validation}
\label{sec:validation}
We now validate the accuracy of DFT-FE on periodic and non-periodic pseudopotential and all-electron benchmark systems. The pseudopotential benchmark systems are validated using plane-wave code QE, while all-electron benchmark systems are validated with \verb|exciting|~\cite{exciting2014}, which uses linearized augmented plane-wave (LAPW) method, and NWChem~\cite{nwchem} using Gaussian basis set.

\subsubsection{Pseudopotential DFT calculations}
\begin{table}[htbp]
\centering
\small
\caption{\label{tab:pspValidation}\small{Validation of DFT-FE with QE on pseudopotential benchmark systems at two different accuracy levels. ${\rm FE}_{\rm ord}, h_{\rm min}$ and $h_{\rm max}$ denote the FE polynomial order, minimum element size and maximum element size (Bohr), respectively, in DFT-FE. ${ E}_{\rm cut}$ denotes the plane-wave basis cut-off used in QE (Hartree). ${ E}_{\rm g}$ denotes ground-state energy (Hartree/atom). $\Delta_{\rm max} {f}=\max\limits_{1\le i \le N_{a}} \left\|{\bf f}_i^{\rm DFT-FE} -{\bf f}_i^{\rm QE}\right\|$ (Hartree/Bohr), where ${\bf f}_i$ denotes the force on the $i^{\rm th}$ atom. $\Delta {\sigma_h}=\abs{\sigma_h^{\rm DFT-FE}-\sigma_h^{\rm QE}}$ (Hartree/${\rm Bohr}^3$), where $\sigma_h$ denotes the hydrostatic cell stress.}}
\begin{subtable}[h]{1.0\textwidth}
\centering
\begin{tabular}{|c|c|c|c|c|c|}
\hline
System & DFT-FE & DFT-FE & QE & QE & Difference in forces \\
       & $\left({\rm FE}_{\rm ord}, h_{\rm min},\, h_{\rm max}\right)$ & $E_g$ &  ${E}_{\rm cut}$ & ${ E_g}$  &  \& stress $\left(\Delta_{\rm max} {f},\,\Delta \sigma_h\right)$  \\
\hline
$\textrm{Mg}_{\textrm{2x2x2}}$ & $4$, $0.46$, $1.92$ & $-54.3195364$  & $45$ & $-54.3195594$& $\Delta_{\rm max} {f}= 2.1 \times 10^{-4}$\\
                            &               &              &    &             & $\Delta \sigma_h= 3.7 \times 10^{-6}$\\
$\textrm{Mg}_{\textrm{4x4x4}}$ & $4$, $0.46$, $1.92$ & $-54.3279442$  & $45$ & $-54.3279638$& $\Delta_{\rm max} {f}= 3.3 \times 10^{-4}$\\
                            &               &              &    &             & $\Delta \sigma_h= 4.6 \times 10^{-6}$\\
$\textrm{Mo}_{\textrm{2x2x2}}$ & $5$, $0.74$, $1.49$ & $-68.5573334$  & $20$ & $-68.5573613$ & $\Delta_{\rm max} {f}=1 \times 10^{-5}$\\
                            &               &              &    &             & $\Delta \sigma_h=1.6 \times 10^{-6}$\\
$\textrm{Mo}_{\textrm{4x4x4}}$ & $5$, $0.74$, $1.49$ & $-68.5811483$  & $20$ & $-68.5811857$ & $\Delta_{\rm max} {f}= 2.1 \times 10^{-5}$\\
                            &               &              &    &             & $\Delta \sigma_h=1.9 \times 10^{-6} $\\
$\textrm{Cu}_{\textrm{3-Shell}}$ & $4$, $0.39$, $12.5$ & $-182.5870759$  & $50$ & $-182.5870221$ & $\Delta_{\rm max} {f}= 7.2 \times 10^{-5}$\\
$\textrm{Cu}_{\textrm{4-Shell}}$ & $4$, $0.39$, $12.2$ & $-182.5908621$ & $50$ &  $-182.5908346$ & $\Delta_{\rm max} {f}=1.4 \times 10^{-4}$\\
\hline
\end{tabular}
		\caption{\small{Medium accuracy level comparisons.}}
		\label{tab:pspValidationLow}
		\vspace{0.3cm}
\end{subtable}

\begin{subtable}[h]{1.0\textwidth}
\centering
\begin{tabular}{|c|c|c|c|c|c|}
\hline
System & DFT-FE & DFT-FE & QE & QE & Difference in forces \\
       & $\left({\rm FE}_{\rm ord}, h_{\rm min},\, h_{\rm max}\right)$ & $E_g$ &  ${E}_{\rm cut}$ & ${ E_g}$  &  \& stress $\left(\Delta_{\rm max} {f},\, \Delta \sigma_h\right)$  \\
\hline
$\textrm{Mg}_{\textrm{2x2x2}}$ & 5, 0.24, 0.96 & $-54.3196337$  & 55 & $-54.3196270$ & $\Delta_{\rm max} {f}=4.6 \times 10^{-6}$\\
                            &               &              &    &             & $\Delta \sigma_h= 2.7 \times 10^{-7}$\\
$\textrm{Mg}_{\textrm{4x4x4}}$ & 5, 0.24, 0.96 & $-54.3280448$  & 55 &  $-54.3280318$& $\Delta_{\rm max} {f}=7.9 \times 10^{-6}$\\
                            &               &              &    &             & $\Delta \sigma_h=2.6 \times 10^{-7}$\\
$\textrm{Mo}_{\textrm{2x2x2}}$ & 5, 0.37, 0.74 & $-68.5574282$  & 50 &  $-68.5574315$ & $\Delta_{\rm max} {f}= 4.5 \times 10^{-6}$\\
                            &               &              &    &             & $\Delta \sigma_h=1.1 \times 10^{-6}$\\
$\textrm{Mo}_{\textrm{4x4x4}}$ & 5, 0.37, 0.74 & $-68.5812495$  & 50 &  $-68.5812527$ & $\Delta_{\rm max} {f}= 6 \times 10^{-6}$\\
                            &               &              &    &             & $\Delta \sigma_h=9.8 \times 10^{-7}$\\
$\textrm{Cu}_{\textrm{3-Shell}}$ & 5, 0.18, 12.5 & $-182.5872871$  & 70 & $-182.5872868$ & $\Delta_{\rm max} {f}=2.5 \times 10^{-5}$\\
$\textrm{Cu}_{\textrm{4-Shell}}$ & 5, 0.18, 12.2 & $-182.5910308$ & 70 &  $-182.5910298$ & $\Delta_{\rm max} {f}=3.5 \times 10^{-5}$\\
\hline
\end{tabular}
		\caption{\small{High accuracy level comparisons.}}
		\label{tab:pspValidationHigh}
		\vspace{0.3cm}
\end{subtable}
\end{table}

We consider three different benchmark systems: (i) hexagonal close packed (hcp) Mg periodic supercells with a mono-vacancy, (ii) body centered cubic (bcc) Mo periodic supercells with a mono-vacancy, and (iii) non-periodic Icosahedron Cu nano-particles. In each of the benchmark systems, we take increasingly refined basis sets in DFT-FE and QE and compare the ground-state energy per atom, ionic forces, and cell stresses at two different accuracy levels: a) medium accuracy---discretization errors of $\sim 10^{-4}$ Ha/atom in ground-state energy, $\sim 10^{-4}$ Ha/Bohr in ionic forces, and $\sim 5 \times 10^{-6}$ Ha/${\rm Bohr}^3$ in cell stress (in periodic benchmark systems), which we consider as chemical accuracy; b) high accuracy---using a more refined basis set in both DFT-FE and QE to demonstrate much closer agreement between them. Further, we note that the validation studies for the aforementioned periodic benchmark systems are conducted using a Gamma point.

In the benchmark system involving Mg, we consider periodic supercells constructed from orthogonal unit cells (containing $4$ atoms) of hcp Mg with lattice constants: $a=5.882$ Bohr and $c=9.586$ Bohr. We consider a mono-vacancy in two supercells: $2\times2\times2$ denoted by $\textrm{Mg}_{\textrm{2x2x2}}$ containing $31$ atoms ($310$ electrons) and $4\times4\times4$ denoted by $\textrm{Mg}_{\textrm{4x4x4}}$ containing $255$ atoms ($2,550$ electrons). The relevant mesh parameters for DFT-FE and cut-off energies for QE are shown in Table~\ref{tab:pspValidation}. Table~\ref{tab:pspValidation} also provides the comparison between DFT-FE and QE at medium and high accuracy levels. At the medium accuracy level, the agreement in ground-state energy is $\mathcal{O}(10^{-5})$ Ha/atom, ionic forces is $\mathcal{O}(10^{-4})$ Ha/Bohr, and hydrostatic stress is $\mathcal{O}(10^{-6})$ Hartree/${\rm Bohr}^3$. Similarly, at high accuracy level the agreement in ground-state energy is $\mathcal{O}(10^{-5})$ Ha/atom, ionic forces is $\mathcal{O}(10^{-5})$ Ha/Bohr, and hydrostatic stress is $\mathcal{O}(10^{-7})$ Hartree/${\rm Bohr}^3$. We additionally remark that Mg has a hard ONCV pseudopotential, which is reflected in basis set parameters.


Next, in the benchmark system involving Mo, we consider periodic supercells constructed from bcc Mo unit cells with lattice constant of $5.95$ Bohr. We consider a mono-vacancy in two supercell sizes---$2\times2\times2$ denoted by $\textrm{Mo}_{\textrm{2x2x2}}$ containing $15$ atoms ($210$ electrons) and $4\times4\times4$ denoted by $\textrm{Mo}_{\textrm{4x4x4}}$ containing 127 atoms (1,778 electrons). Table~\ref{tab:pspValidation} shows the comparison between DFT-FE and QE, which demonstrates a similar excellent agreement as in the case of the Mg benchmark system. We note that the lower plane-wave cut-off or larger $h_{\rm min}$ in DFT-FE in comparison to the Mg benchmark system is attributed to Mo having a softer ONCV pseudopotential than Mg.


Finally, in benchmark system involving Cu, we consider three-dimensional non-periodic Icosahedron Cu nano-particles~\cite{cuNano}. The Icosahedron nano-particles are constructed with nearest neighbour bond length of $6.8$ Bohr and varying the number of shells. We consider two Cu nano-particle sizes: $\textrm{Cu}_{\textrm{3-Shell}}$ containing $147$ atoms ($2,793$ electrons), $\textrm{Cu}_{\textrm{4-shell}}$ containing $309$ atoms ($5,871$ electrons). For the DFT-FE simulations, we choose a non-periodic domain containing the  Cu nano-particle and impose homogeneous Dirichlet boundary conditions on the boundary of the domain. On the other hand, for the QE simulations, we choose an artificial periodic domain containing the Cu nano-particle. The energy and forces are converged with respect to the domain size in both DFT-FE and QE simulations. Table~\ref{tab:pspValidation} shows the comparison between DFT-FE and QE, which demonstrates excellent agreement between the two codes in a non-periodic setting. 


Overall, from Table~\ref{tab:pspValidation}, we show excellent agreement between DFT-FE and QE for both medium and high accuracy calculations, with the difference between the codes systematically reducing with increasing discretization.

\subsubsection{All electron DFT calculations}
We consider three different all-electron benchmark systems: (i) diamond cubic Si periodic unit cell, (ii) diamond cubic Si periodic supercell with a mono-vacancy, and (iii) ${\rm C}_{60}$ (Buckminsterfullerene) molecule. The first two benchmark systems, which are periodic, are validated with the \verb|exciting| code using LAPW basis, while the third non-periodic benchmark system is validated with the NWChem code using Gaussian basis. 

Firstly, we consider a periodic diamond cubic Si unit cell containing $8$ Si atoms ($114$ electrons) and with lattice constant $10.0065225$ Bohr. We conduct the following DFT simulations for the validation study: Gamma point denoted by $\textrm{Si}_{\rm uc}^{\rm Gamma}$, and multiple k-points using a $5 \times 5 \times 5 $ Monkhorst-Pack Grid denoted by $\textrm{Si}_{\rm uc}^{\rm Multi-kpt}$. The mesh parameters for DFT-FE and the relevant parameters for \verb|exciting| are shown in Table~\ref{tab:aeValidationP}. In the case of \verb|exciting|,  \verb|rgkmax| in Table~\ref{tab:aeValidationP} represents the product of the muffin-tin radius and the plane-wave cut-off in the interstitial region, and is the key parameter determining the accuracy of the calculation in the LAPW basis set. Further, we use the default muffin-tin radius for Si. Table~\ref{tab:aeValidationP} shows the comparison between DFT-FE and \verb|exciting| for both $\textrm{Si}_{\rm uc}^{\rm Gamma}$ and  $\textrm{Si}_{\rm uc}^{\rm Multi-kpt}$, where the agreement in ground-state energy is $\sim 1 \times 10^{-4}$ Ha/atom and hydrostatic stress is $\sim 5 \times 10^{-6}$ Hartree/${\rm Bohr}^3$. 


Next, we consider a mono-vacancy in a $2 \times 2 \times 2$ supercell constructed from diamond cubic Si unit cell with lattice constant $10.0065225$ Bohr. This benchmark system is denoted as $\textrm{SiVac}_{2\times2\times2}$ and contains $63$ atoms and $882$ electrons. Table~\ref{tab:aeValidationP} shows the comparison between DFT-FE and \verb|exciting| where the agreement in ground-state energy is $\sim 1 \times 10^{-4}$ Ha/atom, ionic forces is $\sim 5 \times 10^{-4}$ Ha/Bohr, and hydrostatic stress is $\sim 5 \times 10^{-6}$ Hartree/${\rm Bohr}^3$. 

Finally, we consider the non-periodic ${\rm C}_{60}$ Buckminsterfullerene molecule with hexagon-hexagon ring bond length of $2.6258$ Bohr and pentagon-hexagon ring bond length of $2.6197$ Bohr. In both NWChem and DFT-FE we use a non-periodic domain with homogeneous Dirichlet boundary conditions on the boundary of the domain, and the domain size is taken to be large enough for the boundary effects to be negligible. The appropriate mesh parameters for DFT-FE and choice of Gaussian basis for NWChem are shown in Table~\ref{tab:aeValidationNP}. The results demonstrate an excellent agreement between DFT-FE and NWChem in the ground-state energy.



Overall, from Table~\ref{tab:aeValidation}, we show good agreement between DFT-FE and \verb|exciting| and also between DFT-FE and NWChem, underlining a significant capability of DFT-FE to perform highly accurate all-electron calculations with arbitrary boundary conditions.

\begin{table}[htbp]
\centering
\small
\cprotect\caption{\label{tab:aeValidation}\small{Validation of DFT-FE with \verb|exciting| and NWChem on all-electron benchmark systems. ${\rm FE}_{\rm ord}, h_{\rm min}$ and $h_{\rm max}$ denote the FE polynomial order, minimum element size and maximum element size (Bohr), respectively, in DFT-FE. The \verb|rgkmax| parameter in \verb|exciting| determines the number of LAPW basis functions. The polarization consistent Gaussian basis sets in NWChem with increasing levels of accuracy are denoted by: pc2, pc3. ${ E}_{\rm g}$ denotes ground-state energy (Hartree/atom). $\Delta_{\rm max} {f}=\max\limits_{1\le i \le N_{a}} \left\|{\bf f}_i^{\rm DFT-FE} -{\bf f}_i^{\rm QE}\right\|$ (Hartree/Bohr), where ${\bf f}_i$ denotes the force on the $i^{\rm th}$ atom. $\Delta {\sigma_h}=\abs{\sigma_h^{\rm DFT-FE}-\sigma_h^{\rm QE}}$ (Hartree/${\rm Bohr}^3$), where $\sigma_h$ denotes the hydrostatic cell stress.}}
\begin{subtable}[h]{1.0\textwidth}
\centering
\begin{tabular}{|c|c|c|c|c|c|}
\hline
System & DFT-FE & DFT-FE & \verb|exciting| & \verb|exciting| & Difference in forces \\
       & $\left({\rm FE}_{\rm ord}, h_{\rm min},\, h_{\rm max}\right)$ & ${ E_g}$ &  \verb|rgkmax| & ${E_g}$  &  \& stress $\left(\Delta_{\rm max} {f},\, \Delta \sigma_h\right)$  \\
\hline
$\textrm{Si}_{\rm uc}^{\rm Gamma}$ & 5, 0.013, 1.67 & $-289.3500196$  & 7 & $-289.3501763$ & $\Delta \sigma_h=\, 6.1 \times 10^{-6}$ \\
$\textrm{Si}_{\rm uc}^{\rm Multi-kpt}$ & 5, 0.013, 1.67 & $-289.4001025$  & 7 & $-289.4001864$ & $\Delta \sigma_h=\, 4.0 \times 10^{-6}$\\
$\textrm{SiVac}_{2\times2\times2}$ & 5, 0.013, 1.67 &  $-289.3943335$   & 7 &  $-289.3944222$ &  $\Delta_{\rm max} {f}=7.2 \times 10^{-4}$\\
 &  &  &  &  & $\Delta \sigma_h=2.5 \times 10^{-6}$\\
\hline
\end{tabular}
		\cprotect\caption{\small{Validation of periodic calculations with \verb|exciting|.}}
		\label{tab:aeValidationP}
		\vspace{0.3cm}
\end{subtable}

\begin{subtable}[h]{1.0\textwidth}
\centering
\begin{tabular}{|c|c|c|}
\hline
System & DFT-FE (AAMR) &  NWChem \\
       & $\left({\rm FE}_{\rm ord}, h_{\rm min},\, h_{\rm max}: {\rm E_g}\right)$ &  (Gaussian Basis:  ${\rm E_g}$)\\
\hline
$\textrm{C}_{\textrm{60}}$ & 4, 0.037, 20.0: $-38.0701278$  & pc2: $-38.0693824$\\
 & 4, 0.0185, 20.0: $-38.0709949$  & pc3: $-38.0710146$ \\
 & 4, 0.0094, 20.0: $-38.0710439$ & -\\
\hline
\end{tabular}
		\caption{\small{Validation of non-periodic calculations with NWChem.}}
		\label{tab:aeValidationNP}
		\vspace{0.3cm}
\end{subtable}
\end{table}
\subsection{Parallel scaling performance}
\label{sec:scaling}
Here we demonstrate the parallel scalability of DFT-FE on various system sizes. We consider hexagonal close packed (hcp) Mg periodic super cells with a mono-vacancy, and study the strong scaling behavior on three system sizes: (a) small---$\textrm{Mg}_{\textrm{4x4x4}}$ (255 atoms, 2,550 electrons), (b) medium---$\textrm{Mg}_{\textrm{8x8x8}}$ (2,047 atoms, 20,470 electrons), and (c) large---$\textrm{Mg}_{\textrm{10x10x10}}$ (3,999 atoms, 39,990 electrons). In particular, we study the strong scaling behavior by measuring the relative speedup with increasing number of MPI tasks while keeping the discretization fixed for all the three systems. The speedup is measured relative to the wall time taken on 512 MPI tasks, 3,200 MPI tasks, and 12,800 MPI tasks for $\textrm{Mg}_{\textrm{4x4x4}}$, $\textrm{Mg}_{\textrm{8x8x8}}$ and $\textrm{Mg}_{\textrm{10x10x10}}$, respectively. We note that lower number of MPI tasks were not possible due to memory constraints. This is primarily because of low memory per core ($\sim$1.4 GB) of the many-core KNL architecture in the Cori supercomputer.   Further, the FE mesh in the above studies is chosen such that the discretization errors in energy and forces are $\sim 10^{-4}$ Ha per atom and $\sim 10^{-4}$ Ha/Bohr, respectively.

\begin{figure}[t!]
    \centering
    \begin{subfigure}[t]{0.48\textwidth}
        \centering
        \includegraphics[scale=0.25]{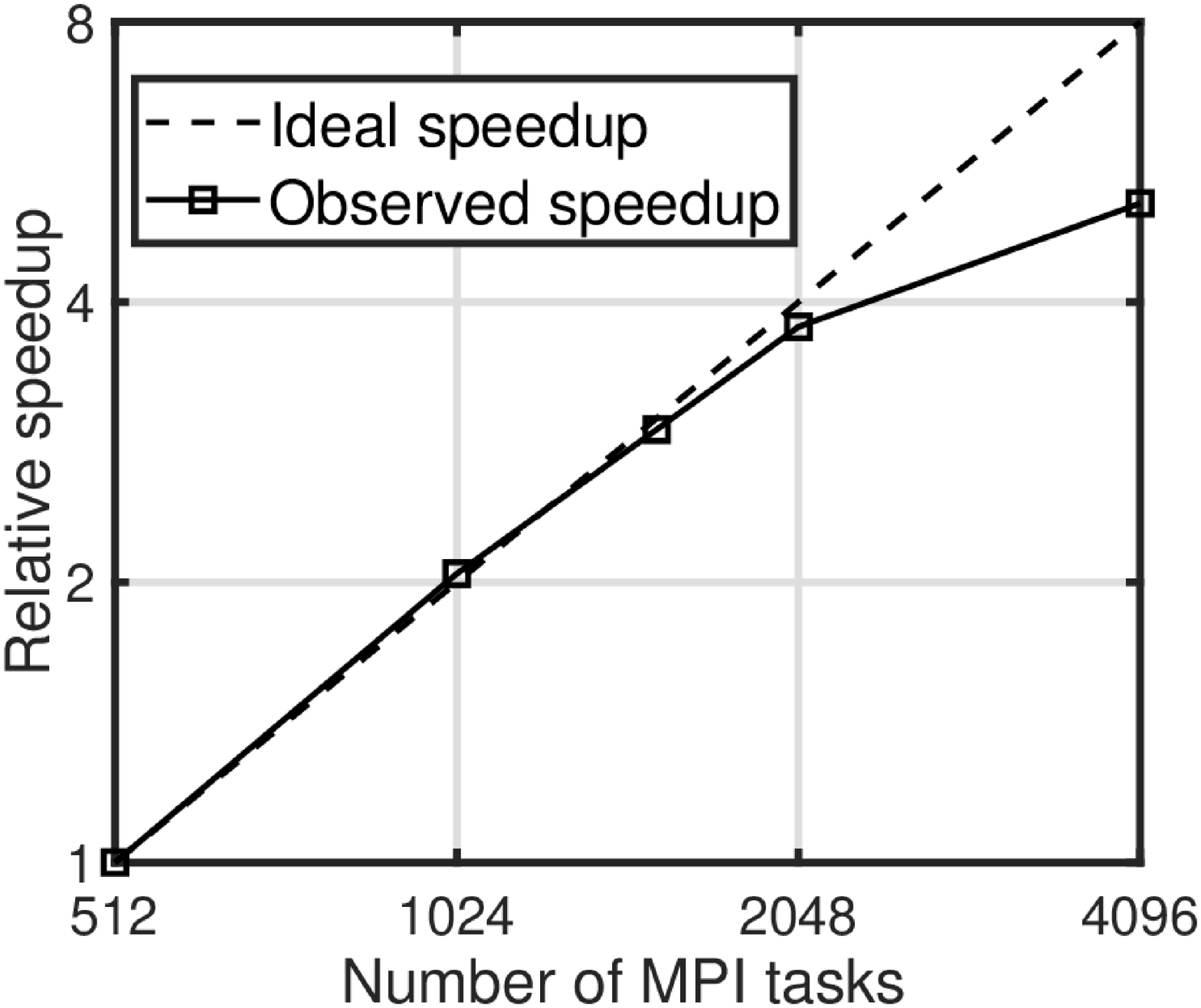}
        \caption{$\textrm{Mg}_{\textrm{4x4x4}}$ (255 atoms, 2,550 electrons)}
        \label{fig:scalingStudy4x4x4}
    \end{subfigure}
    ~ 
    \begin{subfigure}[t]{0.48\textwidth}
        \centering
        \includegraphics[scale=0.25]{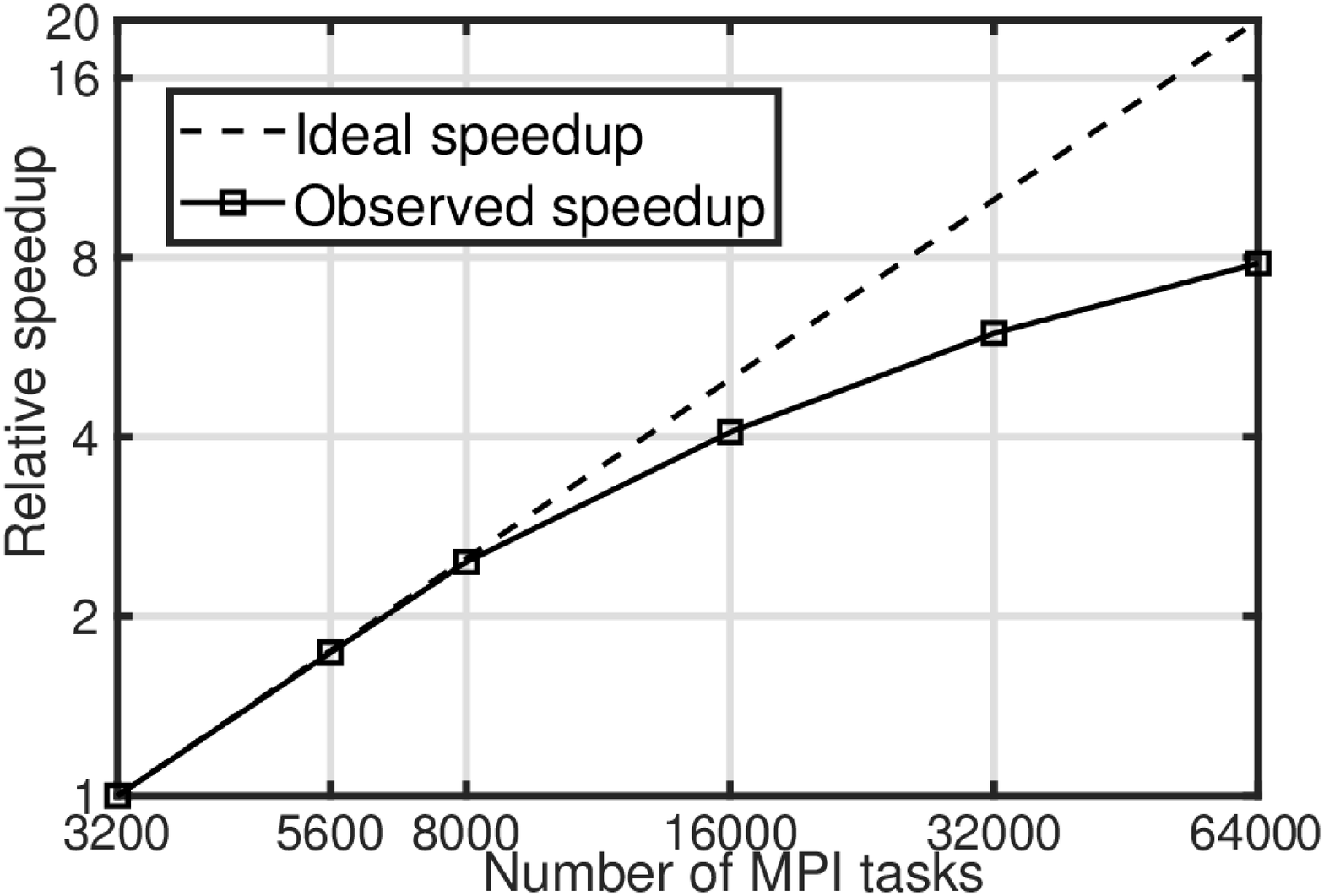}
        \caption{$\textrm{Mg}_{\textrm{8x8x8}}$ (2,047 atoms, 20,470 electrons)}
        \label{fig:scalingStudy8x8x8}
    \end{subfigure}
    
    \begin{subfigure}[t]{0.5\textwidth}
        \centering
        \includegraphics[scale=0.25]{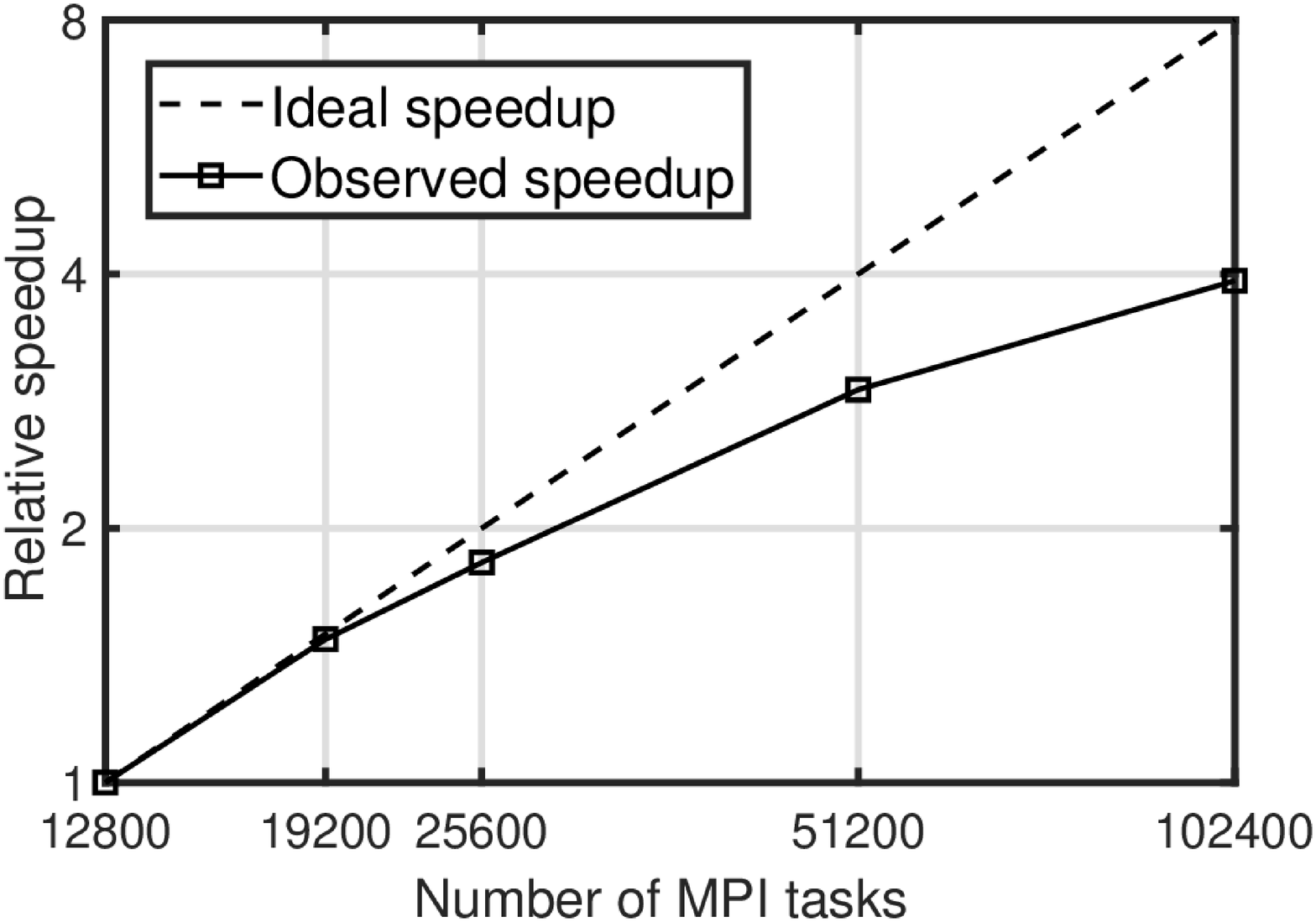}
        \caption{$\textrm{Mg}_{\textrm{10x10x10}}$ (3,999 atoms, 39,990 electrons)}
        \label{fig:scalingStudy10x10x10}
    \end{subfigure}
    \caption{Strong parallel scaling using DFT-FE. Case studies: a) $\textrm{Mg}_{\textrm{4x4x4}}$, b) $\textrm{Mg}_{\textrm{8x8x8}}$, and c) $\textrm{Mg}_{\textrm{10x10x10}}$.}
    \label{fig:scalingStudy}
\end{figure}

First, we consider the parallel scalability of the smaller system size $\textrm{Mg}_{\textrm{4x4x4}}$ as demonstrated in Fig.~\ref{fig:scalingStudy4x4x4}. Here we use only domain decomposition parallelization to scale up to 4,096 MPI tasks at 75$\%$ efficiency, with an average of 1,629 dofs per MPI task. The corresponding wall time for a single SCF iteration step is 28 seconds.  Such excellent parallel scalability is possible due to the aforementioned low communication cost in the FE discretized Hamiltonian matrix and wavefunction vector products involved in Chebyshev filtering, which is the dominant computational cost for small to medium system sizes. 

\begin{figure}[t!]
    \centering
    \begin{subfigure}[t]{0.48\textwidth}
        \centering
        \includegraphics[scale=0.265]{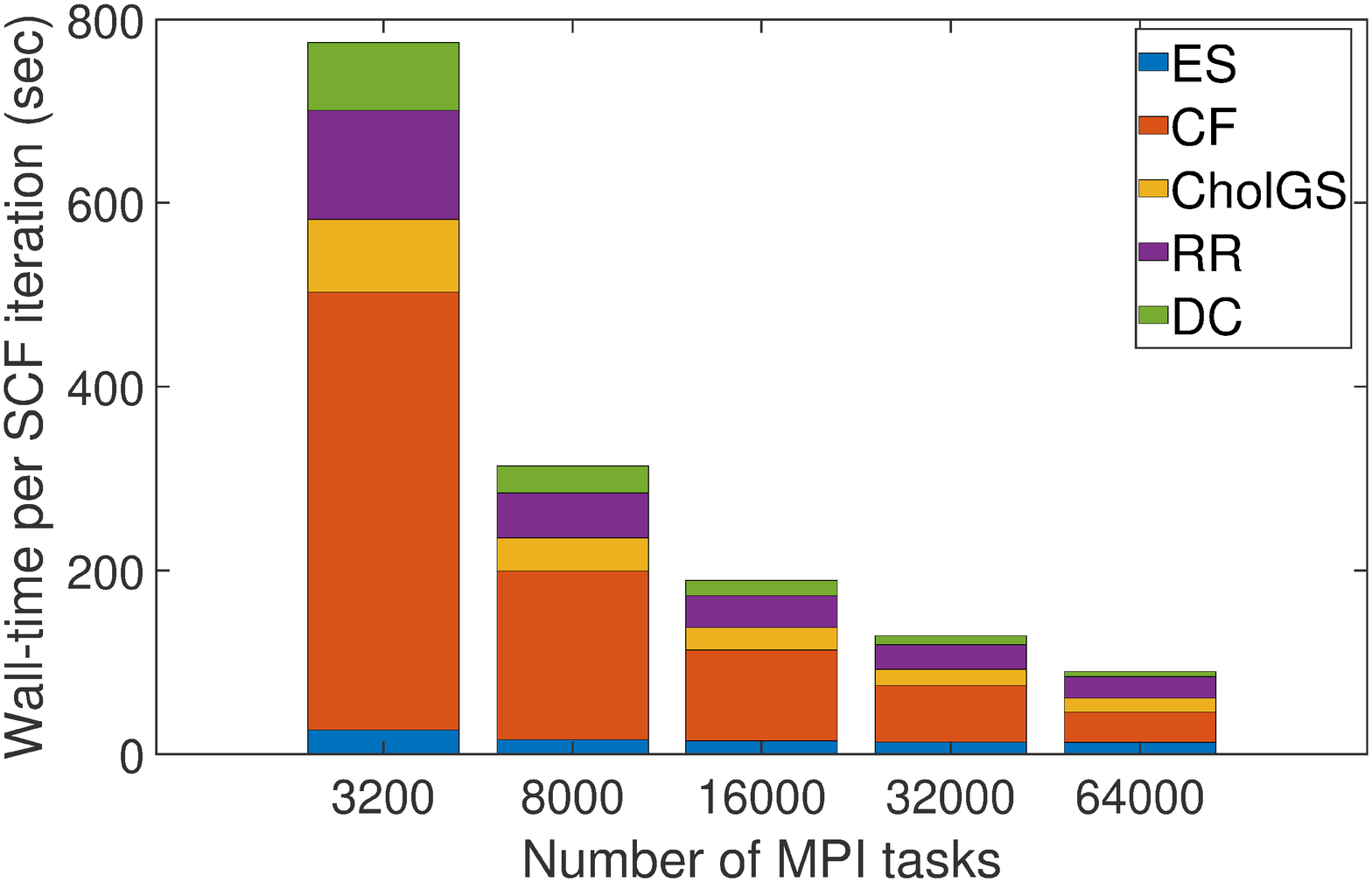}
        \caption{$\textrm{Mg}_{\textrm{8x8x8}}$ (2,047 atoms, 20,470 electrons)}
        \label{fig:scalingStudy8x8x8BreakDown}
    \end{subfigure}
    ~
    \begin{subfigure}[t]{0.5\textwidth}
        \centering
        \includegraphics[scale=0.27]{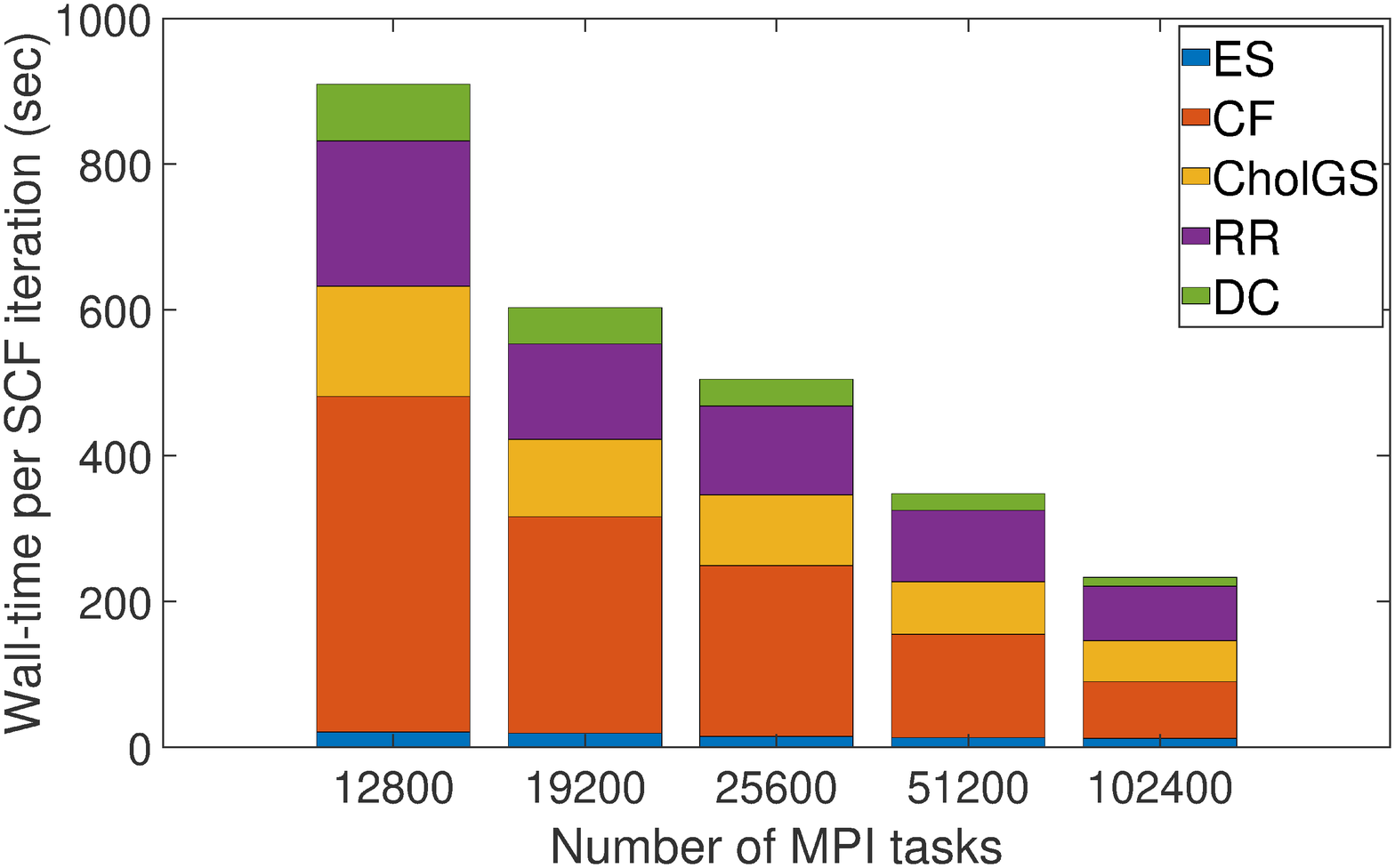}
        \caption{$\textrm{Mg}_{\textrm{10x10x10}}$ (3,999 atoms, 39,990 electrons)}
        \label{fig:scalingStudy10x10x10BreakDown}
    \end{subfigure}
    \caption{Breakdown of total wall-time per SCF iteration into the various computational steps in DFT-FE:  a) ES (Total electrostatic potential solve), b) CF (Chebyshev filtering), c) CholGS (Cholesky-Gram-Schimdt Orthogonalization), d) RR (Rayleigh-Ritz procedure), and e) DC (Electron-density computation). Case studies: a) $\textrm{Mg}_{\textrm{8x8x8}}$, and b) $\textrm{Mg}_{\textrm{10x10x10}}$. The number of MPI tasks correspond to the strong scaling studies in Fig.~\ref{fig:scalingStudy8x8x8} and ~\ref{fig:scalingStudy10x10x10}.}
    \label{fig:scalingStudyBreakdown}
\end{figure}

Next, we consider the parallel scalability of the medium to large system sizes: $\textrm{Mg}_{\textrm{8x8x8}}$ and $\textrm{Mg}_{\textrm{10x10x10}}$, which are shown in Fig.~\ref{fig:scalingStudy8x8x8} and ~\ref{fig:scalingStudy10x10x10}, respectively.
Based on the comparison of three different parallelization strategies in Section~\ref{sec:paral} for achieving maximum parallel scalability for large system sizes, we use the parallelization strategy of combined domain decomposition and band parallelization. In particular, in the case of $\textrm{Mg}_{\textrm{8x8x8}}$, we use domain decomposition parallelization from 3,200 to 32,000 MPI tasks and then use band parallelization with two band parallelization groups to further scale to 64,000 MPI tasks at 43$\%$ efficiency. At 64,000 MPI tasks, we use an average of 1,436 dofs per MPI task, and obtain a wall time of 91 seconds for a single SCF iteration step. Similarly, in the case of $\textrm{Mg}_{\textrm{10x10x10}}$, we use domain decomposition parallelization till 51,200 MPI tasks, and then use band parallelization to further scale to 102,400 MPI tasks at 49$\%$ efficiency. At 102,400 MPI tasks, we use an average of 1,835 dofs per MPI task, and obtain a wall time of 237 seconds for a single SCF iteration step. The parallel scaling of the above medium to large system sizes is dependent on the scalability of the major computational steps: CF, CholGS and RR (section~\ref{sec:numImpl}). We remark that in spite of the computational complexity of CF scaling quadratically in comparison to cubic scaling of CholGS and RR, CF's excellent parallel scalability afforded by FE discretization continues to be crucial for parallel scalability at medium to large system sizes. This is evident from Fig.~\ref{fig:scalingStudyBreakdown} showing the strong scaling of the various computational steps in DFT-FE for $\textrm{Mg}_{\textrm{8x8x8}}$ and $\textrm{Mg}_{\textrm{10x10x10}}$, where we note that  CF is still a significant portion of the total wall-times, and further CF also  demonstrates excellent parallel scalability. Fig.~\ref{fig:scalingStudyBreakdown} also demonstrates good parallel scaling of CholGS and RR steps, where the use of mixed precision arithmetic based algorithms play a key role in reducing communication costs. Overall, DFT-FE's massive parallel scalability, as demonstrated here, is a result of the locality of the FE basis as well as an effective parallel implementation of the various algorithms in DFT-FE, as discussed in section~\ref{sec:numImpl}, that reduce communication costs and latency.

\subsection{Computational efficiency and wall time comparison with plane-wave codes}
\label{sec:timing}
We now consider three different benchmark systems with sizes ranging  from  
$2,550$ to $61,502$ electrons to compare the computational efficiency (CPU-time and minimum wall-time) of the DFT-FE code against the plane-wave codes---QUANTUM ESPRESSO (QE) v6.3~\cite{qe2009,qe2017}, and ABINIT v8.8.4~\cite{gonze2002}. In particular, we consider: (i) hexagonal close packed (hcp) Mg periodic super cells with a mono-vacancy, (ii) body centered cubic (bcc) Mo periodic super cells with a mono-vacancy, and (iii) non-periodic Icosahedron Cu nano-particles.  The details on these benchmark systems are discussed previously in Section~\ref{sec:validation}, with a wider range of system sizes considered here. We note that the FE mesh parameters and plane-wave cut-off energies in all the benchmark calculations are chosen to be commensurate with chemical accuracy, based on the validation studies on smaller system sizes for the same benchmark system types. We do not explicitly measure the discretization errors here as highly refined calculations required to do so will use significant computational resources given the many benchmark systems along with large sizes (up to $61,502$ electrons) considered here.  Further, we note that all QE and ABINIT timings reported below are for an optimal combination of FFT grid parallelization and band parallelization.

The DFT-FE simulations for the above benchmark systems use the following values of Chebyshev polynomial degree $m$ (see Section~\ref{sec:cheby}): $m=45$ for benchmark systems (i) and (ii), and $m=50$ for benchmark system (iii). Further, $N_{\rm fr}$, which is used in the RR step (see Section~\ref{sec:rr}), is chosen to be 15 \% of $N$. Additionally, in all simulations (DFT-FE, QE and ABINIT) $N$ is chosen as $N_e/2+ b$, with $b\sim10\%$ of $N_e/2$ for benchmark systems (i) and (ii), and $b\sim5\%$ of $N_e/2$ for benchmark system (iii).

\paragraph{\small CPU-time comparisons}
In Table~\ref{tab:cpuTime}, we compare the average computational CPU-times per SCF iteration step\footnote{Measured by taking the average of a few SCF iteration steps after the first $2-3$ SCF iteration steps, which are excluded as their timings can be variable depending on the starting wavefunctions guess to the SCF procedure. Furthermore, for few system sizes ($< 10,000$ electrons) in each benchmark system, we verify that the choice of the DFT-FE parameters are adequate to achieve convergence in similar number of SCF iteration steps as taken by QE and ABINIT. We do not use advanced mixing strategies like Kerker preconditioning in QE and ABINIT simulations as such strategies are currently not implemented in DFT-FE.} in the above benchmark systems between DFT-FE and the plane-wave codes QE and ABINIT. The CPU-times are reported in Node-Hrs, which is obtained by multiplying the total number of compute nodes used with the average wall-time per SCF iteration measured in hours. We also compare the number of basis functions used to achieve the desired chemical accuracy in energy and forces. We note that all the simulations for Table~\ref{tab:cpuTime} are run using the minimum number of compute nodes required to fit the peak memory of the simulation, and additionally remark that the dashes in the table corresponding to QE and ABINIT benchmark simulations that are not performed as they are computationally prohibitive. First, we consider the periodic benchmark problems in Tables~\ref{tab:cpuTimesMg} and ~\ref{tab:cpuTimesMo}, where we observe that DFT-FE is more computationally efficient than both QE and ABINIT beyond system sizes of $\sim 3000$ electrons ($300$ atoms)  for hcp Mg supercells, and $\sim 6,000$ electrons ($428$ atoms) for bcc Mo supercells. There are a couple of reasons for DFT-FE's efficiency gains over QE in spite of the number of basis functions advantage of plane-wave basis for periodic problems. Firstly, simulations for medium to large system sizes require more compute nodes to fit the peak memory on many-core architectures like the Cori KNL nodes. This increases the CPU-time of plane-wave codes, relative to DFT-FE, due to the better parallel scaling of DFT-FE. Secondly, the efficient and scalable numerical implementation CF, CholGS and RR in DFT-FE (see Section~\ref{sec:numImpl}) is also a key factor. Next, we consider the non-periodic benchmark problem: Icosahedron nano-particles of varying sizes, in Table~\ref{tab:cpuTimesCu}. Here we observe that that DFT-FE is more computationally efficient than QE beyond a very small system size of $147$ atoms ($2,793$ electrons). We note that the spatial adaptivity of DFT-FE provides a key advantage in non-periodic systems where the FE mesh can be coarse-grained into the vacuum as opposed to a uniform spatial resolution of the plane-wave basis. Furthermore, the spatial adaptivity of the FE basis is also an advantage in systems having hard pseudopotentials such as Cu. 
Overall, from Table~\ref{tab:cpuTime}, we observe that DFT-FE's efficiency gains over QE increases with increasing system size, achieving efficiency gains of $5.7\times$, $12.4\times$, and $11.9\times$ for $\textrm{Mg}_{\textrm{8x8x8}}$, $\textrm{Mo}_{\textrm{10x10x10}}$, and $\textrm{Cu}_{\textrm{5-shell}}$, respectively, which are the largest benchmark systems considered for CPU-time comparison. We note that ABINIT is slower than QE for all benchmark systems considered in Table~\ref{tab:cpuTime}. Finally, another key observation is that due to the efficient numerical implementation of cubic-scaling CholGS and RR steps in DFT-FE, the range of close to quadratic scaling in computational complexity with respect to number of electrons ($N_e$) is extended to much larger system sizes---$\mathcal{O}(N_{e}^{2.12})$ up to $N_e=39,990$ for hcp Mg super cells, $\mathcal{O}(N_{e}^{2.32})$ up to $N_e=27,986$ for bcc Mo super cells, and $\mathcal{O}(N_{e}^{2.04})$ up to $N_e=17,537$ for Cu nano-particles.

\begin{table}[htbp]
\centering
\small
\caption{\label{tab:cpuTime}\small{CPU-time comparison of DFT-FE with QE and ABINIT: Average time per SCF iteration step in Node-Hrs.}}
\begin{subtable}[h]{1.0\textwidth}
\centering
\begin{tabular}{|c|c|c|c|c|c|c|}
\hline
 System & Number of atoms & FE basis & DFT-FE & Plane-wave & QE & ABINIT \\
 & (Number of electrons)  &          &        &    basis   &    &         \\
\hline
$\textrm{Mg}_{\textrm{4x4x4}}$ & 255 (2,550)      & 6,673,513   & 0.3    & 530,051         & 0.1   & 0.3  \\
$\textrm{Mg}_{\textrm{6x6x6}}$ & 863 (8,630)      & 19,852,441  & 3.3    & 1,788,771 & 4.4   & 20.2 \\
$\textrm{Mg}_{\textrm{8x8x8}}$ & 2,047 (20,470)    & 45,954,505  & 21.6   & 4,240,071         & 123.5 & -    \\
$\textrm{Mg}_{\textrm{10x10x10}}$ & 3,999 (39,990) & 93,972,153 & 103.4  &-           & -     & -    \\
\hline
\end{tabular}
		\caption{\small{Benchmark system (i): hcp Mg periodic supercells with a mono-vacancy.}}
		\label{tab:cpuTimesMg}
		\vspace{0.3cm}
\end{subtable}
\begin{subtable}[h]{1.0\textwidth}
\centering
\begin{tabular}{|c|c|c|c|c|c|c|}
\hline
 System & Number of atoms & FE basis & DFT-FE & Plane-wave & QE & ABINIT \\
 & (Number of electrons)  &          &        &    basis   &    &         \\
\hline
$\textrm{Mo}_{\textrm{6x6x6}}$ & 431 (6034)      & 5,475,843  & 0.5  & 194,310  & 0.56 & 0.7 \\
$\textrm{Mo}_{\textrm{8x8x8}}$ & 1,023 (14,322)    & 12,942,743 & 4.2  & 460,725 & 22.1 & 115.7 \\
$\textrm{Mo}_{\textrm{10x10x10}}$ & 1,999 (27,986) & 25,229,995 & 17.7 & 899,849 & 219.5   & -\\
\hline
\end{tabular}
		\caption{\small{Benchmark system (ii): bcc Mo periodic supercells with a mono-vacancy.}}
		\label{tab:cpuTimesMo}
		\vspace{0.3cm}
\end{subtable}
\begin{subtable}[h]{1.0\textwidth}
\centering
\begin{tabular}{|c|c|c|c|c|c|c|}
\hline
 System & Number of atoms & FE basis & DFT-FE & Plane-wave & QE & ABINIT \\
 & (Number of electrons)  &          &        &    basis   &    &         \\
\hline
$\textrm{Cu}_{\textrm{3-Shell}}$ & 147 (2793)  & 6,584,861   & 0.3  & 1,080,751  & 0.2 &0.8\\
$\textrm{Cu}_{\textrm{4-Shell}}$ & 309 (5,871)  & 13,974,767  & 1.7  & 2,110,867  & 5.5 &10.7\\
$\textrm{Cu}_{\textrm{5-Shell}}$ & 561 (10,659) & 26,060,299  & 5.3  & 3,647,655  & 63.4 &-\\
$\textrm{Cu}_{\textrm{6-Shell}}$ & 923 (17,537) & 41,775,101  & 12.7 & 5,792,547   & -&-\\
\hline
\end{tabular}
		\caption{\small{Benchmark system (iii): Cu Icosahedron nano-particles of varying sizes.}}
		\label{tab:cpuTimesCu}
\end{subtable}
\end{table}

\paragraph{\small Wall-time comparisons}
Next, in Table~\ref{tab:wallTimeComparison} we compare the average minimum wall-times per SCF iteration step in the above benchmark systems between DFT-FE and QE, with the restriction that the parallel scaling efficiency is above $40\%$. We observe that DFT-FE wall-times are smaller than QE wall-times for all the benchmark systems considered. Furthermore, the speedups in DFT-FE over QE increases with system size, with substantial speedups of $9\times$, $16.1\times$ and $6.9\times$ for $\textrm{Mg}_{\textrm{8x8x8}}$, $\textrm{Mo}_{\textrm{10x10x10}}$, and $\textrm{Cu}_{\textrm{5-shell}}$, respectively. Even at the smallest system sizes DFT-FE is still significantly faster than QE, with speedups of $1.5\times$, $7.5\times$ and $3.3\times$ for $\textrm{Mg}_{\textrm{4x4x4}}$, $\textrm{Mo}_{\textrm{6x6x6}}$, and $\textrm{Cu}_{\textrm{3-shell}}$, respectively. Additionally, in Table~\ref{tab:wallTimeComparison}, we also report some very large scale simulations conducted using DFT-FE: $\textrm{Mg}_{\textrm{10x10x10}}$, $\textrm{Mo}_{\textrm{13x13x13}}$, and $\textrm{Cu}_{\textrm{6-shell}}$, obtaining very modest minimum wall-times of $203$, $277$ and $70$ seconds, respectively (with parallel scaling efficiencies above $40\%$). We note that such large system sizes are computationally prohibitive using QE, and thereby QE simulations for these systems are not performed. Finally, in Table~\ref{tab:wallTimesLarge} we show the breakdown of DFT-FE wall-times into key computational steps for  the large system sizes in the above benchmark problems. Overall, we have demonstrated that DFT-FE is faster than plane-wave codes QE and ABINIT for system sizes beyond $2,000$ electrons with significant speedups at larger system sizes, and large-scale DFT simulations on generic material systems are practically feasible using DFT-FE for system sizes ranging up to $50,000$--$100,000$ electrons.

\begin{table}[htbp]
\centering
\small
\caption{\label{tab:wallTimeComparison}\small{Minimum wall-time comparison of DFT-FE with QE: Average time per SCF iteration step in seconds (rounded to the nearest whole number).}}
\begin{subtable}[h]{1.0\textwidth}
\centering
\begin{tabular}{|c|c|c|c|c|}
\hline
System & Number of atoms & DFT-FE & QE\\
       &  (Number of electrons) & &\\
\hline
$\textrm{Mg}_{\textrm{4x4x4}}$ & 255 (2,550) & 19 & 29\\
$\textrm{Mg}_{\textrm{6x6x6}}$ & 863 (8,630) & 38 & 165\\
$\textrm{Mg}_{\textrm{8x8x8}}$ & 2,047 (20,470) & 91 &  816\\
$\textrm{Mg}_{\textrm{10x10x10}}$ & 3,999 (39,990) & 203 &  - \\
\hline
\end{tabular}
		\caption{\small{Benchmark system (i): hcp Mg periodic supercells with a mono-vacancy.}}
		\label{tab:wallTimesMg}
		\vspace{0.3cm}
\end{subtable}

\begin{subtable}[h]{1.0\textwidth}
\centering
\begin{tabular}{|c|c|c|c|}
\hline
System & Number of atoms & DFT-FE & QE \\
       & (Number of electrons) & &\\
\hline
$\textrm{Mo}_{\textrm{6x6x6}}$ & 431 (6,034) & 23 & 173 \\
$\textrm{Mo}_{\textrm{8x8x8}}$ & 1,023 (14,322) & 52 & 549 \\
$\textrm{Mo}_{\textrm{10x10x10}}$ & 1,999 (27,986) & 117 &  1883 \\
$\textrm{Mo}_{\textrm{13x13x13}}$ & 4,393 (61,502) & 277 &  - \\
\hline
\end{tabular}
		\caption{\small{Benchmark system (ii): bcc Mo periodic supercells with a mono-vacancy.}}
		\label{tab:wallTimesMo}
		\vspace{0.3cm}
\end{subtable}

\begin{subtable}[h]{1.0\textwidth}
\centering
\begin{tabular}{|c|c|c|c|}
\hline
System & Number of atoms & DFT-FE & QE \\
       & (Number of electrons) & &\\
\hline
$\textrm{Cu}_{\textrm{3-Shell}}$ & 147 (2,793) & 15 & 50 \\
$\textrm{Cu}_{\textrm{4-Shell}}$ & 309 (5,871) & 25 & 183 \\
$\textrm{Cu}_{\textrm{5-Shell}}$ & 561 (10,659) & 44 &  304 \\
$\textrm{Cu}_{\textrm{6-Shell}}$ & 923 (17,537) & 70 &  - \\
\hline
\end{tabular}
		\caption{\small{Benchmark system (iii): Cu Icosahedron nano-particles of varying sizes.}}
		\label{tab:wallTimesCu}
\end{subtable}
\end{table}

\begin{table}[t]
\centering
\small
\caption{\label{tab:wallTimesLarge}\small{Breakdown of average wall-time per SCF iteration step (in seconds, rounded to the nearest whole number) using DFT-FE for large systems into the following computational steps: a) ES (Total electrostatic potential solve), b) CF (Chebyshev filtering), c) CholGS (Cholesky-Gram-Schimdt Orthogonalization), d) RR (Rayleigh-Ritz procedure), e) DC (Electron-density computation) and f) O (other---Discrete Hamiltonian computation, electron density mixing, and computation of Fermi energy). ``NDP'' denotes number of domain decomposition MPI tasks, and ``NBP'' denotes number of band parallelization groups. Total number of MPI tasks is NDP times NBP.}}
\centering
\bgroup
\def\arraystretch{1.5}
\begin{subtable}[h]{1.0\textwidth}
\centering
\begin{tabular}{|c|c|c|c|c|c|}
\hline
System & No. atoms & No. electrons & DOF's  &NDP (NBP) & Total\\
    &  & & per atom &   & MPI tasks\\
\hline
 $\textrm{Mg}_{\textrm{8x8x8}}$ & 2,047 & 20,470  & 22,450 & 32,000 (2) & 64,000\\
\hline
$\textrm{Mg}_{\textrm{10x10x10}}$ & 3,999 & 39,990 & 23,499 & 51,200 (3)& 153,600\\
\hline
 $\textrm{Mo}_{\textrm{10x10x10}}$ & 1,999 & 27,986  &12,621 &16,000 (3) & 48,000\\
\hline
$\textrm{Mo}_{\textrm{13x13x13}}$ & 4,393 & 61,502 &12,594 &48,000 (4) & 192,000\\
\hline
\end{tabular}
		\caption{\small{Setup of the benchmark simulations.}}
		\label{tab:wallTimesLargeA}
\end{subtable}

\begin{subtable}[h]{1.0\textwidth}
\centering
\begin{tabular}{|c|c|c|c|c|c|c|c|}
\hline
System & ES & CF & CholGS & RR & DC & O & Total\\
       &    &    &     &    &    &   &  time  \\
\hline
 $\textrm{Mg}_{\textrm{8x8x8}}$  &11 & 33& 15 & 24 & 6& 2 & 91\\
\hline
$\textrm{Mg}_{\textrm{10x10x10}}$ &12 &63& 53& 63 & 9& 3& 203\\
\hline
 $\textrm{Mo}_{\textrm{10x10x10}}$ &5 & 49 & 23 & 30  & 6 &4& 117\\
\hline
$\textrm{Mo}_{\textrm{13x13x13}}$ &7 & 80 & 80 & 97 & 9 & 4 & 277\\
\hline
\end{tabular}
		\caption{\small{Breakdown of average wall-time per SCF iteration step.}}
		\label{tab:wallTimesLargeB}
\end{subtable}
\egroup
\end{table}

\section{Demonstration of DFT-FE's capabilities}
\label{sec:dftfeCapabilities}
\subsection{Geometry optimization}
\begin{figure}
\includegraphics[scale=0.04]{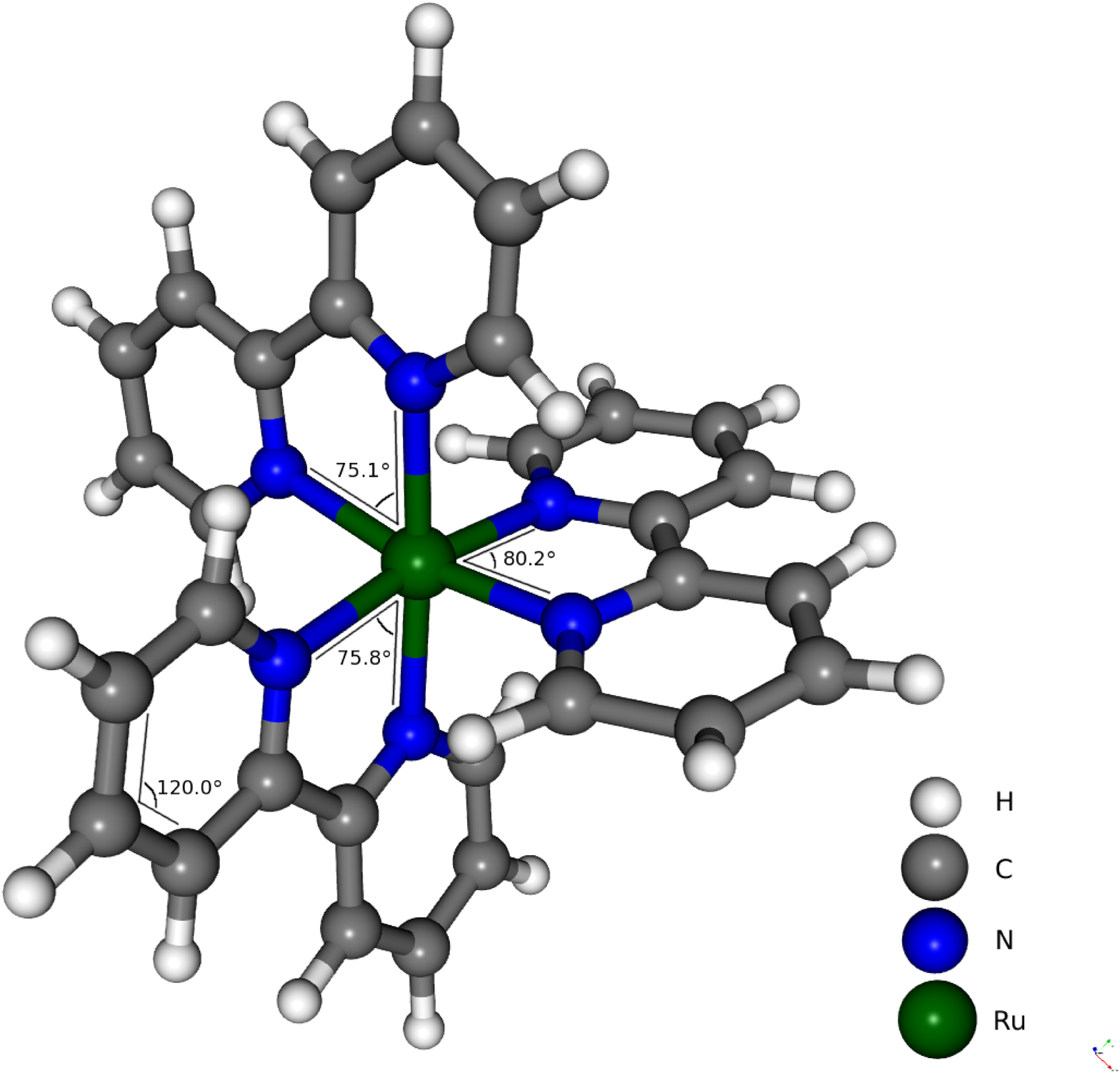}
 \centering
\caption{Schematic of Tris(bipyridine) ruthenium.}
\label{fig:tbr}
\end{figure}

We demonstrate here the capability of DFT-FE to conduct ionic relaxation using the methodology described in Section ~\ref{sec:geoopt}, and, further, we validate the ionic relaxation by comparing with QE. As our benchmark system, we consider tris (bipyridine) ruthenium (TBR), which belongs to a class of transition metal complexes with distinctive optical properties~\cite{campagna2007}. TBR  contains $61$ atoms ($290$ electrons) in total, comprising of $30$ Carbon atoms, $24$ Hydrogen atoms, $6$ Nitrogen atoms and 1 Ruthenium atom. A schematic of the starting structure is shown in Figure~\ref{fig:tbr}. In both the DFT-FE and QE simulations, the discretization errors in ground-state energy and forces are converged to $\sim 10^{-4}$ Ha per atom and $\sim 10^{-4}$ Ha/Bohr, respectively. Furthermore, GGA exchange correlation and ONCV norm conserving pseudopotentials are employed along with use of Fermi-Dirac smearing with temperature $T = 500$ K. In Table~\ref{tab:unrelaxedTBR}, we first consider the un-relaxed TBR structure, and validate that the ground-state energy and forces obtained from DFT-FE are in very good agreement with QE values. Next we conduct ionic relaxation of the TBR structure in both DFT-FE and QE until all ionic force component magnitudes are below $1 \times 10^{-3}$ Ha/Bohr.  The ground-state energy change of the relaxed TBR structure compared to the un-relaxed TBR structure obtained from DFT-FE and QE are in very good agreement as shown in Table~\ref{tab:relaxedTBR}. The maximum difference between relaxed and the initial un-relaxed TBR coordinates obtained from DFT-FE after removing the rigid body modes is $\Delta_{\rm max} {R^{\rm relax}}=0.276$ Bohr. Finally, using a similar metric we compare the relaxed TBR coordinates obtained from DFT-FE and QE, where we obtain $\Delta_{\rm max} {R^{\rm DFT-FE/QE}}=0.023$ Bohr. We note that only the core region of the TBR: $6$ Nitrogen atoms and 1 Ruthenium atom is considered when measuring $\Delta_{\rm max} {R^{\rm DFT-FE/QE}}$, as the outer ligand structure consisting of Carbon and Hydrogen atoms have close to zero energy modes. This coupled with the use of different relaxation solvers (BFGS in QE and CG in DFT-FE) can result in DFT-FE and QE giving significantly different coordinates for the outer ligand structure.

\begin{table}[htbp]
\centering
\small
\caption{\label{tab:geoOpt}\small{Demonstration of geometry optimization in DFT-FE. \textbf{Case study: tris (bipyridine) ruthenium (TBR)}. ${\rm FE}_{\rm ord}, h_{\rm min}$ and $h_{\rm max}$ denote the FE polynomial order, minimum element size and maximum element size (Bohr), respectively. ${\rm E}_{\rm cut}$ denotes plane-wave basis cut-off (Hartree), and ${ E}_{\rm g}$ denotes ground-state energy per atom (Hartree). $f_{\rm max}=\max\limits_{1\le i \le N_{a}} \left\|{\bf f}_i\right\|$ (Hartree/Bohr), where ${\bf f}_i$ denotes the force on the $i^{\rm th}$ atom, and maximum difference in forces is measured by $\Delta_{\rm max} {f}=\max\limits_{1\le i \le N_{a}} \left\|{\bf f}_i^{\rm DFT-FE} -{\bf f}_i^{\rm QE}\right\|$ (Hartree/Bohr). $\Delta {\rm E}_{\rm relax}={\rm E}_{\rm g, relax}-{\rm E}_{\rm g, un-relaxed}$ denotes the difference in the ground-state energy per atom between the relaxed and un-relaxed TBR structures (Hartree). Maximum difference between relaxed and the un-relaxed TBR coordinates in DFT-FE is measured by $\Delta_{\rm max} {R^{\rm relax}}=\max\limits_{1\le i \le N_{a}} \left\|{\bf R}_i^{\rm relaxed} -{\bf R}_i^{\rm un-relaxed}\right\|$ (Bohr).}}
\begin{subtable}[h]{1.0\textwidth}
\centering
\begin{tabular}{|c|c|c|c|c|c|c|}
\hline
 DFT-FE & DFT-FE & DFT-FE & QE & QE & QE & $\Delta_{\rm max} {f}$\\
$\left({\rm FE}_{\rm ord}, h_{\rm min},\, h_{\rm max}\right)$ & ${E}_{\rm g}$ & $f_{\rm max}$ & ${\rm E}_{\rm cut}$ & ${\rm E}_{\rm g}$ & $f_{\rm max}$ &    \\
\hline
4, 0.42, 13.33 & -5.5566947  & 0.1626628 & 50 &  -5.5566265 & 0.1626344 & $1.1 \times 10^{-4}$\\
\hline
\end{tabular}
		\caption{\small{Comparison of energies and forces in the un-relaxed TBR structure between DFT-FE and QE.}}
		\label{tab:unrelaxedTBR}
		\vspace{0.3cm}
\end{subtable}

\begin{subtable}[h]{1.0\textwidth}
\vspace{0.25in}
\centering
\begin{tabular}{|c|c|c|}
\hline
DFT-FE & QE & $\Delta_{\rm max} {R^{\rm relax}}$ \\
 $\Delta {\rm E}_{\rm relax}$ & $\Delta {\rm E}_{\rm relax}$      &\\
\hline
-0.0015992  &  -0.0016009 &  0.276 \\
\hline
\end{tabular}
		\caption{\small{Comparison between relaxed and un-relaxed TBR structrues.}}
		\label{tab:relaxedTBR}
		\vspace{0.3cm}
\end{subtable}

\end{table}

\subsection{Molecular dynamics}
We demonstrate the capability of DFT-FE to conduct Born-Oppenheimer molecular dynamics (MD) calculations~\cite{marx_hutter_2009}. In particular, we choose fcc Al $3 \times 3 \times 3$ supercell with fcc lattice constant  $7.571944$ Bohr and containing 108 atoms and conduct a NVE molecular dynamics simulation. We use an initial ionic temperature of $T = 1,500$ K and employ a time step of $0.5$ fs using a velocity Verlet time integration algorithm. This simulation is conducted using norm-conserving Troullier-Martins pseudopotential and  LDA exchange correlation. We assign initial velocities using a Maxwell-Boltzmann distribution corresponding to the initial temperature of $1,500$ K. Further, we choose the electronic temperature to be equal to the ionic temperature at every step and the MD simulation is conducted until $800$ fs.

\begin{figure}[htp]
\includegraphics[scale=0.25]{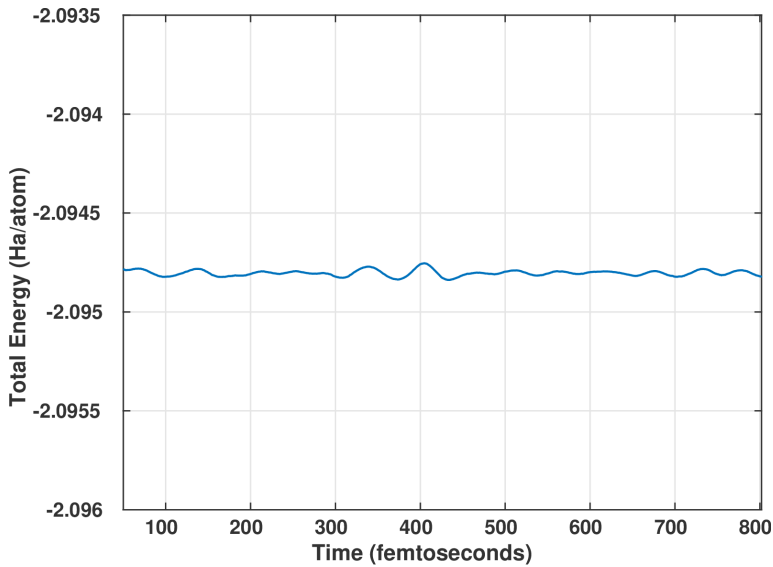}
 \centering
\caption{Variation of total energy during NVE molecular dynamics simulation using DFT-FE. Case study: fcc Al $3\times3\times3$ super-cell.}
\label{fig:nvemd}
\end{figure}

Fig.~\ref{fig:nvemd} shows the variation of the total energy of the system up to $800$ fs. The mean and the standard deviation of the total energy is computed to be $-2.0948$ Ha/atom and $1.482 \times 10^{-5}$ Ha/atom, respectively. We also compute the drift in the total energy by evaluating the slope of linear fit, which is found to be $1.8045 \times 10^{-9}$ Ha/atom-fs.

\subsection{Band-structure calculations}
\begin{figure}[t]
\includegraphics[width=1.0\textwidth, trim={2cm 0cm 0cm 0cm},clip]{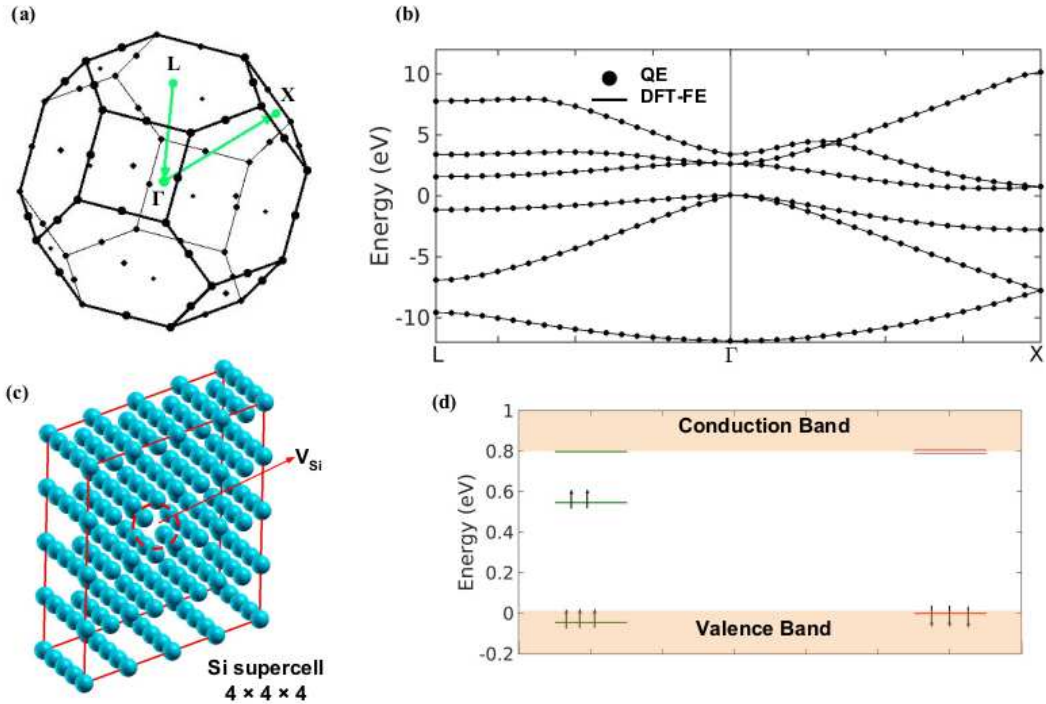}
 \centering
\caption{(a) Brillouin zone (BZ) of diamond unit cell of silicon. (b) Band-structure of diamond unit cell of silicon computed using DFT-FE. (c) $4\times 4\times 4$ supercell of diamond-Si with a silicon vacancy (V\textsubscript{Si}). (d) Spin-polarized midgap states $4\times 4\times 4$ supercell of diamond-Si with a silicon vacancy (V\textsubscript{Si}) computed using DFT-FE.}
\label{fig:sibands}
\end{figure}

We demonstrate the bandstructure calculation capability in DFT-FE with a diamond unit cell of silicon (lattice constant 10.26 Bohr). The Brillouin zone (BZ) is shown in Fig.~\ref{fig:sibands} (a) along with two high-symmetry lines. ONCV pseudopotential is used along with PBE~\cite{pbe} functionals to carry out the DFT calculation. The BZ is sampled by a $8\times 8\times 8$ Monkhorst-Pack grid. The KS eigenvalues along the two high-symmetry lines are extracted from the converged Hamiltonian and are shown in Fig.~\ref{fig:sibands} (b). For the sake of comparison, a similar calculation is done using QE, and the corresponding bandstructure is overlaid in Fig.~\ref{fig:sibands} (b). 

Defects in crystalline solids often give rise to midgap states that act as recombination centers and have drastic consequences on many applications in electronics and photonics. DFT-FE can be used to explore novel defects and their energy locations within the optical gap. The excellent scaling capability of DFT-FE with increasing system size helps to probe the physics of a system with very low defect concentrations, for example unintentional doping in semiconductors. We demonstrate a simple example of a $4\times 4\times 4$ supercell of diamond-Si with a silicon vacancy (V\textsubscript{Si}) as illustrated in Fig.~\ref{fig:sibands} (c). This gives rise to spin-polarized midgap states shown on Fig.~\ref{fig:sibands} (d). While the majority spin channel has a relatively deep state the minority spin has a very shallow state almost at the conduction band minima. The occupied majority spin states in the midgap lie at energy of $0.54700$ eV ($0.54714$ eV) and the corresponding empty minority spin states lie at $0.79010$ eV ($0.79016$ eV), where the top of the valence band is taken as the energy reference in both cases. The numbers within the parenthesis represent data obtained from a similar calculation in QE.   

\section{Conclusions}
\label{sec:conclusions}
In this work, we have developed DFT-FE (Density Functional Theory with Finite-Elements), an accurate, computationally  efficient and scalable finite-element (FE) based code for large-scale first-principles based materials modeling using Kohn-Sham DFT. The DFT-FE code can conduct both pseudopotential and all-electron calculations on non-periodic, semi-periodic and periodic systems, which is a unique feature that has been possible due to the real-space formulation employed in this work in conjunction with the versatility of the FE basis. Besides the systematic convergence afforded by the FE basis, the spatial adaptivity of the FE basis (realized through `p4est' library in deal.II package) and the higher-order spectral finite-elements employed in DFT-FE play an important role in the computational efficiency of the code. DFT-FE offers spectral FE basis up to ${12}^{\text{th}}$ order, and the spatial adaptivity can be realized either via user defined mesh parameters, or by employing the automatic adaptive mesh refinement algorithm implemented in DFT-FE that is based on estimates of local error. 

The computational efficiency and scalability of the DFT-FE code can largely be attributed to the locality of the FE basis, the algorithms employed in the solution of the discrete Kohn-Sham problem, and a careful numerical implementation of the algorithms---minimizing floating point operations, communication costs and latency---some of which leverage the attributes of the FE basis. In particular, the solution to the Kohn-Sham problem is efficiently computed by: (i) employing Chebyshev filtered subspace iteration technique to compute the eigensubspace of interest; (ii) using Cholesky factorization based Gram-Schmidt orthonormalization (CholGS) to compute an orthonormal basis spanning the subspace; (iii) employing Rayleigh-Ritz (RR) procedure to diagonalize the Hamiltonian in the projected subspace and compute the electron density to continue the self-consistent field (SCF) iteration. We developed and implemented mixed precision arithmetic based algorithms for the various steps in the solution procedure, and also employed spectrum splitting technique, that significantly reduced the computational prefactors for CholGS and RR procedure by factors of around $2$ and $3$, respectively, for large-scale systems. Since the computational complexity of CholGS and RR procedure scale cubically with system size, these efficiency gains delay the onset of cubic computational complexity in DFT-FE to very large system sizes. Our numerical studies on benchmark problems demonstrate that DFT-FE exhibits close to quadratic-scaling in system size even for those as large as $40,000$ electrons. The scalability of DFT-FE has been tested on up to $192,000$ MPI tasks, with a parallel efficiency of $42\%$ realized on a $\sim60,000$ electron system. The locality of the FE basis is an important factor in the excellent parallel scalability of the DFT-FE code. However, a careful implementation of the various aspects of the algorithms---such as the use of elemental level matrix-matrix products and blocked approach in Chebyshev filtering, the use mixed precision arithmetic based algorithms in the CholGS and RR procedure---have been instrumental in reducing communication costs and latency, thus resulting in better scalability and reduced wall-times.

In order to assess the accuracy of DFT-FE with respect to state-of-the-art DFT codes, we have conducted a comprehensive comparison study. For validation of pseudopotential calculations we compared with QUANTUM ESPRESSO (QE), a widely used pseudopotential plane-wave code. The benchmark systems considered included periodic metallic systems and non-periodic nano-particles of various sizes. Importantly, the agreement between DFT-FE and QE on ground-state energies, ionic forces and stresses was excellent, with the differences being significantly lower than the discretization errors of the codes. In order to validate all-electron calculations, we compared with \verb|exciting| for periodic calculations and NWChem for non-periodic calculations. The ground-state energies, forces and stresses obtained from DFT-FE on benchmark all-electron systems were again in excellent agreement with these codes.

In order to compare the computational efficiency and scalability afforded by DFT-FE with respect to widely used DFT codes, we conducted an extensive comparison study with QE using ONCV pseudopotentials. We used the CPU-time per SCF iteration on a wide range of benchmark problems as a metric to assess the computational efficiency, and the minimum wall-time (with at least $40-45\%$ parallel efficiency) as a metric to also assess the effectiveness of the scalability. These benchmark studies suggest that DFT-FE is more efficient than QE for periodic system sizes beyond 3,000 electrons. Furthermore, for larger systems ($10,659-20,470$ electrons), we find DFT-FE to be substantially more efficient than QE by $4.5-12\times$ in terms of CPU-times. We used the same benchmark systems to also compare the minimum  wall-time per SCF iteration. Importantly, DFT-FE was significantly faster than QE for all the systems considered, with $1.5-7.5\times$ speedups for smaller system sizes ($2,550-6,034$ electrons) to $7-16\times$ speedups for larger system sizes ($14,322-20,470$ electrons). We also considered three very large periodic metallic systems with $27,986$, $39,990$ and $61,502$ electrons, where we achieve very modest average minimum wall-times per SCF iteration of $1.9$, $3.4$ and $4.6$ minutes, respectively.

DFT-FE code development has adopted best software development practices like unit and regression tests, code modularity, code documentation (using Doxygen and comments), and strict procedures for code review and merging of new changes. Such steps are very critical for sustainable and inclusive software development of DFT-FE and maintaining reproducibilty of the accuracy and performance benchmarks. We have compiled and tested DFT-FE on the following supercomputers: Flux at University of Michigan, XSEDE Comet, TACC Stampede2, NERSC Cori (both Intel KNL and Haswell nodes), ALCF Theta, OLCF Summit (IBM Power 9 CPUs), and BSC MareNostrum (Intel Xeon Platinum).

Overall, DFT-FE provides a practical capability to perform accurate massively parallel large-scale pseudopotential DFT calculations (reaching 100,000 electrons) in generic material systems with arbitrary boundary conditions and complex geometries. An attractive feature of DFT-FE is also the ability to perform all-electron calculations in the same framework, which can aid transferability studies on pseudopotentials. Further, the framework, in principle, can support mixed pseudopotential and all-electron calculations, where some atoms are treated at the all-electron level and others are treated at the pseudopotential level. This can be useful in a wide range of applications from using all-electron calculations for certain atoms with unreliable pseudopotentials to the computation of spin Hamiltonian parameters that require an all-electron treatment around the defect states (e.g. NV center in diamond)~\cite{Ghosh2019}. Implementation of enriched finite-element basis~\cite{kanungo2017} in DFT-FE, which can enable large-scale efficient all-electron DFT calculations, is currently being pursued. Further, implementation of advanced exchange-correlation functionals (hybrid and dispersion corrected), advanced mixing schemes, spin-orbit coupling, and implementation of polarizability and dielectric calculations in DFT-FE are other efforts that are being pursued.

\section*{Acknowledgements}
We thank Bikash Kanungo, Nelson Rufus and Shukan Parekh for independently testing the DFT-FE code, providing useful feedback at many stages of the code development, and providing the reference data for validating the all-electron calculations reported in this manuscript.  We gratefully acknowledge the support from the Department of Energy, Office of Basic Energy Sciences, under Award numbers DE-SC0008637 and DE-SC0017380 for supporting the pseudopotential and all-electron development, respectively. We are also grateful for the support from Toyota Research Institute that funded the development efforts in the later stages of this effort. This work used the Extreme Science and Engineering Discovery Environment (XSEDE), which is supported by National Science Foundation Grant number ACI-1053575. This research used resources of the National Energy Research Scientific Computing Center, a DOE Office of Science User Facility supported by the Office of Science of the U.S. Department of Energy under Contract No. DE-AC02-05CH11231. This research also used resources of Argonne Leadership Computing Facility (ALCF)  resources, a DOE Office of Science User Facility. V.G. also gratefully acknowledges the support of the Army Research Office through the DURIP grant W911NF1810242, which also provided computational resources for this work. D.D. acknowledges the financial support of the German Research Foundation (DFG) grant DA 1664/2-1, that made this collaboration possible. 




\clearpage
\nocite{*}
\bibliographystyle{elsarticle-num}
\bibliography{references}

\begin{thebibliography}{100}
\expandafter\ifx\csname url\endcsname\relax
  \def\url#1{\texttt{#1}}\fi
\expandafter\ifx\csname urlprefix\endcsname\relax\def\urlprefix{URL }\fi
\expandafter\ifx\csname href\endcsname\relax
  \def\href#1#2{#2} \def\path#1{#1}\fi

\bibitem{kohn65}
W.~Kohn, L.~J. Sham, Self-consistent equations including exchange and
  correlation effects, Phys. Rev. 140~(4A) (1965) A1133.

\bibitem{kohn96}
W.~Kohn, Density functional and density matrix method scaling linearly with the
  number of atoms, Phys. Rev. Lett. 76 (1996) 3168--3171.

\bibitem{XCReview2005}
G.~E. Scuseria, V.~N. Staroverov, Progress in the development of
  exchange-correlation functionals (2005).

\bibitem{rodney2017}
D.~Rodney, L.~Ventelon, E.~Clouet, L.~Pizzagalli, F.~Willaime, Ab initio
  modeling of dislocation core properties in metals and semiconductors, Acta
  Mater. 124 (2017) 633--659.

\bibitem{Arias2000}
S.~Ismail-Beigi, T.~A. Arias, Ab initio study of screw dislocations in mo and
  ta: A new picture of plasticity in bcc transition metals, Phys. Rev. Lett. 84
  (2000) 1499--1502.

\bibitem{Trinkle2005}
D.~R. Trinkle, C.~Woodward, The chemistry of deformation: How solutes soften
  pure metals, Science 310~(5754) (2005) 1665--1667.

\bibitem{Trinkle2008}
C.~Woodward, D.~R. Trinkle, L.~G. Hector, D.~L. Olmsted, Prediction of
  dislocation cores in aluminum from density functional theory, Phys. Rev.
  Lett. 100 (2008) 045507.

\bibitem{Clouet2009}
E.~Clouet, L.~Ventelon, F.~Willaime, Dislocation core energies and core fields
  from first principles, Phys. Rev. Lett. 102 (2009) 055502.

\bibitem{shin2011}
I.~Shin, E.~A. Carter, Orbital-free density functional theory simulations of
  dislocations in magnesium, Modelling and Simulation in Materials Science and
  Engineering 20~(1) (2011) 015006.

\bibitem{shin2013}
I.~Shin, E.~A. Carter, Possible origin of the discrepancy in peierls stresses
  of fcc metals: First-principles simulations of dislocation mobility in
  aluminum, Phys. Rev. B 88 (2013) 064106.

\bibitem{iyer2015}
M.~Iyer, B.~Radhakrishnan, V.~Gavini, Electronic-structure study of an edge
  dislocation in {Aluminum} and the role of macroscopic deformations on its
  energetics, J. Mech. Phys. Solids 76 (2015) 260--275.

\bibitem{radhakrishnan2016}
B.~Radhakrishnan, V.~Gavini, Orbital-free density functional theory study of
  the energetics of vacancy clustering and prismatic dislocation loop
  nucleation in {Aluminium}, Philos. Mag. 96 (2016) 2468--2487.

\bibitem{das2017}
S.~Das, V.~Gavini, Electronic structure study of screw dislocation core
  energetics in {Aluminum} and core energetics informed forces in a dislocation
  aggregate, J. Mech. Phys. Solids 104 (2017) 115--143.

\bibitem{dawson2018}
J.~A. Dawson, H.~Chen, M.~S. Islam, Composition screening of {Lithium}- and
  {Sodium}-rich anti-perovskites for fast-conducting solid electrolytes, J.
  Phys. Chem. C 122~(42) (2018) 23978--23984.

\bibitem{dive2018}
A.~Dive, C.~Benmore, M.~Wilding, S.~W. Martin, S.~Beckman, S.~Banerjee,
  Molecular dynamics modeling of the structure and {${\textrm{Na}}^{+}$}-ion
  transport in {${\textrm{Na}}_2{\textrm S}$ + ${\textrm{Si}}\textrm{S}_2$}
  glassy electrolytes, J. Phys. Chem. B 122~(30) (2018) 7597--7608.

\bibitem{cole2016}
D.~J. Cole, N.~D.~M. Hine, Applications of large-scale density functional
  theory in biology, J. Phys.: Condens. Matter 28~(39) (2016) 393001.

\bibitem{qe2009}
P.~Giannozzi, S.~Baroni, N.~Bonini, M.~Calandra, R.~Car, C.~Cavazzoni,
  D.~Ceresoli, G.~L. Chiarotti, M.~Cococcioni, I.~Dabo, A.~{Dal Corso},
  S.~de~Gironcoli, S.~Fabris, G.~Fratesi, R.~Gebauer, U.~Gerstmann,
  C.~Gougoussis, A.~Kokalj, M.~Lazzeri, L.~Martin-Samos, N.~Marzari, F.~Mauri,
  R.~Mazzarello, S.~Paolini, A.~Pasquarello, L.~Paulatto, C.~Sbraccia,
  S.~Scandolo, G.~Sclauzero, A.~P. Seitsonen, A.~Smogunov, P.~Umari, R.~M.
  Wentzcovitch, {QUANTUM ESPRESSO}: a modular and open-source software project
  for quantum simulations of materials, J. Phys. Condens. Matter 21~(39) (2009)
  395502.

\bibitem{gonze2002}
X.~Gonze, J.-M. Beuken, R.~Caracas, F.~Detraux, M.~Fuchs, G.-M. Rignanese,
  L.~Sindic, M.~Verstraete, G.~Zerah, F.~Jollet, M.~Torrent, A.~Roy, M.~Mikami,
  P.~Ghosez, J.-Y. Raty, D.~Allan, First-principles computation of material
  properties: the \{ABINIT\} software project, Comput. Mater. Sci. 25~(3)
  (2002) 478--492.

\bibitem{VASP}
G.~Kresse, J.~Furthm{\"u}ller, Efficient iterative schemes for ab initio
  total-energy calculations using a plane-wave basis set, Phys. Rev. B 54~(16)
  (1996) 11169--11186.

\bibitem{exciting2014}
A.~Gulans, S.~Kontur, C.~Meisenbichler, D.~Nabok, P.~Pavone, S.~Rigamonti,
  S.~Sagmeister, U.~Werner, C.~Draxl, exciting: a full-potential all-electron
  package implementing density-functional theory and many-body perturbation
  theory, J. Phys.: Condens. Matter 26~(36) (2014) 363202.

\bibitem{Pople}
W.~J. Hehre, R.~F. Stewart, J.~A. Pople, Self-consistent molecular-orbital
  methods. i. {U}se of {G}aussian expansions of {S}later-type atomic orbitals,
  J. Chem. Phys. 51~(6) (1969) 2657--2664.

\bibitem{jensen2002}
F.~Jensen, {P}olarization consistent basis sets. ii. estimating the
  {K}ohn--{S}ham basis set limit, J. Chem. Phys. 116~(17) (2002) 7372--7379.

\bibitem{cp2k2014}
J.~Hutter, M.~Iannuzzi, F.~Schiffmann, J.~VandeVondele, cp2k: atomistic
  simulations of condensed matter systems, Wiley Interdiscip. Rev.: Comput.
  Mol. Sci. 4.

\bibitem{blum2009}
V.~Blum, R.~Gehrke, F.~Hanke, P.~Havu, V.~Havu, X.~Ren, K.~Reuter,
  M.~Scheffler, Ab initio molecular simulations with numeric atom-centered
  orbitals, Comput. Phys. Commun. 180~(11) (2009) 2175--2196.

\bibitem{nwchem}
M.~Valiev, E.~Bylaska, N.~Govind, K.~Kowalski, T.~Straatsma, H.~V. Dam,
  D.~Wang, J.~Nieplocha, E.~Apra, T.~Windus, W.~de~Jong, Nwchem: A
  comprehensive and scalable open-source solution for large scale molecular
  simulations, Comput. Phys. Commun. 181~(9) (2010) 1477--1489.

\bibitem{tsuchida1995}
E.~Tsuchida, M.~Tsukada, Electronic-structure calculations based on the
  finite-element method, Phys. Rev. B 52 (1995) 5573--5578.

\bibitem{tsuchida1996}
E.~Tsuchida, M.~Tsukada, Adaptive finite-element method for
  electronic-structure calculations, Phys. Rev. B 54 (1996) 7602--7605.

\bibitem{tsuchida1998}
E.~Tsuchida, M.~Tsukada, Large-scale electronic-structure calculations based on
  the adaptive finite-element method, Journal of the Physical Society of Japan
  67~(11) (1998) 3844--3858.

\bibitem{pask1999}
J.~E. Pask, B.~M. Klein, C.~Y. Fong, P.~A. Sterne, Real-space local polynomial
  basis for solid-state electronic-structure calculations: A finite-element
  approach, Phys. Rev. B 59 (1999) 12352--12358.

\bibitem{pask2005}
J.~E. Pask, P.~A. Sterne, Finite element methods in ab initio electronic
  structure calculations, Modell. Simul. Mater. Sci. Eng. 13~(3) (2005) R71.

\bibitem{sukumar2009}
N.~Sukumar, J.~E. Pask, Classical and enriched finite element formulations for
  bloch-periodic boundary conditions, Int. J. Numer. Methods Eng. 77~(8) (2009)
  1121--1138.

\bibitem{suryanarayana2010}
P.~Suryanarayana, V.~Gavini, T.~Blesgen, K.~Bhattacharya, M.~Ortiz,
  Non-periodic finite-element formulation of {K}ohn--{S}ham density functional
  theory, J. Mech. Phys. Solids 58 (2010) 256--280.

\bibitem{zhou2011}
H.~Chen, L.~He, A.~Zhou, Finite element approximations of nonlinear eigenvalue
  problems in quantum physics, Comput. Methods in Appl. Mech. Eng. 200~(21)
  (2011) 1846--1865.

\bibitem{motamarri2013}
P.~Motamarri, M.~Nowak, K.~Leiter, J.~Knap, V.~Gavini, Higher-order adaptive
  finite-element methods for {K}ohn-{S}ham density functional theory, J.
  Comput. Phys. 253 (2013) 308--343.

\bibitem{SCHAUER2013644}
V.~Schauer, C.~Linder, All-electron {K}ohn--{S}ham density functional theory on
  hierarchic finite element spaces, J. Comput. Phys. 250 (2013) 644--664.

\bibitem{zhou2014}
H.~Chen, X.~Dai, X.~Gong, L.~He, A.~Zhou, Adaptive finite element
  approximations for {K}ohn--{S}ham models, Multiscale Model. Simul. 12~(4)
  (2014) 1828--1869.

\bibitem{denis2016}
D.~Davydov, T.~D. Young, P.~Steinmann, On the adaptive finite element analysis
  of the {Kohn--Sham} equations: methods, algorithms, and implementation, Int.
  J. Numer. Methods Eng. 106~(11) (2016) 863--888.

\bibitem{kanungo2017}
B.~Kanungo, V.~Gavini, Large-scale all-electron {density functional theory}
  calculations using an enriched finite-element basis, Phys. Rev. B 95 (2017)
  035112.

\bibitem{Davydov2018}
D.~Davydov, T.~Heister, M.~Kronbichler, P.~Steinmann, Matrix-free locally
  adaptive finite element solution of density-functional theory with
  nonorthogonal orbitals and multigrid preconditioning, Physica Status Solidi
  B: Basic Solid State Physics 255~(9).
\newblock \href {http://dx.doi.org/10.1002/pssb.201800069}
  {\path{doi:10.1002/pssb.201800069}}.

\bibitem{parsec2006}
L.~Kronik, A.~Makmal, M.~L. Tiago, M.~M.~G. Alemany, M.~Jain, X.~Huang,
  Y.~Saad, J.~R. Chelikowsky, {PARSEC} --- the pseudopotential algorithm for
  real-space electronic structure calculations: recent advances and novel
  applications to nano-structures, Phys. Status Solidi B 243~(5) (2006)
  1063--1079.

\bibitem{rescu2016}
V.~Michaud-Rioux, L.~Zhang, H.~Guo, {RESCU}: A real space electronic structure
  method, J. Comput. Phys. 307 (2016) 593--613.

\bibitem{sparc2017a}
S.~Ghosh, P.~Suryanarayana, {SPARC}: Accurate and efficient finite-difference
  formulation and parallel implementation of density functional theory:
  Isolated clusters, Comput. Phys. Commun. 212 (2017) 189--204.

\bibitem{sparc2017b}
S.~Ghosh, P.~Suryanarayana, {SPARC}: Accurate and efficient finite-difference
  formulation and parallel implementation of density functional theory:
  Extended systems, Comput. Phys. Commun. 216 (2017) 109--125.

\bibitem{octopus2015}
X.~Andrade, D.~Strubbe, U.~De~Giovannini, A.~H. Larsen, M.~J.~T. Oliveira,
  J.~Alberdi-Rodriguez, A.~Varas, I.~Theophilou, N.~Helbig, M.~J. Verstraete,
  L.~Stella, F.~Nogueira, A.~Aspuru-Guzik, A.~Castro, M.~A.~L. Marques,
  A.~Rubio, Real-space grids and the {O}ctopus code as tools for the
  development of new simulation approaches for electronic systems, Phys. Chem.
  Chem. Phys. 17 (2015) 31371--31396.

\bibitem{gpaw}
J.~Enkovaara, C.~Rostgaard, J.~J. Mortensen, J.~Chen, M.~Du{\l}ak, L.~Ferrighi,
  J.~Gavnholt, C.~Glinsvad, V.~Haikola, H.~A. Hansen, H.~H. Kristoffersen,
  M.~Kuisma, A.~H. Larsen, L.~Lehtovaara, M.~Ljungberg, O.~Lopez-Acevedo, P.~G.
  Moses, J.~Ojanen, T.~Olsen, V.~Petzold, N.~A. Romero, J.~Stausholm-M{\o}ller,
  M.~Strange, G.~A. Tritsaris, M.~Vanin, M.~Walter, B.~Hammer, H.~HÃ¤kkinen,
  G.~K.~H. Madsen, R.~M. Nieminen, J.~K. N{\o}rskov, M.~Puska, T.~T. Rantala,
  J.~Schi{\o}tz, K.~S. Thygesen, K.~W. Jacobsen, Electronic structure
  calculations with {GPAW}: a real-space implementation of the projector
  augmented-wave method, Journal of Physics: Condensed Matter 22~(25) (2010)
  253202.

\bibitem{luigi2008}
L.~Genovese, A.~Neelov, S.~Goedecker, T.~Deutsch, S.~A. Ghasemi, A.~Willand,
  D.~Caliste, O.~Zilberberg, M.~Rayson, A.~Bergman, R.~Schneider, Daubechies
  wavelets as a basis set for density functional pseudopotential calculations,
  J. Chem. Phys. 129~(1) (2008) 014109.

\bibitem{skylaris2005}
C.-K. Skylaris, P.~D. Haynes, A.~A. Mostofi, M.~C. Payne, Introducing {ONETEP}:
  Linear-scaling density functional simulations on parallel computers, J. Chem.
  Phys. 122~(8) (2005) 084119.

\bibitem{dgdft2015}
W.~Hu, L.~Lin, C.~Yang, {DGDFT}: A massively parallel method for large scale
  {density functional theory} calculations, J. Chem. Phys. 143~(12) (2015)
  124110.

\bibitem{motamarri2015}
P.~Motamarri, V.~Gavini, Tucker-tensor algorithm for large-scale {K}ohn-{S}ham
  density functional theory calculations, Phys. Rev. B 93 (2017) 035111.

\bibitem{godecker99}
S.~Goedecker, Linear scaling electronic structure methods, Rev. Mod. Phys. 71
  (1999) 1085--1123.

\bibitem{bowler2012}
D.~R. Bowler, T.~Miyazaki, $\mathcal{O(N)}$ methods in electronic structure
  calculations, Rep. Prog. Phys. 75~(3) (2012) 036503.

\bibitem{fattebert2006}
J.-L. Fattebert, F.~Gygi, Linear-scaling first-principles molecular dynamics
  with plane-waves accuracy, Phys. Rev. B 73 (2006) 115124.

\bibitem{ls3df2008}
L.-W. Wang, Z.~Zhao, J.~Meza, Linear-scaling three-dimensional fragment method
  for large-scale electronic structure calculations, Phys. Rev. B 77 (2008)
  165113.

\bibitem{motamarri2014}
P.~Motamarri, V.~Gavini, Subquadratic-scaling subspace projection method for
  large-scale {K}ohn-{S}ham density functional theory calculations using
  spectral finite-element discretization, Phys. Rev. B 90 (2014) 115127.

\bibitem{skylaris2018}
J.~Aarons, C.-K. Skylaris, Electronic annealing {F}ermi operator expansion for
  {DFT} calculations on metallic systems, J. Chem. Phys. 148~(7) (2018) 074107.

\bibitem{mohr2018}
S.~Mohr, M.~Eixarch, M.~Amsler, M.~J. Mantsinen, L.~Genovese, Linear scaling
  {DFT} calculations for large tungsten systems using an optimized local basis,
  Nucl. Mater. Energy 15 (2018) 64 -- 70.

\bibitem{saad2006}
Y.~Zhou, Y.~Saad, M.~L. Tiago, J.~R. Chelikowsky, Parallel
  self-consistent-field calculations via {C}hebyshev-filtered subspace
  acceleration, Phys. Rev. E 74~(6) (2006) 066704.

\bibitem{saad2012}
G.~Schofield, J.~R. Chelikowsky, Y.~Saad, A spectrum slicing method for the
  {K}ohn--{S}ham problem, Comp. Phys. Comm 183~(3) (2012) 497--505.

\bibitem{zhou2006}
Y.~Zhou, Y.~Saad, M.~L. Tiago, J.~R. Chelikowsky, Self-consistent-field
  calculations using {C}hebyshev-filtered subspace iteration, J. Comput. Phys.
  219~(1) (2006) 172--184.

\bibitem{motamarri2018}
P.~Motamarri, V.~Gavini, Configurational forces in electronic structure
  calculations using {K}ohn-{S}ham density functional theory, Phys. Rev. B. 97
  (2018) 165132.

\bibitem{qe2017}
P.~Giannozzi, O.~Andreussi, T.~Brumme, O.~Bunau, M.~B. Nardelli, M.~Calandra,
  R.~Car, C.~Cavazzoni, D.~Ceresoli, M.~Cococcioni, N.~Colonna, I.~Carnimeo,
  A.~D. Corso, S.~de~Gironcoli, P.~Delugas, R.~A.~D. Jr, A.~Ferretti,
  A.~Floris, G.~Fratesi, G.~Fugallo, R.~Gebauer, U.~Gerstmann, F.~Giustino,
  T.~Gorni, J.~Jia, M.~Kawamura, H.-Y. Ko, A.~Kokalj, E.~KÃ¼Ã§Ã¼kbenli,
  M.~Lazzeri, M.~Marsili, N.~Marzari, F.~Mauri, N.~L. Nguyen, H.-V. Nguyen,
  A.~O. de-la Roza, L.~Paulatto, S.~PoncÃ©, D.~Rocca, R.~Sabatini, B.~Santra,
  M.~Schlipf, A.~P. Seitsonen, A.~Smogunov, I.~Timrov, T.~Thonhauser, P.~Umari,
  N.~Vast, X.~Wu, S.~Baroni, Advanced capabilities for materials modelling with
  {QUANTUM ESPRESSO}, J. Phys. Condens. Matter 29~(46) (2017) 465901.

\bibitem{rmartin}
R.~M. Martin, Electronic structure: basic theory and practical methods,
  Cambridge university press, Cambridge, UK, 2004.

\bibitem{gga1}
D.~C. Langreth, M.~J. Mehl, Beyond the local-density approximation in
  calculations of ground-state electronic properties, Phys. Rev. B 28 (1983)
  1809--1834.

\bibitem{becke}
A.~D. Becke, Density-functional exchange-energy approximation with correct
  asymptotic behavior, Phys. Rev. A. 38 (1988) 3098--3100.

\bibitem{pw}
J.~P. Perdew, Y.~Wang, Accurate and simple analytic representation of the
  electron-gas correlation energy, Phys. Rev. B. 45 (1992) 13244--13249.

\bibitem{pbe}
J.~P. Perdew, K.~Burke, M.~Ernzerhof, {Generalized Gradient Approximation} made
  simple, Phys. Rev. Lett. 77 (1996) 3865--3868.

\bibitem{kb82}
L.~Kleinman, D.~M. Bylander, Efficacious form for model pseudopotentials, Phy.
  Rev. Lett. 48~(20) (1982) 1425.

\bibitem{tm91}
N.~Troullier, J.~L. Martins, Efficient pseudopotentials for plane-wave
  calculations, Phys. Rev. B 43 (1991) 1993--2006.

\bibitem{oncv2013}
D.~R. Hamann, Optimized norm-conserving {V}anderbilt pseudopotentials, Phys.
  Rev. B 88 (1995) 239906.

\bibitem{science2016}
K.~Lejaeghere, et~al., Reproducibility in density functional theory
  calculations of solids, Sci. 351.

\bibitem{gavini2007}
V.~Gavini, J.~Knap, K.~Bhattacharya, M.~Ortiz, Non-periodic finite-element
  formulation of orbital free density functional theory, J. Mech. Phys. Solids
  55 (2007) 669--696.

\bibitem{das2015}
S.~Das, M.~Iyer, V.~Gavini, Real-space formulation of orbital-free density
  functional theory using finite-element discretization: The case for {Al},
  {Mg}, and {Al-Mg} intermetallics, Phys. Rev. B 92 (2015) 014104.

\bibitem{mermin}
N.~W. {Ashcroft}, N.~D. {Mermin}, {Solid State Physics}, Hartcourt College
  Publishers, San Diego, US, 1976.

\bibitem{mpgrid}
H.~J. Monkhorst, J.~D. Pack, Special points for brillouin-zone integrations,
  Phys. Rev. B 13 (1976) 5188--5192.

\bibitem{brenner2002}
S.~C. Brenner, L.~R. Scott, The Mathematical Theory of Finite-element Methods,
  Springer, New York, 2002.

\bibitem{dealII90}
G.~Alzetta, D.~Arndt, W.~Bangerth, V.~Boddu, B.~Brands, D.~Davydov,
  R.~Gassm{\"o}ller, T.~Heister, L.~Heltai, K.~Kormann, M.~Kronbichler,
  M.~Maier, J.-P. Pelteret, B.~Turcksin, D.~Wells, The \texttt{deal.II}
  {L}ibrary, {V}ersion 9.0, Journal of Numerical Mathematics 26~(4) (2018)
  173--183.

\bibitem{bylaska}
E.~J. Bylaska, M.~Holst, J.~H. Weare, Adaptive finite element method for
  solving the exact {K}ohn-{S}ham equation of density functional theory, J.
  Chem. Theory. Comput. 5~(4) (2009) 937--948.

\bibitem{lehtovaara}
L.~Lehtovaara, V.~Havu, M.~Puska, All-electron density functional theory and
  time-dependent density functional theory with high-order finite elements, J.
  Chem. Physics 131~(5) (2009) 054103.

\bibitem{qcofdft}
V.~Gavini, K.~Bhattacharya, M.~Ortiz, Quasi-continuum orbital-free
  density-functional theory: A route to multi-million atom non-periodic {DFT}
  calculation, J. Mech. Phys. Solids 55~(4) (2007) 697--718.

\bibitem{choly2005}
N.~Choly, G.~Lu, W.~E, E.~Kaxiras, Multiscale simulations in simple metals: A
  density-functional-based methodology, Phys. Rev. B 71 (2005) 094101.

\bibitem{gang2006}
G.~Lu, E.~B. Tadmor, E.~Kaxiras, From electrons to finite elements: A
  concurrent multiscale approach for metals, Phys. Rev. B 73 (2006) 024108.

\bibitem{p4est2011}
C.~Burstedde, L.~C. Wilcox, O.~Ghattas, {\texttt{p4est}}: Scalable algorithms
  for parallel adaptive mesh refinement on forests of octrees, SIAM J. Sci.
  Comput 33~(3) (2011) 1103--1133.

\bibitem{zhou2008}
X.~Dai, J.~Xu, A.~Zhou, Convergence and optimal complexity of adaptive finite
  element eigenvalue computations, Numerische Mathematik 110~(3) (2008)
  313--355.

\bibitem{chen2011}
H.~Chen, L.~He, A.~Zhou, Finite element approximations of nonlinear eigenvalue
  problems in quantum physics, Computer Methods in Applied Mechanics and
  Engineering 200~(21) (2011) 1846 -- 1865.

\bibitem{bao2012}
G.~Bao, G.~Hu, D.~Liu, An h-adaptive finite element solver for the calculations
  of the electronic structures, Journal of Computational Physics 231~(14)
  (2012) 4967 -- 4979.

\bibitem{chen2014}
H.~Chen, X.~Dai, X.~Gong, L.~He, A.~Zhou, Adaptive finite element
  approximations for kohn-sham models, Multiscale Modeling and Simulation
  12~(4) (2014) 1828--1869.

\bibitem{shen2018}
Y.~Shen, Y.~Kuang, G.~Hu, An asymptotics-based adaptive finite element method
  for kohn--sham equation, Journal of Scientific Computing.

\bibitem{radio}
R.~Radovitzky, M.~Ortiz, Error estimation and adaptive meshing in strongly
  nonlinear dynamic problems, Comput. Methods Appl. Mech. Engrg. 172~(1) (1999)
  203--240.

\bibitem{ciarlet2002}
P.~G. {Ciarlet}, {The Finite Element Method for Elliptic Problems}, SIAM,
  Philadelphia, 2002.

\bibitem{anderson1965}
D.~G. Anderson, Iterative procedures for nonlinear integral equations, J.
  Assoc. Comput. Mach. 12~(4) (1965) 547--560.

\bibitem{broyden1965}
C.~G. Broyden, A class of methods for solving nonlinear simultaneous equations,
  Math. Comput. (1965) 577--593.

\bibitem{kronbichler2012}
M.~Kronbichler, K.~Kormann, A generic interface for parallel cell-based finite
  element operator application, Comput. Fluids 63 (2012) 135--147.

\bibitem{bekas2010}
C.~Bekas, A.~Curioni, Very large scale wavefunction orthogonalization in
  density functional theory electronic structure calculations, Comput. Phys.
  Commun. 181~(6) (2010) 1057--1068.

\bibitem{scalapack1997}
L.~S. Blackford, J.~Choi, A.~Cleary, E.~D'Azevedo, J.~Demmel, I.~Dhillon,
  J.~Dongarra, S.~Hammarling, G.~Henry, A.~Petitet, K.~Stanley, D.~Walker,
  R.~C. Whaley, {ScaLAPACK} Users' Guide, SIAM, Philadelphia, PA, 1997.

\bibitem{tsuchida2012}
E.~Tsuchida, Y.-K. Choe, Iterative diagonalization of symmetric matrices in
  mixed precision and its application to electronic structure calculations,
  Comput. Phys. Commun. 183~(4) (2012) 980--985.

\bibitem{amartya2018}
A.~S. Banerjee, L.~Lin, P.~Suryanarayana, C.~Yang, J.~E. Pask, Two-level
  chebyshev filter based complementary subspace method: Pushing the envelope of
  large-scale electronic structure calculations, J. Chem. Theory Comput. 14~(6)
  (2018) 2930--2946.

\bibitem{motamarri2017}
P.~Motamarri, V.~Gavini, K.~Bhattacharya, M.~Ortiz, Spectrum-splitting approach
  for {F}ermi-operator expansion in all-electron {K}ohn-{S}ham {DFT}
  calculations, Phys. Rev. B 95 (2017) 035111.

\bibitem{elpa2014}
A.~Marek, V.~Blum, R.~Johanni, V.~Havu, B.~Lang, T.~Auckenthaler, A.~Heinecke,
  H.-J. Bungartz, H.~Lederer, The {ELPA} library: scalable parallel eigenvalue
  solutions for electronic structure theory and computational science, J.
  Phys.: Condens. Matter 26~(21) (2014) 213201.

\bibitem{elpaCray}
B.~Cook, T.~Kurth, J.~Deslippe, P.~Carrier, N.~Hill, N.~Wichmann, Eigensolver
  performance comparison on {Cray XC} systems, Concurr. Comp.-Pract. E. 0~(0)
  (2018) e4997.

\bibitem{elpaOpt}
P.~Kus, A.~Marek, S.~S. Koecher, H.~Kowalski, C.~Carbogno, C.~Scheurer,
  K.~Reuter, M.~Scheffler, H.~Lederer, Optimizations of the eigensolvers in the
  {ELPA} library, arXiv:1811.01277.

\bibitem{Kerker1981}
G.~P. Kerker, Efficient iteration scheme for self-consistent pseudopotential
  calculations, Phys. Rev. B 23 (1981) 3082--3084.

\bibitem{eyert1996}
V.~Eyert, A comparative study on methods for convergence acceleration of
  iterative vector sequences, J. Comp. Phys 124~(2) (1996) 271--285.

\bibitem{kudin2002}
K.~N. Kudin, G.~E. Scuseria, E.~Canc$\grave{e}$s, A black-box self-consistent
  field convergence algorithm: One step closer, J. Chem. Phys. 116~(19) (2002)
  8255--8261.

\bibitem{linlin2013b}
L.~Lin, C.~Yang, Elliptic preconditioner for accelerating the self-consistent
  field iteration in {K}ohn--{S}ham density functional theory, SIAM J. Sci.
  Comput 35~(5) (2013) S277--S298.

\bibitem{zhou2018}
Y.~Zhou, H.~Wang, Y.~Liu, X.~Gao, H.~Song, Applicability of {K}erker
  preconditioning scheme to the self-consistent density functional theory
  calculations of inhomogeneous systems, Phys. Rev. E 97 (2018) 033305.

\bibitem{BangerthBursteddeHeisterEtAl11}
W.~Bangerth, C.~Burstedde, T.~Heister, M.~Kronbichler, Algorithms and data
  structures for massively parallel generic adaptive finite element codes, ACM
  Trans. Math. Software 38~(2) (2011) 14:1--14:28.

\bibitem{shewchuk1994}
J.~R. Shewchuk, An introduction to the conjugate gradient method without the
  agonizing pain, Tech. rep., Carnegie Mellon University, Pittsburgh, PA, USA
  (1994).

\bibitem{brent1971}
R.~P. Brent, An algorithm with guaranteed convergence for finding a zero of a
  function, Comput. J. 14~(4) (1971) 422--425.

\bibitem{liu1989}
D.~C. Liu, J.~Nocedal, On the limited memory {BFGS} method for large scale
  optimization, Math. Program. 45~(1-3) (1989) 503--528.

\bibitem{bitzek2006}
E.~Bitzek, P.~Koskinen, F.~G\"ahler, M.~Moseler, P.~Gumbsch, Structural
  relaxation made simple, Phys. Rev. Lett. 97 (2006) 170201.

\bibitem{oncv2015}
M.~Schlipf, F.~Gygi, Optimization algorithm for the generation of {ONCV}
  pseudopotentials, Comput. Phys. Commun. 196 (2015) 36--44.

\bibitem{cuNano}
J.~M. Rahm, P.~Erhart, Beyond magic numbers: Atomic scale equilibrium
  nanoparticle shapes for any size, Nano Lett. 17~(9) (2017) 5775--5781.

\bibitem{campagna2007}
S.~Campagna, F.~Puntoriero, F.~Nastasi, G.~Bergamini, V.~Balzani,
  Photochemistry and photophysics of coordination compounds: Ruthenium, Topics
  in current chemistry (2007) 117--214.

\bibitem{marx_hutter_2009}
D.~Marx, J.~Hutter, Ab Initio Molecular Dynamics: Basic Theory and Advanced
  Methods, Cambridge University Press, 2009.

\bibitem{Ghosh2019}
K.~Ghosh, H.~Ma, V.~Gavini, G.~Galli, All-electron density functional
  calculations for electron and nuclear spin interactions in molecules and
  solids, arXiv:1902.07377.

\bibitem{vb90}
D.~Vanderbilt, Soft self-consistent pseudopotentials in a generalized
  eigenvalue formalism, Phys. Rev. B 41 (1990) 7892--7895.

\end{thebibliography}







\end{document}